\documentclass[twocolumn]{aastex631}
\usepackage{CJK, graphicx, amsmath, placeins, xcolor, multirow, physics, xspace}

\newcommand{\st}[1]{_\mathrm{#1}}

\definecolor{indigo}{HTML}{332288}
\newcommand{\colesyn}[1]{#1}
\definecolor{olive}{HTML}{999933}
\newcommand{\colpsyn}[1]{#1}
\definecolor{wine}{HTML}{882255}
\newcommand{\colpizero}[1]{#1}
\definecolor{purple}{HTML}{AA4499}

\definecolor{cyan}{HTML}{BBCCEE}
\newcommand{\colpgamma}[1]{#1}

\newcommand{\xrt}{\textit{Swift}-XRT\xspace}
\newcommand{\bat}{\textit{Swift}-BAT\xspace}
\newcommand{\gbm}{\textit{Fermi}-GBM\xspace}
\newcommand{\lat}{{\textit{Fermi}-LAT}\xspace}
\newcommand{\magic}{MAGIC\xspace}
\newcommand{\hess}{H.E.S.S.\xspace}
\newcommand{\lhaaso}{LHAASO\xspace}

\newcommand{\timedep}{\textit{time-dependent}\xspace}
\newcommand{\steadystate}{\textit{steady-state}\xspace}

\newcommand{\xrayband}{\textit{X-ray}\xspace}
\newcommand{\vheband}{\textit{VHE}\xspace}

\newcommand{\ssc}{\textit{SSC}\xspace}
\newcommand{\extsyn}{\textit{Extended-syn}\xspace}
\newcommand{\psyn}{\textit{Proton-syn}\xspace}
\newcommand{\ppcasc}{\textit{pp-cascade}\xspace}
\newcommand{\pgcasc}{$p\gamma$\textit{-cascade}\xspace}

\newcommand{\AM}{\textsc{AM$^3$}\xspace}

\newcommand{\Ounity}{\mathcal{O}(1)\xspace}

\newcommand{\pram}{p_{\mathrm{ram}}'}
\newcommand{\tdyn}{t_{\mathrm{dyn}}'}
\newcommand{\tadi}{t_{\mathrm{adi}}'}
\newcommand{\tesc}{t_{\mathrm{esc}}'}
\newcommand{\texp}{t_{\mathrm{exp}}'}
\newcommand{\tacc}{t_{\mathrm{acc}}'}
\newcommand{\tsyn}{t_{\mathrm{syn}}'}

\newcommand{\Ekiniso}{E\st{kin,iso}}
\newcommand{\Euhecriso}{E\st{UHECR,iso}}

\newcommand{\Bce}{B^\mathrm{crit}_{e}}
\newcommand{\Bcp}{B^\mathrm{crit}_{p}}
\newcommand{\Bcpi}{B^\mathrm{crit}_{\pi}}
\newcommand{\Bcmu}{B^\mathrm{crit}_{\mu}}

\newcommand{\Eep}{E_e'}
\newcommand{\Epp}{E_p'}
\newcommand{\Emup}{E_\mu'}

\newcommand{\Egp}{E_\gamma'}
\newcommand{\Eemaxp}{E_e^{\prime\mathrm{max}}}
\newcommand{\Epmaxp}{E_p^{\prime\mathrm{max}}}
\newcommand{\Eminp}{E_{\mathrm{min}}^{\prime}}

\newcommand{\Eg}{E_\gamma}
\newcommand{\Egcb}{E_\gamma^{\mathrm{cb}}}
\newcommand{\Egt}{E_\gamma^{\mathrm{t}}}
\newcommand{\EgKN}{E_\gamma^{\mathrm{KN}}}

\newcommand{\photonindex}{\gamma}
\newcommand{\electronindex}{s}

\newcommand{\injindex}{s\st{inj}}

\newcommand{\nhat}{\hat{n}_E'}

\newcommand{\uesyn}{u_{e\mathrm{syn}}'}
\newcommand{\uessc}{u_{e\mathrm{SSC}}'}

\newcommand{\eV}{\mathrm{eV}}

\newcommand{\refsec}{Section~}
\newcommand{\reffig}{Figure~}
\newcommand{\reftab}{Table~}
\newcommand{\refap}{Appendix~}
\newcommand{\refeq}{Equation~}

\newcommand{\winered}{Bordeaux-red\xspace}

\begin{document}
\begin{CJK*}{UTF8}{gbsn}

\title{Lepto-Hadronic Scenarios for TeV Extensions of Gamma-Ray Burst Afterglow Spectra}

\correspondingauthor{Marc Klinger}
\email{marc.klinger@desy.de}

\author[0000-0002-4697-1465]{Marc Klinger}
\affiliation{Deutsches Elektronen-Synchrotron DESY, Platanenallee 6, 15738 Zeuthen, Germany}

\author[0000-0003-0327-6136]{Chengchao Yuan (袁成超)}
\affiliation{Deutsches Elektronen-Synchrotron DESY, Platanenallee 6, 15738 Zeuthen, Germany}

\author[0000-0001-9473-4758]{Andrew M. Taylor}
\affiliation{Deutsches Elektronen-Synchrotron DESY, Platanenallee 6, 15738 Zeuthen, Germany}



\author[0000-0001-7062-0289]{Walter Winter}
\affiliation{Deutsches Elektronen-Synchrotron DESY, Platanenallee 6, 15738 Zeuthen, Germany}



\begin{abstract}
Recent multi-wavelength observations of gamma-ray burst afterglows observed in the TeV energy range challenge the simplest Synchrotron Self-Compton (SSC) interpretation of this emission and are consistent with a single power-law component spanning over eight orders of magnitude in energy. 
To interpret this generic behaviour in the single-zone approximation without adding further free parameters, we perform an exhaustive parameter space study using the public, time-dependent, multi-messenger transport software \AM. This description accounts for the radiation from non-thermal protons and the lepto-hadronic cascade induced by $pp$- and $p\gamma$-interactions. 
We summarise the main scenarios which we have found (\ssc, \extsyn, \psyn, \ppcasc, and \pgcasc) and discuss their advantages and limitations.
We find that possible high-density environments, as may be typical for surrounding molecular cloud material, offer an alternative explanation for producing flat hard (source) spectra up to and beyond energies of 10~TeV.
\end{abstract}



\section{Introduction} \label{sec:intro}

Recent detections of gamma-ray bursts (GRBs) with photons in the very-high-energy (VHE; 0.1-100~TeV) regime have opened a new opportunity to test the capability of particle acceleration at relativistic shocks \citep{HESS_180720B,MAGIC_190114_data19,HESS_190829,LHAASO_221009_WCDA}. 

The typical observational signature of a GRB consists of an early (few to few hundred seconds) bright, prompt emission phase during which complex temporal behaviour is observed. This is followed by an afterglow emission phase, during which the light curve decays in a smooth (sometimes broken) power-law shape, which can be observed for much longer (up to months) depending on the sensitivity of the observing instrument. While the origin of the prompt emission is still under debate --- see, e.g., \citet{MochkovitchDaigne98,Kobayashi_internalshocks97,GianniosSpruit_prompt_pynting05,Zhang_ICMART11,BeloborodovMeszaros17,PeerRyde18,Ghisellini_ProtonSynPrompt20} --- the afterglow emission is commonly believed to emerge from a relativistic shock set up at the interface between a fast outflow, typically called blast wave, and its surrounding medium \citep[for reviews see, e.g.,][]{Piran_review04,2006RPPh...69.2259M,Zhang_GRBbook}. The simultaneous generation of turbulent magnetic fields and the acceleration of charged electrons that cool via synchrotron radiation results in a plausible explanation for the observations of afterglows from sub-eV up to multi-GeV energies.  

A widely established but constraining assumption in the modelling is to assume that the dissipation of bulk kinetic energy into radiation can be described in a single, homogeneous, and isotropic region \citep[the one-zone approximation; see, e.g.,][]{Zhang_GRBbook}. This has the advantage of only a handful ($\sim 5-10$) of free parameters and allows for a broken power-law description of the spectral energy distributions (SEDs) observed from often sparse eV to GeV data (below these energies, typically a less certain reverse shock component is included). However, this synchrotron component is commonly expected to not reach far above GeV energies (so-called synchrotron burn-off), limited by the interplay of acceleration and radiation by the same magnetic field.

The new clear detection of VHE emission of multiple long GRBs offers a novel level of spectral information up to unprecedentedly high photon energies. This allows one to critically address the question of the production mechanism of the emission in the VHE band, particularly if a new spectral component is required.

A common interpretation of the VHE emission is that these are synchrotron photons which are inverse-Compton up-scattered to the VHE band by the same population of electrons \citep[the one-zone synchrotron self-Compton (SSC) scenario, e.g.,][]{MeszarosReesPapathanassiou_94,ChiangDermer99,SariEsin01}. The observational picture, namely hard (spectral) photon indices extending up to energies of 10~TeV (see \refsec\ref{sec:obs_picture}), raises problems with this interpretation due to the spectral softening by the Klein-Nishina effects in the cross-section (see \refsec\ref{sec:SSC}). 

Besides the simple SSC interpretation, a variety of other mechanisms have been proposed, departing from the single-zone approximation to multi-zone models. This includes for example multiple/decaying magnetic fields \citep[e.g.,][]{Vanthieghem_Micro_20,KhangulyanEtAl_clumpyB_21,KhangulyanEtAl_2zoneSSC_23,GroseljEtAl2022}, external Compton scenarios \citep{2018ApJ...854...60M,ZhangPetropoulou_190114C_ecternalIC20,2021ApJ...908L..36Z,2022ApJ...932...80Y}, a reverse shock component \citep[e.g.,][]{Zhang_reverseShockproton_23}, structured jets \citep[e.g.,][]{GillGranot,SatoEtAl,OConnor_structuredjet_221009_23}, a significant role of the precursor \citep{Vanthieghem_Micro_20} and far upstream regions \citep{Beloborodov_pairloading_02,Derishev_pairbalance16}, as well as cascades along the propagation to Earth \citep[e.g.,][]{Das_intergalCasc_221009_23}. We highlight that many of these ideas require new parameters that could be hard to justify by the current quality of the observational data. Instead, it is necessary to first systematically explore the possibilities of the one-zone model to reproduce the emerging single power-law observations.

Besides the non-thermal electrons, there are multiple reasons to expect ions to be accelerated at the relativistic shock as well. GRBs are a common candidate of ultra-high-energy (UHE; $\gtrsim 10^{18}~\eV$) cosmic-ray accelerators \citep{Waxman_GRB_CR_95,Vietri95,Bell_RelShock18}, their surrounding interstellar or stellar-wind medium is expected to be ion-dominated, and ion acceleration is also seen in short-time-scale microphysics simulations \citep[e.g.,][]{MarcowithEtAl2016}. 
The presence of these particles leads to a range of additional channels to dissipate the kinetic energy to radiation.
We, therefore, explore the potential of lepto-hadronic signatures in the one-zone approximation to explain the VHE emission and limit ourselves for simplicity to the injection of protons (and electrons).

One possible channel, the synchrotron radiation of non-thermal protons, has been discussed in the literature without a coherent conclusion \citep{HESS_180720B,MAGIC_190114C_MWL19,HESS_190829,Isravel_190114_psyn_23,Isravel_221009_psyn_23,LHAASO_221009_KM2}. 

As a different channel, electromagnetic cascades inside the source initiated at energies above the VHE regime via photo-hadronic or proton-proton interactions are other ideas to generate SEDs up to energies $> 10$~TeV \citep[for a $p\gamma$ model see, e.g.,][]{Wang_pgammacasc_23}. From the observational perspective, see \refsec\ref{sec:obs_picture}, this is particularly interesting, as cascades naturally yield power-law spectra with photon indices of $\photonindex\approx 2$ \citep[similar to][]{Berezinsky_cascade_16}. Note that in this paper, we use the notation $EF_E\propto E^{2-\photonindex}$, see \reftab\ref{tab:notation}.

We note that these lepto-hadronic ideas would also lead to neutrino signatures from the afterglow. However, this has been recently found to be observationally not very promising for some cases \citep{Guarini_neutrinos23}.

These conceptual ideas necessitate a systematic exploration of the potential of lepto-hadronic one-zone models to explain the emerging observational picture of hard ($\photonindex\approx 2$) photon SEDs that also resemble a single power-law component spanning from keV to TeV energies. We give a summary of the observational picture in \refsec\ref{sec:obs_picture}. We then discuss our model for the emission in \refsec\ref{sec:methods} and present the representative scenarios we found in \refsec\ref{sec:results}. Finally, we discuss their implications in \refsec\ref{sec:discussion}, and conclude in \refsec\ref{sec:conclusions}.

\begin{table}[]
\centering
\begin{tabular}{|l|l|}
\multicolumn{2}{l}{Observed fluxes and comoving densities}                             \\ \hline
$EF_E \propto E^2 \frac{\dd N}{\dd E}$ & Observed photon energy flux \\ \hline
$\nhat \propto \frac{\dd N'}{\dd E'} $ & Particle spectral density \\ \hline
\multicolumn{2}{l}{Power-law indices}                             \\ \hline
$\photonindex$ with $EF_E \propto E^{2-\photonindex}$ & Photon index \\ \hline
$\electronindex$ with $\nhat \propto E'^{-\electronindex}$ & Spectral index (electrons, protons)\\ \hline
$\injindex$ with $q_E' \propto E'^{-\injindex}$ & Injection spectral index \\  \hline
\multicolumn{2}{l}{}                             \\ \hline
Primed & Frame momentarily comoving \\
 &with blast wave \\ \hline
Un-primed & Observed frame \\ \hline
Star $\star$ & Progenitor rest frame \\ \hline
\end{tabular}
    \caption{Notation used in this paper}
    \label{tab:notation}
\end{table}

\newpage
\section{Observational Picture at the highest energies} \label{sec:obs_picture}
We focus in this work on the window from keV to TeV energies, where the observed energy flux is not severely absorbed, see \refsec\ref{sec:propagation} for details. 

Up to which energies the afterglow radiation can be observed has been a topic of speculation for a long while. A first observational hint was given by the detection of the unprecedentedly bright GRB~130427A up to $\sim$100~GeV energies with a photon index of $\photonindex \sim 2$ ($EF_E\propto E^{2-\photonindex}$), however resulting in no clear detection at VHE \citep{Fermi_130427_14, Kouveliotou_130427_13}.

\begin{figure}
    \centering
    \includegraphics[width=1.1\linewidth]{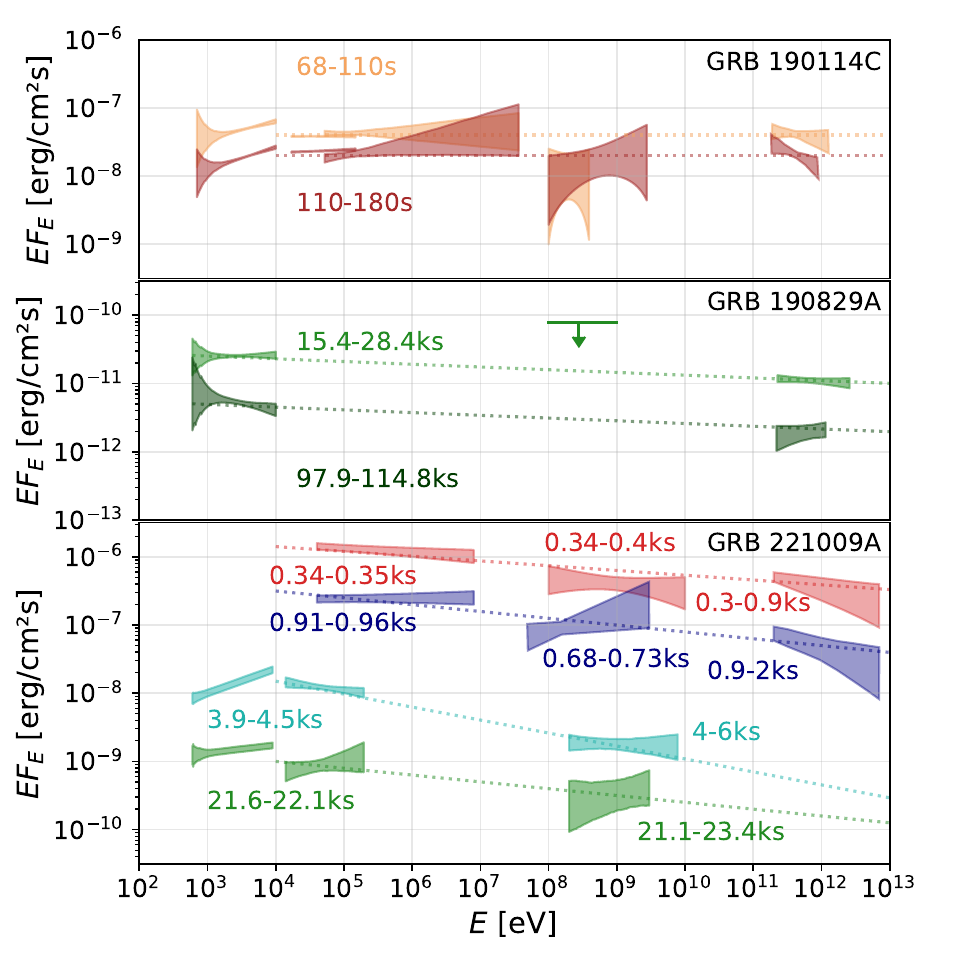}
    \caption{Observational picture of the three GRB afterglows detected at VHE with the highest level of spectral information: GRB~190114C, GRB~190829A, and GRB~221009A. The dotted lines indicate the emerging power-law behaviour discussed in the text. The data is compiled from \cite{KlingerEtAl_GRB190114C,FermiSwift_190114_20,MAGIC_190114C_MWL19,HESS_190829,LHAASO_221009_WCDA,Zhang_GBM_221009_23,Liu_LAT_221009_23,Tavani_AGILE_221009_23,Klinger_221009_24}. All times refer to the corresponding \gbm trigger time. }
    \label{fig:observational_picture_VHEGRBs}
\end{figure}

More recently, the afterglows of multiple GRBs in the VHE band have been detected with high significance. Here, we focus on the three GRBs with the best measurements of the VHE photon index and broad multi-wavelength (MWL) coverage. 
We summarise their contemporaneous keV--TeV SEDs in \reffig\ref{fig:observational_picture_VHEGRBs} and discuss the observational picture for each of them in the following.

\paragraph{GRB~190114C} 
In the VHE regime, this GRB was detected particularly early, around 1~min after the initial trigger, by the Major Atmospheric Gamma Imaging Cherenkov (\magic) telescopes \citep{MAGIC_190114_data19}. However, the challenging observational conditions with bright moonlight and the extragalactic-background-light (EBL) absorption due to the intermediate distance (redshift $z=0.43$) limited an estimation of the de-absorbed photon index at TeV energies to $\photonindex = 1.9\pm0.4\mathrm{(stat)}^{+0.2}_{-0.1}\mathrm{(sys)}$ for 62--100~s after the trigger, and $\photonindex = 2.2^{+0.5}_{-0.4}\mathrm{(stat)}\pm0.3\mathrm{(sys)}$ for 100--140~s after trigger \citep[see Extended Data Table~1 of][]{MAGIC_190114_data19}. Besides the VHE data, the contemporaneous MWL coverage for this GRB spans a broad range from keV to GeV energies, including data from the \textit{Swift} X-ray Telescope ({\xrt}), \textit{Swift} Burst Alert Telescope ({\bat}), \textit{Fermi} Gamma-Ray Burst Monitor ({\gbm}), and the \textit{Fermi} Large Area Telescope ({\lat}); see \citet{FermiSwift_190114_20}, \citet{MAGIC_190114C_MWL19}, and \citet{KlingerEtAl_GRB190114C}. It should be noted that the photon statistics in \lat are only 5 and 6 photons in the envelopes shown in the top panel of \reffig\ref{fig:observational_picture_VHEGRBs}. \lat, therefore, does not significantly constrain the combined MWL SED \citep{KlingerEtAl_GRB190114C}. Remarkably, this SED appears to be consistent with a single power-law component with photon index $\photonindex\approx2$ decaying consistently in time from keV to TeV energies (indicated by the dotted lines in \reffig\ref{fig:observational_picture_VHEGRBs}, see also \citet{FermiSwift_190114_20}). Even though widely attributed to a two-component synchrotron self-Compton (SSC) origin \citep[e.g.,][]{MAGIC_190114C_MWL19,AsanoEtAl2020,DerishevPiran21}, a more sophisticated statistical test on the counts-level shows no robust preference over a single component origin, in particular when including the cross-calibration uncertainties from the multiple instruments \citep{KlingerEtAl_GRB190114C}. This leads us to conclude that the large MWL data set is indicative of a single power-law photon spectrum but remains inconclusive about the origin of the VHE emission.

\paragraph{GRB~190829A} 
The afterglow of this GRB was detected relatively late in the VHE band with the High Energy Stereoscopic System (\hess), at about 4--8~hours (first night corresponding to 15.4--28.4~ks) and 27--32~hours (second night corresponding to 97.9--11.8~ks) \citep{HESS_190829}. However, the proximity of the event ($z=0.08$) allowed for a relatively precise measurement of the de-absorbed photon index at VHE, yielding $\photonindex = 2.1\pm0.1\mathrm{(stat)}\pm0.3\mathrm{(sys)}$ (night 1) and $\photonindex = 1.9\pm0.3\mathrm{(stat)}\pm0.2\mathrm{(sys)}$ (night 2). Unfortunately, at such late times, the contemporaneous MWL coverage is limited to only \xrt at keV energies. Remarkably, extrapolation of both individual spectra by a power-law of photon index $\photonindex \approx 2.1$ is consistent with each other (indicated by the dotted lines in \reffig\ref{fig:observational_picture_VHEGRBs}), leading to a similar picture as for GRB~190114C above. A statistical test on the counts-level basis prefers a single synchrotron component at $>5\sigma$\citep{HESS_190829}. Assuming a single radiation zone, this tension is mainly driven by the hard spectrum observed in the VHE band. This contrasts with the expectation of a softer SSC component, softened by the Klein-Nishina effects. We conclude that this MWL data set is consistent with a single power-law photon spectrum and in tension with a one-zone SSC scenario as the origin of the VHE emission.

\paragraph{GRB~221009A}
This event is the brightest GRB detected so far and was followed up by a large amount of MWL observations. It triggered \gbm (we use this as $T_0$) about 225~s before the main burst saturated the detector. The event was detected by the Large High Altitude Air Shower Observatory (\lhaaso), where the combined data sets of the Water Cherenkov Detector Array (WCDA) and the Kilometer Squared Array (KM2A) show a single power-law with de-absorbed photon index $\photonindex\approx 2.35 \pm 0.03 \mathrm{(stat)}$ for 230-300~s after trigger, and $\photonindex\approx 2.26 \pm 0.02\mathrm{(stat)}$ for 300-900~s after trigger, which extends from 0.2~TeV up to at least 10~TeV \citep{LHAASO_221009_WCDA,LHAASO_221009_KM2}. With a redshift of $z=0.15$, the EBL absorption introduces systematic uncertainties to the de-absorbed photon index of at least $0.1$ and limits robust conclusions towards the intrinsic spectral shape at energies above 10~TeV. Due to the smoothly varying light curve detected with the WCDA, the emission is commonly attributed to the early afterglow phase. This hard VHE spectrum itself is again challenging to describe within an SSC scenario, again due to the Klein-Nishina spectral softening effects expected. 
In addition, a time and energy-dependent analysis of the contemporaneous \gbm data indicates a hard afterglow component showing up in light-curve valleys between the prompt signatures \citep{FermiGBM_grb221009a,Zhang_GBM_221009_23}. Remarkably, the spectra of \gbm and \lhaaso are compatible with a single power-law component with a photon index of $\photonindex \approx 2.1-2.2$ (see again indicative dotted lines in \reffig\ref{fig:observational_picture_VHEGRBs}). Taking into account the observational challenges introduced by the viewing direction through the Galactic plane \citep[e.g.,][]{Klinger_221009_24}, this conclusion is also consistent with the observations by \lat and the instruments on board the AGILE satellite \citep{Liu_LAT_221009_23,Tavani_AGILE_221009_23}, although a prompt origin of this emission cannot be excluded. A comparison of this picture to later observations (4~ks, 22~ks) of \xrt, \bat, and \lat also shows consistency with a power-law of photon index $\photonindex \approx 2.2$ \citep[see][for a counts-level MWL fit]{Klinger_221009_24}. We conclude that this large MWL data set is indicative of a single power-law photon spectrum and is in tension with a one-zone SSC scenario as the origin of the VHE emission.
\\

In summary, we find an emerging observational picture of photon spectra resembling a single power-law component extending from roughly keV to sometimes even tens of TeV energies and with a hard photon index of $\photonindex\approx 2-2.2$.

\paragraph{Population Picture}
Putting these observations into the context of the population of GRBs with detection up to only GeV energies (though without constraining VHE upper limits), one finds that the picture inferred from the VHE-detected GRB is not particularly unique. Looking at the 2nd GRB catalogue from \lat \citep{LAT_GRB_cat_19}, one finds for their afterglow emission (``EXT") window an average photon index of $\gamma \approx 2$ too, although with a large spread (10\%/90\% percentile of 1.6/2.5). Taking all \xrt-detected GRBs, and the subset also detected by \lat, the photon index distribution at keV energies peaks around $\photonindex\approx 2$ and $\photonindex\approx 1.8$, respectively \citep{SwiftFermi18}.
This indicates that a spectral break typically occurs somewhere in this keV-MeV energy range or below, and that the hard power-law spectrum commonly observed at the highest energies extends also down to X-ray energies on average.

\section{Methods} \label{sec:methods}
In this section, we describe how we model the observed afterglow emission. We do this by performing the blast wave modelling independently from the radiation modelling and ignoring the feedback from the radiation on the blast wave. This has been done in many works; see, e.g., \cite{Zhang_GRBbook} for a review.

We discuss in \refsec\ref{sec:BW_description} our description of the circumstances in the blast wave. These provide the conditions for our multi-messenger radiation modelling in this radiation zone (\refsec\ref{sec:MM_modeling}), where we use the open-source software \AM (version 1.0.0) for the first time also to perform GRB afterglow modelling. Then, we discuss in \refsec\ref{sec:propagation} how we convert our radiation densities to observed spectra and the expected propagation effects during their way to Earth. Finally, we summarise our performed parameter space exploration and qualitative scenario type selection in \refsec\ref{sec:scan_and_selection}.

For better reproducibility of our results, we publish the source code for our scenarios on GitHub\footnote{ \url{https://github.com/maklinger/LepHadGRBAfterglows}, \\ preserved on Zenodo at \citet{zenodo_ref}}.

\subsection{Shock Description, Blast Wave Dynamics, One-Zone approximation} \label{sec:BW_description}

In our definition, the blast wave refers to a thin shell immediately downstream of the relativistic shock. We approximate this shock as an infinitely thin discontinuity and ignore the effects of magnetic fields, non-thermal particles, and radiation pressure on the jump conditions.
Motivated by micro-physical simulations, we also assume that the shock discontinuity amplifies turbulent magnetic fields and accelerates particles, producing a non-thermal (power-law) spectrum between a minimum and maximum energy.
We highlight that this set of assumptions --- commonly used in the GRB literature --- is only accurate to factors of order unity, $\Ounity$, see also \refsec\ref{sec:limitations}. Consequently, we focus on the dominant scaling relations and ignore details of order unity factors for the rest of the paper.

The instantaneous conditions in the blast wave, relevant for the radiation modelling, are defined via the Lorentz factor of the shock $\Gamma$, 
observation time $t$, and proton number density just upstream of the shock $n$. 
We define the (momentarily comoving) dynamical timescale $\tdyn = \Gamma t$ (Doppler factor $\mathcal{D} \approx \Gamma$, we ignore the redshift factor $(1+z)\approx \Ounity$), and adopt $\pram = \Gamma^2 n m_{p} c^2$ for the upstream ram pressure (ie. we assume the average upstream particle mass to be the proton mass $m_p$). With this, we can define the timescale for the free-streaming escape of uncharged particles $\tesc$, the adiabatic cooling $\tadi$ of charged particles (coupled via the magnetic fields), and the timescale of volume dilution $\texp$ introduced by the expansion of the blast wave:
\begin{align}
    &\tesc \approx \tadi \approx \texp \approx \tdyn \approx \Gamma t \\
    & = 3\times 10^4 \mathrm{s}\qty(\frac{\Gamma}{30})\qty(\frac{t}{10^3 \mathrm{s}})  \; .
\end{align}

As a $pp$ target, we estimate the thermal proton density in the blast wave as $n'$ from the shock jump conditions and ignore their momentum compared to non-thermal protons (cold target approximation).
\begin{equation}
    n' \approx \Gamma n = 30\mathrm{cm}^{-3} \qty(\frac{\Gamma}{30})\qty(\frac{n}{1\mathrm{cm}^{-3}})  \; .
\end{equation}
We parameterise the isotropic magnetic field pressure  $p_B'$ and the non-thermal proton/electron injection in terms of the ram pressure $\pram$ via the fractions\footnote{For the hydrodynamic shock approximation these pressures need to be negligible to the dynamics of the system (e.g., compared to the proton thermal pressure)} $\varepsilon_{B/p/e} \ll 1$:
\begin{equation}
    p_{B/e/p} = \varepsilon_{B/e/p} \Gamma^2 n m_p c^2  \; .
\end{equation}
This defines our magnetic field strength $B'$:
\begin{align}
    B' = 0.6 \,\mathrm{G} \qty(\frac{n}{1\mathrm{cm}^{-3}})^{1/2} \qty(\frac{\Gamma}{30}) \qty(\frac{\varepsilon_B}{10^{-2}})^{1/2}  \; .
\end{align}

We furthermore assume the non-thermal particle injection spectra $q_{e/p}'$ to be of a power-law form with spectral index $\injindex$ for electrons and protons, an exponential cut-off at $\Epmaxp$ (protons) and $\Eemaxp$ (electrons), and a step-like turn-on at the minimum energy $E_e^{\prime\mathrm{min}} = E_p^{\prime\mathrm{min}} \equiv \Eminp$ (same free parameter for both\footnote{We note that, motivated by the large uncertainty, we use the injection energy scale, $\Eminp$, as a free parameter, rather than the often adopted constant fraction parameters \citep[e.g.,][]{SarietAl96}: the non-thermal particle number ($\zeta$) and energy injection ($\varepsilon$) rates \citep[e.g.,][]{VDHorstEtAl2014,MisraEtAl2021,WarrenEtAl2015,WarrenEtAl2018,ResslerLaskar2017,AsanoEtAl2020}.}). Thus:
\begin{align}
    q_{E,e/p}' &\equiv \frac{\dd^3 N_{e/p}'}{\dd t' \dd V' \dd E_{e/p}'} \nonumber\\
    &= q_{E,e/p}^{\prime \mathrm{min}} \qty(\frac{E_{e/p}'}{\Eminp})^{-\injindex} \exp \qty(-\frac{E_{e/p}'}{E^{\prime\mathrm{max}}_{e/p}}) \\
    &\hspace{4cm} \mathrm{for} \; E_{e/p}' \geq E\st{min}'  \; ,\nonumber
\end{align}
with $q_{E,e/p}^{\prime \mathrm{min}} $ defined via
\begin{equation}
    \int \dd E_{e/p}' \: E_{e/p}' \, q_{E, e/p}' = \frac{\varepsilon_{e/p} \pram}{\tdyn}  \; .
\end{equation}

We re-parameterise the maximum energy in terms of the Bohm parameter $\eta$ \citep{Bohm_1949}, by balancing the acceleration time $\tacc = \eta E_{e/p}'/(eB'c)$ (with electron/proton charge $e$, speed of light $c$, and $\tacc$ dominated by the downstream component of the acceleration cycle) to the fastest cooling time at a given instance (computed via \AM, see \refsec\ref{sec:MM_modeling}).

In order to perform time-dependent radiation modelling, i.e. including the evolution of the blast wave parameters with time, the blast wave dynamics (i.e. $\Gamma$ as a function of $t$) are needed. This requires further assumptions on the density profile through which the blast wave passes. 
For simplicity, we consider a constant density case in the power-law deceleration regime \citep{BlandfordMcKee1976}:
\begin{align}
    \Gamma  &\approx \qty( \frac{3}{2^{8} \pi m_p c^5}  \frac{\Ekiniso}{n \, t^3})^{1/8} \label{eq:LF_vs_time}\\
    &\approx 56 \qty(\frac{n}{1\mathrm{cm}^{-3}})^{-1/8} \qty(\frac{\Ekiniso}{10^{54}\mathrm{erg}})^{1/8} \qty(\frac{t}{10^3\mathrm{s}})^{-3/8} \; . \nonumber
\end{align}
Our normalisation is obtained by the conservation of the initial kinetic energy $\Ekiniso$: 
\begin{align}
    \Ekiniso \approx \Gamma(t)^2 m\st{sw}(r_\star(t)) c^2  \; ,
\end{align}
with the isotropic-equivalent swept-up mass $m\st{sw}/4\pi = m_p \int \dd r_\star r_\star^2 n(r_\star)$, $r_\star(t) = 4 \Gamma(t)^2 t c$ and $\mathcal{D} \approx \Gamma$.

\subsection{Multimessenger Modelling} \label{sec:MM_modeling}
The evolution of the spectral and spatial density $\hat{n}_{E,i}' = \dd^2 N_i'/ \dd E_i' \dd V'$ of each considered particle species $i$ involves the temporal evolution of a system of coupled transport equations of the following form:
\begin{equation}
    \label{eq:transport_eq}
    \frac{\partial \nhat}{\partial t} = - \frac{\partial}{\partial E'}\qty( \frac{E \nhat}{\tau\st{cool}'} ) - \frac{\nhat}{\tau\st{sink}'} + q_E'  \; .
\end{equation}
This requires the estimation of the timescales of the relevant cooling processes $\tau\st{cool}'$, and sink/loss/escape processes $\tau\st{sink}'$, as well as the injection terms $q_E'$\footnote{We relate our timescales to the \AM notation in \citet{AM3_paper} via $\tau\st{cool}'= E/\dot{E}$ and $\tau\st{sink}' = 1/\alpha $.}. All three can be, in principle, a function of time and energy and can depend on the density of other species. 
We use \AM to calculate these terms and to evolve in time the densities of protons, neutrons, electrons, positrons, pions, muons, neutrinos and photons as a combined system.
As explained in detail in \citet{GaoEtAl_AM3_17} and \citet{AM3_paper}, we include synchrotron and inverse Compton cooling and emission of all charged particles, synchrotron self-absorption of electrons and positrons, photon-photon annihilation ($\gamma\gamma\to e^+e^-$) including pair feedback, escape of neutral particles and adiabatic cooling of charged particles, photohadronic ($p\gamma\to \pi...$) and proton-proton ($pp\to \pi...$) induced pion production, the subsequent decays of pions and muons and the electron-positron pair production via the Bethe-Heitler process ($p\gamma\to p e^+e^-$). 
We also use an effective sink term for the particle density dilution in the case of an expanding volume.

\begin{figure*}
    \centering
    \includegraphics[width=0.9\linewidth]{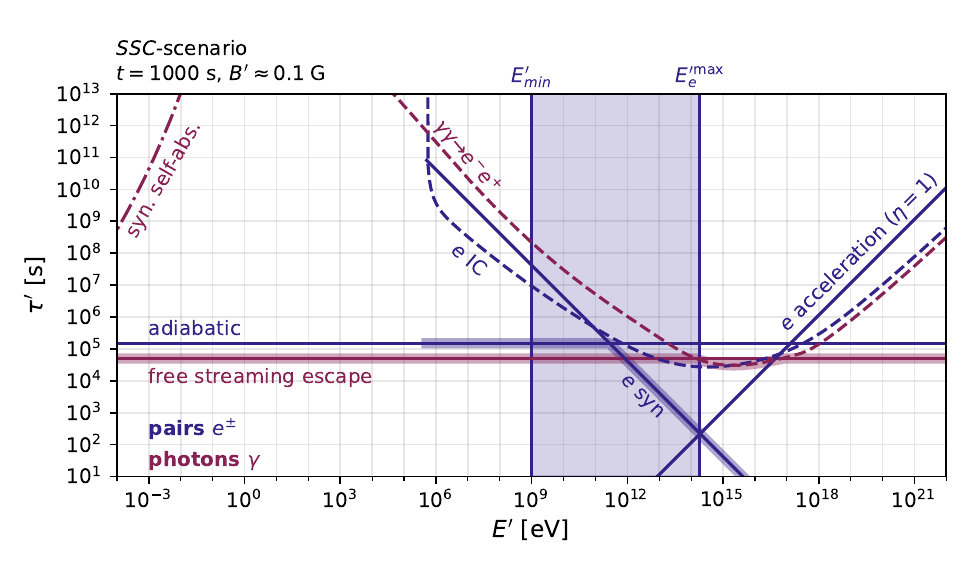}
    \caption{Comoving timescales for the \ssc-scenario of this work (see \reffig\ref{fig:SED_SSC} for the SED and parameters used): Indigo-blue corresponds to the electrons/positrons, \winered to photons. The plot shows the electron/positron timescales of adiabatic, synchrotron and inverse Compton (IC) scattering, as well as the acceleration timescale used to estimate the maximum energy, which, combined with the minimum energy, spans the power-law injection range. For photons, the timescales for free-streaming escape, synchrotron self-absorption, and internal annihilation of electron-positron pairs are shown. In addition, the dominant timescale at each energy is highlighted for intuition of the steady-state spectra.}
    \label{fig:SSC_timescales}
\end{figure*}
 
An example of some of these timescales is given in \reffig\ref{fig:SSC_timescales} for our \ssc-scenario, in which the SED is dominated by the synchrotron and inverse Compton radiation of primary injected electrons.

We consider two methods for the temporal evolution in this paper: In the \steadystate method, we keep all parameters fixed and run the system into the steady state. We use a time step of approximately $10^{-2}\tdyn$ and stop our evolution after $5\times \tdyn$. This truncation time only alters the results at the few per cent levels. This method is often applied in analytic estimates when fitting afterglow data \citep{SariEsin01,VanEertenZhangMacFadyen10,Ryan_Afterglowpy20}. Secondly, in the \timedep method, we change $\Gamma$ over time (see equation~\ref{eq:LF_vs_time}), with the system subsequently reaching a quasi-steady state, in which the blast wave parameters evolve slowly, and the system continuously catches up with the change of the blast wave parameters. This method has also been applied in a couple of works \citep{ChiangDermer99,PeerWaxman05,FanPiran08,PetropoulouMastichiadis09,PennanenVurmPoutanen14,FukushimaToAsano17}. For the \timedep method, we use a logarithmically growing time step of typically $10^{-2}\tdyn(t)$ and start our evolution typically a few hundred steps before $t$.

It can be shown that in the afterglow cases, both methods yield essentially $\nhat \propto q_E' \tau'$, with $\tau'$ being the dominant (smallest) cooling timescale, and agree up to a factor $\alpha\st{corr}$ of order unity and a convergence term $C(t)$ (when starting from empty initial conditions $\nhat(t_0')=0$):
\begin{align}
    \label{eq:n_q_tau_steadystate}
    \nhat \approx q_E' \tau' \times \underbrace{\alpha\st{corr}}_{\Ounity} \times \underbrace{C(t')}_{\to 1}  \; .
\end{align}
 
\begin{figure*}
    \centering
    \includegraphics[width=0.9\linewidth]{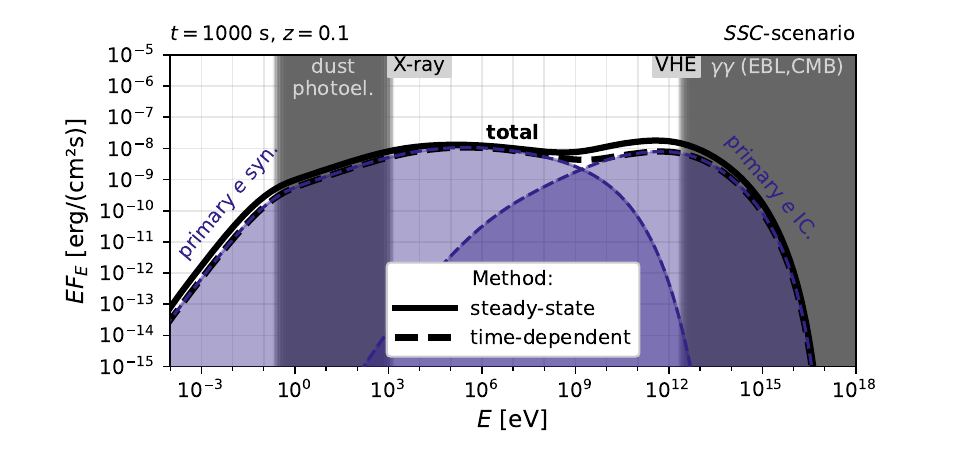}
    \caption{Comparison of time-dependent result to steady-state approximation for our \ssc scenario. Observed energy flux as a function of observed energy for a representative parameter choice at an observed time of 1000~s. Here, we show the total photon spectrum (thick black lines), with solid corresponding to the \steadystate and dashed corresponding to the \timedep method. We also indicate the contributions to the spectrum of the synchrotron (low energy component) and inverse Compton emission (high energy component) of the primary injected electrons for the \timedep method. The source is placed at a representative redshift of $z=0.1$. Furthermore, the grey shaded energy bands correspond to representative ranges of strong absorption by dust and photo-electric absorption ($\sim$ eV-keV) and EBL absorption ($\gtrsim $TeV), see \refsec\ref{sec:propagation}. We also sketch at the top the energy ranges (X-ray and VHE) where we filter for the solutions to be ``flat"; see \refsec\ref{sec:scan_and_selection}. }
    \label{fig:QSS_comparison}
\end{figure*}

We show a comparison of a typical \textit{SSC} scenario for the \textit{steady-state} method and the \textit{time-dependent} method with a constant density profile in \reffig\ref{fig:QSS_comparison}.
It can be seen that the synchrotron components of both spectra differ only within $\Ounity$, with the upscattered SSC component for the \textit{steady-state} method being only twice that obtained for the \textit{time-dependent} method (still well within an $\Ounity$ difference). In \refap\ref{ap:qss_approx}, we show that this is also true for the lepto-hadronic scenarios presented in our work.

We conclude from this that within $\Ounity$ the \steadystate method is a good approximation to the \timedep results and that both methods lead conceptually to the same conclusions. We, therefore, choose the slightly faster \steadystate method for our parameter space exploration, while we show representative scenarios obtained with the \timedep method in \refsec\ref{sec:results}.

\subsection{Conversion to Observed Flux and Propagation Effects}\label{sec:propagation}
We convert the comoving number density of escaping photons and neutrinos to the observed flux as follows (again up to $\Ounity$ and with $\mathcal{D}\approx \Gamma$; compare, e.g., \citet{PetropoulouMastichiadis09}):
\begin{equation}
    \label{eq:obs_flux_def}
    EF_E \equiv E^{2} \frac{\dd^3 N }{\dd E \dd t \dd A }  = \frac{\Gamma^2 V' E'^2 \nhat}{4\pi d_L^2 \tesc}   \; .
\end{equation}
Here, we define the comoving volume to be $V' = 4\pi r_\star^3/\Gamma$, approximating the shell thickness to $r_\star/\Gamma$, where $r_\star = 4\Gamma^2 t c$. Also, we use the cosmological model from \cite{Planck18} to convert the representative redshift $z=0.1$ to the luminosity distance $d_L\approx 1.5\times 10^{27}$~cm. Lastly, we convert the comoving energy $E'$ to the observed one via $E = \Gamma E'$.

In order to demonstrate the observability of the SEDs, we show in our figures two grey-shaded energy ranges for typical parameters. At these energies, the observed flux is exponentially suppressed during the propagation, compared to the emitted flux\footnote{Note that synchrotron self-absorption and photon-photon-annihilation are included in the in-source modelling in \AM via sink terms in the transport equation (\refeq\ref{eq:transport_eq}).}. In particular, the uncertainties in the absorption model become very large and hardly allow for a reliable de-absorption, and thus we also do not attempt to include a specific absorption scenario for each process. We describe the specifics of these absorption processes in the following, leading to our choice for the energy range of the bands.

\paragraph{Dust absorption} The low energy (optical to ultraviolet) window is limited by scattering on dust grains of uncertain size and location. For representative parameters, the commonly used empirical model of \cite{Pei92} starts to become optically thick around 1~eV, with a large systematic uncertainty depending on the environment type chosen (Milky Way or Small/Large Magellanic Cloud) and the underlying dust size distribution. 

\paragraph{Photo-electric absorption} With increasing energy, the absorption mechanism is taken over by the ionisation of material along the line of sight via the photo-electric effect \citep[e.g.,][]{Wilms2000}. This effect depends strongly on the number of atoms and the metalicity, but the general scaling of the cross-section leads to an exponential suppression towards lower energies, starting from around $\sim$keV energies. We note that besides absorption due to the material in our own Galaxy, which can be calibrated from other known sources, typically an additional absorption component local to the emission is needed \citep{Willingale_abs13}, whose metalicity must be guessed. 

\paragraph{EBL/CMB absorption} At higher energies, roughly above a few TeV for a typical redshift of 0.1, the pair production of high-energy photons from the source with low-energy photons of the extragalactic background light (EBL) and cosmic microwave background (CMB, relevant above PeV energies) during the propagation causes again an exponential suppression. The systematic uncertainty of the EBL target field and the typical energy resolution of 10\% limit our observational capabilities at higher energies \citep[e.g.,][]{HESS_EBL17}.

\subsection{Parameter scan and model selection} \label{sec:scan_and_selection}
For computational speed, we adopted the \steadystate method to perform a grid scan over the parameters. We summarise the parameters varied and their ranges in \reftab\ref{tab:parameter_range}. We note that in order to explore extended synchrotron scenarios, we also explored the parameter space with $\eta < 1$.

Throughout the paper, we focus on the spectrum at an observation time of 1000~s and $z=0.1$, as representative for VHE observations.

\begin{table}
    \caption{Summary of the parameters which we varied from minimum (min.) to maximum (max.) values with a step size (step). The same values of $\Eminp, \eta$, and $p$ are used for electrons and protons}. We note that we did not explore the full parameter space for the $\eta < 1$ case.
    \label{tab:parameter_range}
    \begin{tabular}{|l|c|c|c|}
        \hline
        \textbf{Parameter} & \textbf{min.} & \textbf{max.} & \textbf{step} \\ \hline
        $\log_{10}\Gamma$ & $0.7$ & $1.7$& $0.25$ \\
        $\log_{10}(n/1\mathrm{cm}^{-3})$ & $-3$ & $5$& $1$ \\
        $\log_{10}\varepsilon_B$ & $-9$ & $0$& $0.5$ \\
        $\log_{10}\varepsilon_e$ & $-9$ & $0$& $0.5$ \\
        $\log_{10}\varepsilon_p$ & $-5$ & $0$& $0.5$ \\
        $\log_{10}((\Eminp/1\eV)$ & $9$ & $12$& $0.5$ \\
        $\log_{10}\eta$ & $-4$ & $4$& $0.5$ \\
        $p$ & $2.0$ & $2.6$& $0.1$ \\ \hline
    \end{tabular}
\end{table}

\begin{figure*}
    \centering
    \includegraphics[width=0.7\linewidth]{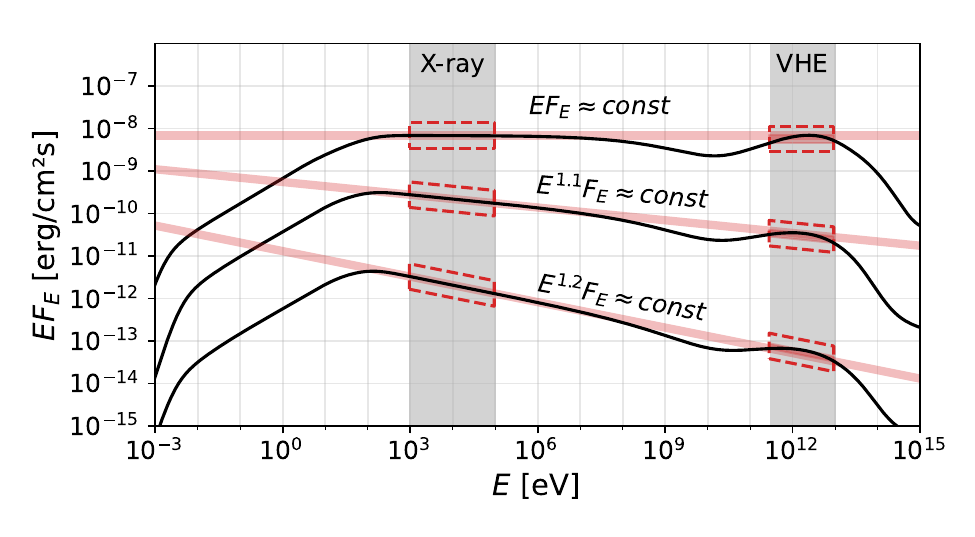}
    \caption{Conceptual sketch of the selection process for scenarios that follow our definition of ``flat". Three observed \steadystate SEDs are shown corresponding to $E^{1+\Delta \photonindex} F_E \approx const$ for $\Delta \photonindex \in \{0, 0.1, 0.2\}$, corresponding to the red power-law lines. We define constant as being confined to the red dashed boxes in the two observed energy regimes $10^{3}-10^{5}$~eV (\xrayband), and $10^{10.5}-10^{13}$~eV (\vheband), both visualised by the grey vertical bands. We emphasise the aspect ratio of 1, which allows the best visualisation of the level of flatness.}
    \label{fig:conceptual_sketch}
\end{figure*}

We calculate the total photon SED for each point of the parameter space. 
From this we select models with a flat spectrum, namely when the photon SED follows approximately a single power-law, by requiring the SED to not vary by more than a factor (in log space $\Delta \log_{10} EF_E$) in two bands with spectrally strong observations, see \reffig\ref{fig:conceptual_sketch}: \xrayband (from 1~keV to 100~keV), and \vheband (from 0.3~TeV to 10~TeV). We motivate this factor by the typical spread, $\sigma_\gamma$, of photon indices, $\photonindex$, from \xrt, $\sigma_{\photonindex,\mathrm{XRT}}\approx 0.2$ \citep{Willingale_abs13,SwiftFermi18}, as well as $\sigma_{\photonindex,\mathrm{VHE}}\approx 0.3$ from GRB~190114C \citep{MAGIC_190114_data19}, GRB~190829A \citep{HESS_190829}, and GRB~221009A \citep{LHAASO_221009_KM2}. We convert this into the factor $\Delta \log_{10} EF_E$ by fully containing a power-law with slope variation within $ [-\sigma/2, \sigma/2]$. We add an additional systematic uncertainty of about 0.1 and end up with a box height of about $\Delta \log_{10} EF_E \approx 0.3$, i.e. variation of the flux in each band by less than 30\% of an order of magnitude. We use these criteria to select models with approximate flatness in each of the energy ranges and require additionally that the average height ($ 0.5\log_{10} (EF_E|\st{max} \times EF_E|\st{min})$) of both components is contained within a factor $10^{0.2}\approx 1.6$.

Since the observational data also indicates slightly softer power-laws than photon index $2$ ($\photonindex=2$ corresponds to completely horizontal in an $EF_E$ plot), we generalise our criteria to softer spectra by using $E^{1+\Delta \photonindex} F_E$ instead. We consider three values for the tilt, $\Delta \photonindex = 0$ as for GRB~190114C, $\Delta \photonindex = 0.1$ as for GRB~190829A, and $\Delta \photonindex = 0.2$ as for GRB~221009A. Three examples are given in a conceptual sketch in \reffig\ref{fig:conceptual_sketch}.
As all scenarios work for all three cases of $\Delta\photonindex$, we choose to show for this paper only figures of the case $\Delta \photonindex = 0$ and discuss the validity of the scenario for the other cases in the corresponding section. 

Besides the total spectrum, we additionally track the following photon components: synchrotron and inverse Compton scattered photons from electrons from primary injection, annihilation feedback pairs ($\gamma \gamma \to e^- e^+$), proton-proton cascade secondaries, photo-pion cascade secondaries and pairs from the Bethe-Heitler process, as well as from protons, pions, and muons, and also the photons from the decay of neutral pions produced in proton-proton and photo-pion production. More precisely, we co-evolve (without feedback into the main solution) the transport equation of secondaries using only the source term from a certain channel from the list above (in combination with the sink and cooling terms based on the main solution). Based on this, we can classify the type of solution by checking which component contributes most to the energy-integrated flux in each of the bands. For example, an \ssc-scenario would correspond to the primary injected electron synchrotron component dominating the \xrayband band, whereas the inverse Compton component of the same particles dominates the \vheband band.
We note that as soon as the radiation from the cascade electrons contributes significantly, also the radiation of the pairs from photon annihilation becomes similarly strong, and classification via a dominant component is hard to make.

\section{Results} \label{sec:results}

Following our parameter space scan and model selection (\refsec\ref{sec:scan_and_selection}), we find five families of scenarios with the required flat SED behaviour. For these, we find four photon-emission channels to be dominant:
\begin{enumerate}
    \item synchrotron radiation of primary injected electrons, relevant for all scenarios
    \item inverse Compton up-scattered synchrotron radiation of primary injected electrons
    \item proton-proton induced cascade emission: $\pi^0$ decay photons at high energies, synchrotron radiation of secondary electrons/positrons at low energies
    \item photo-pion induced cascade emission, i.e. synchrotron radiation of secondary electrons/positrons
\end{enumerate}
This leads to five conceptually different types of flat scenarios: \ssc, \extsyn, \psyn, \ppcasc, and \pgcasc. We summarise the properties, advantages, and problems of these scenarios in \reftab\ref{tab:summary} and discuss a representative parameter set for each scenario in the following sections. For the sake of comparability, the y-range (energy flux) is kept the same in the figures of all scenarios.

\begin{table*}
\centering 
\caption{Summary table of the discussed scenarios\label{tab:summary}}
\begin{tabular}{|l|c|c|l|l|l|}
\hline
Name    & Low $E$ Comp.                 & High $E$ Comp.                        & Specifications & Advantages      & Limitations  \\ \hline
\ssc    & \colesyn{primary} & \colesyn{primary}           & - leptonic dominated   & - bright & KN suppression \\ 
(\refsec\ref{sec:SSC})&\colesyn{electron syn.}&\colesyn{electron IC}& - slow cooling electrons & & $\to$ VHE slope\\ 
&&& - low $\Eminp$ & & $\to$ fine-tuned height ratio\\ 
&&& - efficient acceleration ($\eta \sim 1$)& & \\ 
&&& - moderate $n,\Gamma $ & & \\ \hline

\extsyn & \multicolumn{2}{|c|}{\colesyn{primary electron syn.}}  & - leptonic dominated & - bright& - requires $\eta\ll1$ \\ 
(\refsec\ref{sec:extsyn})&\multicolumn{2}{|c|}{}& - extreme acceleration ($\eta \ll 1)$ & - directly yields& (challenging for 1 zone) \\ 
&\multicolumn{2}{|c|}{}& - higher $\Eminp$ (less IC targets)&  single power-law&  \\ 
&\multicolumn{2}{|c|}{}& - moderate $n,\Gamma$& &  \\  \hline

\psyn   & \colesyn{primary} & \colpsyn{proton syn.} & - lepto-hadronic& - bright & - $p$-syn. component   \\ 
(\refsec\ref{sec:psyn})&\colesyn{electron syn.}&& - high $n,\Gamma$& &  at exp. cut-off (fine-tuning) \\ 
&&& - efficient acceleration ($\eta \sim 1$)& &  $\to$ flux level\\ 
&&& - moderate baryonic loading & & $\to$ peak energy \\ 
&&&     ($\varepsilon_p/\varepsilon_e \sim 10^{2}$)& & $\to$ shape of UHE $p$ cut-off \\ \hline

\ppcasc & \colesyn{primary} & \colpizero{$pp$ created} & - lepto-hadronic& - flat VHE comp.& - inefficient \\
(\refsec\ref{sec:ppcascade})&\colesyn{electron syn.}& \colpizero{$\pi^0 $-decay}& - very high $n,\Gamma$&   extending &  \\ 
&&& - extreme baryonic loading & beyond 10~TeV &  \\ 
&&&     ($\varepsilon_p/\varepsilon_e \sim 10^{6}$)& &  \\ \hline

\pgcasc & \colesyn{primary } & \colpgamma{$p\gamma$ created} & - lepto-hadronic& - bright & - extreme energy and\\ 
(\refsec\ref{sec:pgammacascade})&\colesyn{electron syn.}&\colpgamma{electron ($e^{\pm}$)}& - extreme $n,\Gamma$& & density requirements \\ 
&& syn.& - high baryonic loading& &  \\ 
&&&     ($\varepsilon_p/\varepsilon_e \sim 10^{4}$)& &  \\ \hline 

\end{tabular}

\end{table*}

\subsection{The \ssc scenario} \label{sec:SSC}

\begin{figure*}
    \centering
    \includegraphics[width=\linewidth]{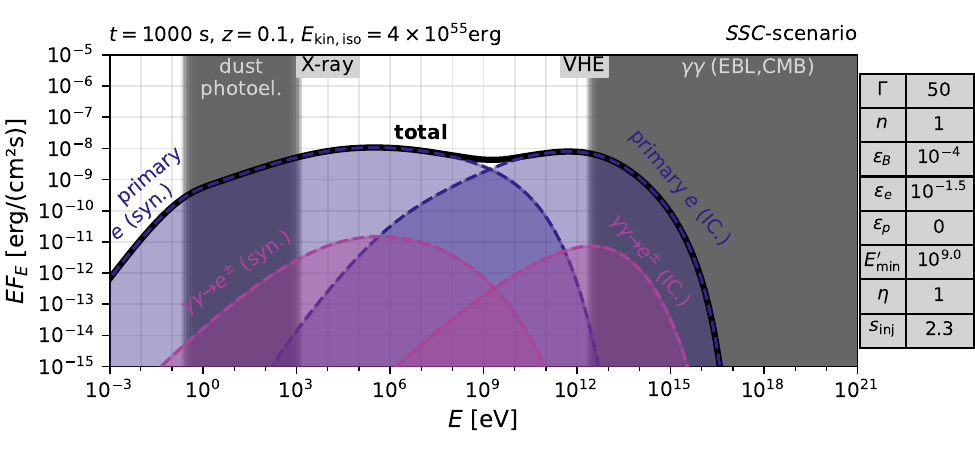}
    \caption{\textbf{\ssc scenario (\timedep method):} Observed energy flux as a function of observed energy for a representative parameter choice (see table at the side with symbols as in text, note $n$ in cm$^{-3}$ and $\Eminp$ in eV) at an observed time of $t=1000$~s. The source is placed at a redshift of $z=0.1$. Shown here are the total photon spectrum and its contributions as described next to the dashed lines: the synchrotron and inverse Compton emission of the primary injected electrons; and the synchrotron and inverse Compton emission of the electron/positron pairs created by annihilated photons ($\gamma\gamma\to e^{-}e^{+}$). See \reffig\ref{fig:QSS_comparison} for a more detailed description of the SED plots and \reffig\ref{fig:SSC_timescales} for the corresponding timescales.}
    \label{fig:SED_SSC}
\end{figure*}

In the \ssc scenario, the photon emission in our \xrayband and \vheband bands are dominated by the synchrotron and inverse Compton components, respectively, of the primary injected electrons. 
We show in \reffig\ref{fig:SED_SSC} the observed SED for a representative set of parameters. The figure is similar to \reffig\ref{fig:QSS_comparison}, though we focus here on the \timedep method.
We refer to \reffig\ref{fig:SSC_timescales} for a plot of the corresponding timescales.

\reffig\ref{fig:SED_SSC} shows the total photon SED and the contributions from the different radiation channels. In indigo-blue, the synchrotron (dominant at low energies) and inverse Compton up-scattered synchrotron emission (at high energies) of the primary shock-accelerated electrons are shown. The subdominant purple lines show the contribution of the secondary pairs from photon annihilation ($\gamma\gamma\to e^{-}e^{+}$), more specifically, their synchrotron emission and their inverse Compton emission (with the total photon spectrum as a target).

We note that the parameters in the table next to the plot are rather typical for GRB modelling. The choice $\varepsilon_p=0$ is for simplicity, and we note that even for $\varepsilon_p=1$, the hadronic components stay subdominant in the regime of the parameter space occupied by the \ssc scenario.

In order to understand the photon spectrum, it is necessary to understand the shape of the primary injected electron spectrum. For this, in the steady-state approximation $\nhat\sim q_E' \tau'$ (cf. eq.~\ref{eq:n_q_tau_steadystate}), the dominant timescales and the injection energy range in \reffig\ref{fig:SSC_timescales} (shaded in indigo-blue) are instructive. Three regimes emerge, which is also known as the slow cooling case, with cooled electron spectral indices ($\nhat \propto \Eep^{-\electronindex}$):
\begin{enumerate}
    \item \textbf{adiabatic cooling tail:} $\Eep < \Eminp$ with $\electronindex = 1$
    \item \textbf{adiabatic cooling regime:} \\$\Eminp < \Eep \lesssim 3\times 10^{11}~\eV$ with $\electronindex \approx \injindex$
    \item \textbf{synchrotron cooling regime:} \\$3\times 10^{11}~\eV \lesssim \Eep < \Eemaxp $ with $\electronindex \approx \injindex + 1 $
\end{enumerate}

The resulting synchrotron photon spectrum resembles the electron spectrum with these critical energies in the electron spectrum mapping to 
\begin{equation}
    \Eg^{\prime\mathrm{syn}} \approx \frac{B'}{\Bce} \frac{\Eep^2}{m_e c^2}  \; ,
\end{equation}
in the photon spectrum. Here, $\Bce\approx 4\times 10^{13}~\mathrm{G}$ is the critical magnetic field for electrons, and $m_e$ is the electron mass. Thus, the cooling break at $\Eep\approx 3\times 10^{11}~\eV$ maps to an observed photon energy of 
\begin{equation}
    \Egcb \approx \Gamma \frac{B'}{\Bce} \frac{\Eep^2}{m_e c^2} \approx 3\times10^{4}~\eV  \; .
\end{equation}
Additionally, the slopes of the electron spectra get stretched ($\Delta \ln \Eg^{\prime\mathrm{syn}} = 2 \Delta \ln \Eep$) and one finds for the photon indices (defined as $EF_E\propto E^{2-\photonindex}$): $\photonindex = (\injindex+1)/2\approx 1.7$ for $\Egp<\Egcb$ and $\photonindex= (\injindex+2)/2\approx 2.2$ for $\Egp>\Egcb$. This results in a roughly flat synchrotron spectrum in the \xrayband band. The choice of $\eta=1$ leads to a maximum synchrotron photon energy $\Eg^{\mathrm{syn,max}} \approx \Gamma \Eg^{\prime\mathrm{syn,burnoff}} \approx 10~\mathrm{GeV}$ \citep[here $\Eg^{\prime\mathrm{syn,burnoff}} \approx 150~\mathrm{MeV}$ is the (comoving) burn-off limit; see, e.g.,][for a detailed discussion]{KhangulyanEtAl_clumpyB_21}. 

The IC spectrum can be understood in a similar way to the synchrotron spectrum, although the break features and the power-law slopes are washed out stronger through the broader kernel from the synchrotron target, which extends over many (5-10) orders of magnitude in energy. Additionally, for observed energies in/above the \vheband band, the Klein-Nishina effects on the IC kernel have to be taken into account (note also the discussion of the validity of this approximation in the supplementary materials of \citet{HESS_190829}, Equation~S21):

\begin{equation}
    \label{eq:KN_effects}
    \EgKN \approx \Gamma^2 \frac{(m_e c^2)^2}{\Egt} \approx 1~\mathrm{TeV}\qty(\frac{\Gamma}{50})^2 \qty(\frac{\Egt}{10^3~\eV})  \; ,
\end{equation}
with photon target energy $\Egt$.
Here, we assume that the electron gives most of its energy to the up-scattered photon. The suppression of the IC cross-section above these energies requires the existence of a target photon spectrum extending \textit{at least down to} energies $\Egt$ in order to maintain flatness in the IC component. In our \ssc scenario, the spectrum starts to decrease, however, slowly already below the \xrayband band, as also typical for the observational picture (see, e.g., \reffig\ref{fig:observational_picture_VHEGRBs}). This leads to a washed-out turnover at TeV energies and has two limiting implications in light of flat photon spectra extending to even higher energies.

First, in the extreme Klein-Nishina regime ($\Eep \Egp \gg m_e^2 c^4 \approx 3\times10^{11}~\mathrm{eV}^2$) the observed photon index is $\photonindex = \injindex +1 \approx 3.3$ (adiabatically cooled electrons) or $\photonindex = \injindex +2 \approx 4.3$ (synchrotron-cooled electrons). This is in conflict with the much harder observed values of $\photonindex \approx 2-2.2$. The broad target spectrum can alter the slope slightly but cannot resolve this conflict.

Secondly, the height (i.e. energy density) of the IC component no longer scales with the energy density in the electrons but sits below this level. This implies that the well-known result for the height ratio of the synchrotron ($\uesyn$) and the IC ($\uessc$) component no longer scale as $\uesyn/\uessc \propto \varepsilon_B/\varepsilon_e$. Thus, an equal height ratio requires a special choice of $\varepsilon_B$ and $\varepsilon_e$.

The timescale plot in \reffig\ref{fig:SSC_timescales} shows two other interesting aspects. 
First, the IC cooling timescale for the electrons is sufficiently large so as to not affect the electron spectral shape beyond a smoothing of the break at $\Ounity$ level. Indeed, at the cooling break, all three (IC, synchrotron, adiabatic) timescales are at the same level. This is linked to the fact that the energy density in both photon components, synchrotron and IC, is tuned to a similar value. For electron spectra with a spectral index between 2 and 3, the power ($\propto \Eep^{2}$ for synchrotron and IC) sits mainly at the cooling break energy. On the other hand, the cooling time is linked to exactly the power lost by an electron. Thus, a comparable energy density will always result in roughly comparable cooling times, such as that in \reffig\ref{fig:SSC_timescales} \citep[see also the discussion in \refap{A} of][]{KlingerEtAl_GRB190114C}. 

The second interesting aspect is the effect of photon annihilation into electron/positron pairs. Its importance can be read off from the dashed \winered curve (also for the photons $\nhat \approx q_E' \tau'$), showing that for this scenario, the effect would be only very small around energies of $1-10~\mathrm{PeV}$. For this set of parameters, this is just above the maximum energy of injected electrons and not relevant for their SED. However, stronger $\gamma\gamma\to e^+e^-$ absorption can lead to additional softening in the photon spectrum.

We note that all the above-discussed aspects also qualitatively hold for the case of slightly tilted spectra ($\Delta \photonindex \gtrsim 0$).

We conclude that the \ssc scenario allows for similar energy flux levels in the \xrayband and \vheband bands. However, a more detailed inspection of the spectral shape reveals significant curvature, especially at the highest energies ($>1$~TeV) due to the Klein-Nishina effects.
Additionally, we find \ssc scenarios to exist only for reasonably large values of $\Gamma \gtrsim 10$ (compare eq.~\ref{eq:KN_effects}) and $\eta \approx 1$ (to fill up the GeV range with synchrotron photons). Both points, curvature and large $\Gamma$, create conflicts for the cases of GRB~190829A with a low $\Gamma$ \citep[][]{HESS_190829} and GRB~221009A extending with an unbroken power-law to at least 10~TeV \citep[][]{LHAASO_221009_KM2}.

\subsection{The \extsyn scenario} \label{sec:extsyn}
Before turning to scenarios with dominant hadronic components, we first discuss the conceptually simple idea of a single electron synchrotron component extending from the \xrayband to the \vheband band.

\begin{figure*}
    \centering
    \includegraphics[width=\linewidth]{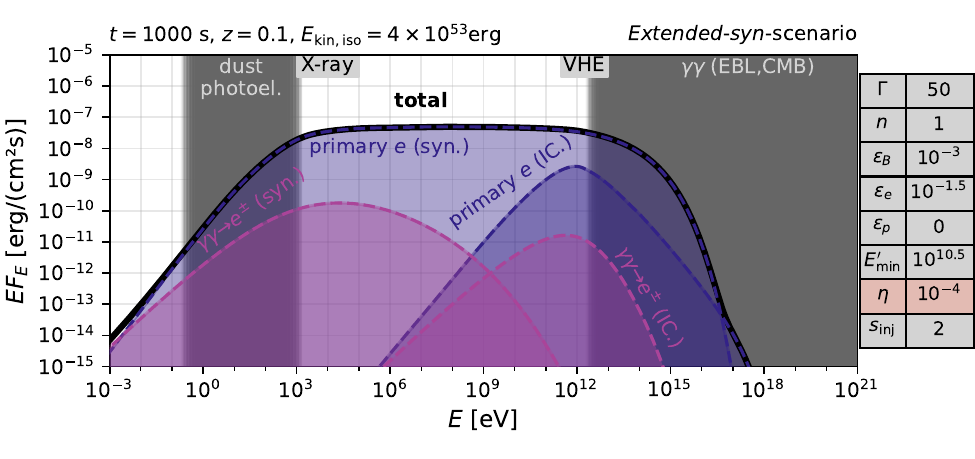}
    \includegraphics[width=0.9\linewidth]{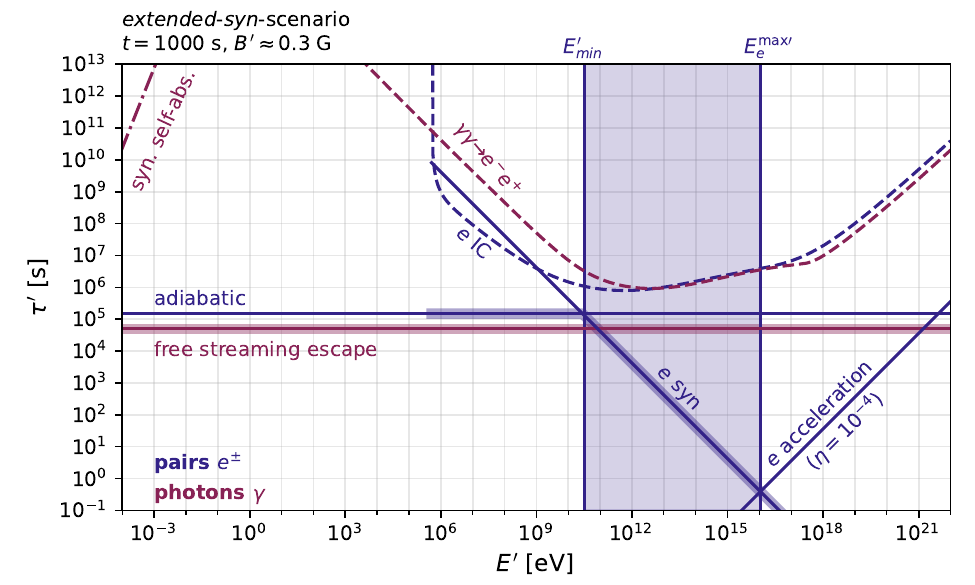}
    \caption{\textbf{\extsyn scenario  (\timedep method) --- Top:} Observed energy flux as a function of observed energy for a representative parameter choice ($n$ in cm$^{-3}$ and $\Eminp$ in eV). We highlighted the parameter $\eta$ due to its extreme value. For a detailed description of the figure, see \reffig\ref{fig:SED_SSC}.
    \textbf{Bottom:} Comoving timescales similar to \reffig\ref{fig:SSC_timescales}.}
    \label{fig:extsyn_scenario}
\end{figure*}

\reffig\ref{fig:extsyn_scenario} combines the SED and the timescales for a representative set of parameters. 
We note that we choose for simplicity $\varepsilon_p=0$, but this does not affect our conclusions.

The SED can be understood in a similar manner to the \ssc scenario. In the \extsyn scenario, slightly larger values of $\varepsilon_B$ (which places $\Egcb\lesssim 10^3\eV$ for flatness in the \xrayband band) and $\Eminp$ (which reduces the IC component) are required. From the timescale plot, we can see that this leads to all electrons being cooled (no adiabatically cooled regime, compare to regime 2 of the \ssc scenario). Furthermore, the IC target for $\Egt<10^3~\eV$ is reduced, leading to a subdominant SSC component and an almost constant IC cooling time for $10^{10}~\eV \lesssim \Eep \lesssim 10^{16}~\eV$.

Most crucial is, however, the parameter $\eta \ll 1$, which increases the maximum synchrotron energy into the \vheband band. This can be seen as follows:
Particle acceleration requires an electric field $\mathcal{E}'$, such that in the simplest case the relativistic particle energy $E\st{particle}' \approx e \mathcal{E}' c \tacc$ after an acceleration time $\tacc$. However, in an ideal magneto-hydrodynamic (MHD), plasma currents short out these electric fields on a much shorter timescale. Instead, one can consider a moving magnetic field resulting in an effective electric field with some efficiency factor depending on the details of the setup, like the geometry. We thus define here:
\begin{equation}
    \tacc = \frac{E'}{e\mathcal{E}'c} = \eta \frac{E'}{eB'c} = \eta \frac{r\st{L}'}{c}  \;.
\end{equation}
$\eta$ can thus be interpreted as the ratio of the magnetic and effective electric field or the number of Larmor radii $r\st{L}'$ needed to increase the particle energy by an e-fold.
Thus, in earlier works the parameter $\eta$ was rather expected to be $\gg 1$ \citep[e.g., recently][]{HuangReville22}.

A value of $\eta\ll 1$ in the one-zone approximation would require that the ideal MHD conditions are no longer valid, such as in the case of reconnecting field lines \citep[for details see][] {KhangulyanEtAl_clumpyB_21}. An alternative explanation is that $\eta$ acts only as an effective parameter for one-zone modelling. In this case, it incorporates the effect of two (or multi) zone models with different magnetic field strengths; see, e.g., \cite{KhangulyanEtAl_clumpyB_21}. For example, a weak magnetic field fills up most of the space and dictates the particle acceleration rate, whilst a strong magnetic field in tiny blob regions yields catastrophic energy losses and produces the observed synchrotron radiation.

Obviously, all these results are also true for the case of slightly tilted spectra ($\Delta \photonindex \gtrsim 0$).

We conclude that the \extsyn scenario can produce flat, extended SEDs in a straightforward manner. However, the necessary assumption of $\eta\ll 1$ requires careful explanation going beyond the scope of this paper.

\subsection{The proton synchrotron scenario} \label{sec:psyn}

\begin{figure*}
    \centering
    \includegraphics[width=\linewidth]{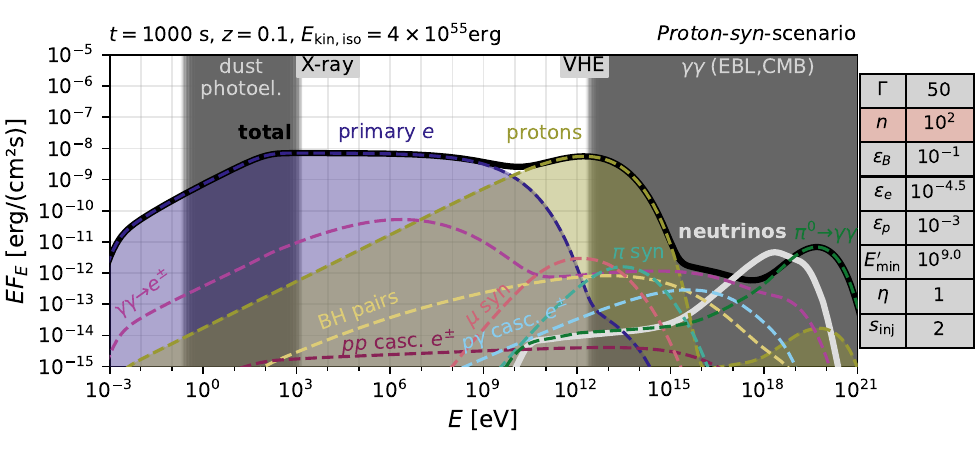}
    \includegraphics[width=0.9\linewidth]{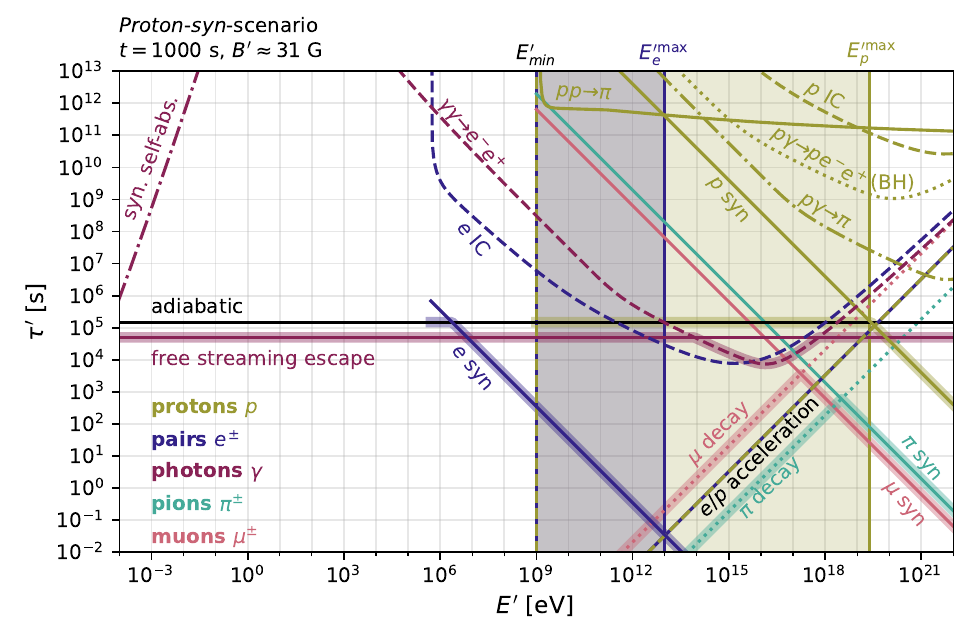}
    \caption{\textbf{\psyn scenario (\timedep method) --- Top:} Observed energy flux as a function of observed energy for a representative parameter choice (for a detailed description of the figure see \reffig\ref{fig:SED_SSC}, note $n$ in cm$^{-3}$ and $\Eminp$ in eV). Additionally, all the hadronic contributions are shown: proton synchrotron and inverse Compton (protons); synchrotron and inverse Compton from the pairs created in the Bethe-Heitler process (BH pairs), photo-pion cascade ($p\gamma$ casc. $e^{\pm}$) and proton-proton cascade ($pp$ casc. $e^{\pm}$); synchrotron emission from charged pions ($\pi$ syn) and muons ($\mu$ syn) and the photons from neutral pion decays ($\pi^0\to\gamma\gamma$). We also show the total flux of neutrinos of all flavours in light grey (neutrinos). We highlighted the density $n$ due to its extreme value.\\
    \textbf{Bottom:} Comoving timescales: Olive-green corresponds to protons, indigo-blue to electrons/positrons, \winered to photons, cyan to charged pions and pink to muons. The plot shows the timescales of adiabatic (black, same for all charged particles), synchrotron ($e/p/\pi/\mu$ syn), inverse Compton ($e/p$ IC) scattering, inelastic proton-proton collisions ($pp\to\pi$), photo-pion production ($p\gamma\to \pi$), Bethe-Heitler process ($p\gamma \to p e^{-}e^{+}$) and charged pion and muon decay. It also shows the acceleration timescale used to estimate the maximum energy for proton and electrons, which, combined with the minimum energy, spans the power-law injection range shaded in dark-blue/olive. For the photons, the timescales for free-streaming escape, synchrotron self-absorption and internal annihilation into electron-positron pairs ($\gamma\gamma\to e^{-}e^{+}$) are shown. In addition, the dominant timescale at each energy is highlighted for each species for intuition of the steady-state spectra.}
    \label{fig:psyn_scenario}
\end{figure*}

Another idea recently suggested to explain the \vheband band photons is the \psyn scenario \citep[e.g.,][]{Isravel_190114_psyn_23,Isravel_221009_psyn_23}. In this case, the proton synchrotron emission dominates the \vheband band, whereas the \xrayband band is dominated by the electron synchrotron component. We show the SED and the corresponding timescales for a representative parameter set in \reffig\ref{fig:psyn_scenario}. As an example, we also added for this scenario the comoving particle energy distributions in \refap\ref{ap:psyn_particles} in \reffig\ref{fig:psyn_particles}.

The SED shows, besides the proton synchrotron component (olive-green, including an IC contribution relevant only above $\Eg \approx 10^{16}~\eV$), the other hadronic components. However, these other components are subdominant below $\Eg < 10^{15}~\eV$. Also, the timescales plot shows the relevant timescales for protons, pions and muons.

We briefly discuss these different SED components. The primary electron synchrotron spectrum can be understood in analogy to the \ssc scenario in \refsec\ref{sec:SSC}. In the \psyn scenario, the magnetic field is stronger, such that the timescale plot shows a fast cooling case, with two relevant regimes (electron spectral index $\nhat \propto \Eep^{-\electronindex}$): 

\begin{enumerate}
    \item \textbf{synchrotron cooling tail:} \\$\Eep < \Eminp$ with $\electronindex \approx 2$
    \item \textbf{synchrotron cooling regime:} \\$\Eep \geq \Eminp$ with $\electronindex \approx \injindex + 1 $
\end{enumerate}
This results in the two power-law regimes below and above $\Eg \approx 10^2~\eV$ with photon indices $\photonindex=1.5$ (below) and $\photonindex= (\injindex+2)/2=2$ (above). The other break at $\Eg\approx 3\times 10^{-3}~\eV$ originates from synchrotron self-absorption, as can be seen from the photon timescales.
The maximum energy of the electron synchrotron component is, as in the \ssc scenario, at around 10~GeV corresponding to $\eta=1$.

The proton synchrotron component can be understood analogously by replacing the mass $m_e \to m_p$ and thus also the critical magnetic field $\Bce = m_e^2 c^3/e\hbar \to \Bcp \approx \Bce (m_p/m_e)^2 \approx 1.5\times 10^{20}~\mathrm{G}$. In the quasi-steady state approximation, the proton cooling is dominated by adiabatic cooling for all injected energies, leading to a single power-law spectrum with the original injected spectral index and an exponential cut-off at $\Epmaxp\approx 2\times 10^{19}~\eV$. The resulting synchrotron spectrum therefore has the photon index $\photonindex = (\injindex+1)/2=1.5$ and a cut-off position at 
\begin{equation}
    \Eg^{\mathrm{syn,max},p} \approx \Gamma \frac{B}{\Bcp} \frac{\qty(\Epmaxp)^2}{m_p c^2} \approx 10^{13}~\eV  \; ,
\end{equation}
in agreement with the peak energy in \reffig\ref{fig:psyn_scenario}.
The presence of UHE protons requires comparably strong magnetic fields (here $\approx$~30~G), leading to either large upstream density requirements (highlighted here in the table), Lorentz factors, or values of $\varepsilon_B$ (which is bound to be smaller than 1). This large magnetic field is also needed to cool the primary electrons to yield a flat synchrotron spectrum down to the \xrayband band.

Besides the synchrotron radiation, the UHE protons also induce a secondary cascade via photo-pion production on the target field provided by the electron synchrotron component. Approximating that the neutral pions obtain about 20\% of the maximum proton energy and then decay into two photons with half of that energy results in the green $\pi^0\to \gamma \gamma$ peak at $\Eg^{\pi^0\to\gamma\gamma}\approx \Gamma \Epmaxp / 10 \approx 10^{20}~\eV$. 

The charged pions predominantly decay into muons and neutrinos, as can be seen from the pion synchrotron cooling curve only being dominant above about $3\times 10^{18}~\eV$.
The resulting muons decay slower than the pions, such that they are affected stronger by synchrotron cooling, roughly by one order of magnitude down to energies of $\Emup \approx 10^{17}~\eV$ (compare dominant timescale highlighted in pink). They subsequently decay further into neutrinos with a complex spectral shape consisting of a plateau, a peak and a cut-off at $E_\nu^{\mathrm{max}}$. We can estimate roughly $E_\nu^{\mathrm{max}} \approx \Gamma \Epmaxp / 20 \approx 10^{19}~\eV$ and a peak position originating from the peak in the muon spectrum at $E_\nu \approx \Gamma \times 10^{17}~\eV / 3 \approx 2\times10^{18}~\eV$. The low energy plateau is dominated by neutrinos from $pp$-interactions, with the flat shape due to the flat proton spectrum. We note that the neutrino flux is below the current and planned detector limits; see \refap\ref{ap:neutrinos} for details.

The synchrotron radiation for both charged pions and muons is also shown in the top panel of \reffig\ref{fig:psyn_scenario}. We can estimate the peak frequencies from the peak positions of the charged pion and muon spectra at $E_\pi^{\prime\mathrm{peak}}\approx 3\times10^{18}~\eV$ and $E_\mu^{\prime\mathrm{peak}}\approx 10^{17}~\eV$, similar to the proton case by replacing $\Bcpi\approx 3\times 10^{18}~\mathrm{G}$, $\Bcmu\approx 2\times 10^{18} ~\mathrm{G}$. This is in good agreement with the pion synchrotron peak at around $\Eg^{\mathrm{syn,max},\pi} \approx 3\times 10^{13} ~\eV$ and the slightly cooling-broadened muon peak observed at $10^{12}-10^{13}~\eV$.

Furthermore, the muons also decay into secondary electrons and positrons, whose synchrotron and inverse Compton emission is also shown in the SED ($p\gamma$ casc. $e^\pm$, light-blue). This component follows qualitatively the power-law shape of the cooled muons, but softer by a factor of $\Eep^2$ due to the energy dependent decay rate and synchrotron cooling (see also \reffig\ref{fig:psyn_particles}), leading to a relevant contribution in the range of $\Eg \approx 10^{15}-10^{18}~\eV$. As can be seen from the photon loss timescale for $\gamma \gamma \to e^- e^+$ pair annihilation, at these energies, the feedback into secondary pairs is very efficient, triggering the electromagnetic cascade that dominates the photon spectrum in this energy range ($\gamma \gamma \to e^\pm$, purple). We can also see that the pair annihilation does not affect the proton synchrotron component sufficiently to impact the shape in the \vheband band. Inspecting the $\gamma \gamma \to e^- e^+$ induced cascade emission curve in the SED, we can additionally see two other features. 
The broad peak in the observed MeV range is due to the synchrotron emission radiated by secondary pairs, produced through $\gamma \gamma$ annihilation of primary electron synchrotron photons with the proton synchrotron photons. At the highest energies (up to $\Eg \approx 10^{20}~\eV$), the $\gamma \gamma \to e^- e^+$ cascade is instead triggered by the $\pi^0\to \gamma \gamma $ peak, and suppressed in the quantum synchrotron regime (\textit{Klein-Nishina} cut-off for synchrotron radiation).

Finally, we also see a subdominant component of the secondary electrons from the $pp$ induced cascade, which follows the shape of the parent proton spectrum (power-law with spectral index 2 in this case).

In light of our search for extended and flat spectra, we find that by tweaking the parameters (mainly $\Gamma, n, \varepsilon_B, \varepsilon_p$) towards high magnetic fields (here $\approx$~30~G), the proton synchrotron component can be fine-tuned to ensure that the total SED is approximately flat up into the \vheband band. However, similar to the \ssc scenario, the proton synchrotron component is not genuinely creating flat spectra in the \vheband band due to the presence of the exponential cut-off. This implies that a certain level of fine-tuning is required in order to ensure that the energy of the peak, as well as its flux level, are both at the right positions. It should also be noted that the proton cut-off is likely more complex than in our simplified exponential cut-off due to the transition from adiabatic cooling to free streaming escape.

We additionally note that this scenario is able to create flat spectra in our slightly tilted cases ($\Delta \photonindex \gtrsim 0$) too.

We conclude that in a situation of large magnetic fields and efficient proton acceleration, the proton synchrotron component can dominate the emission, but its exponential cut-off struggles to genuinely explain the observed flatness of the power-law spectra.

\subsection{The $pp$ cascade scenario} \label{sec:ppcascade}

\begin{figure*}
    \centering
    \includegraphics[width=\linewidth]{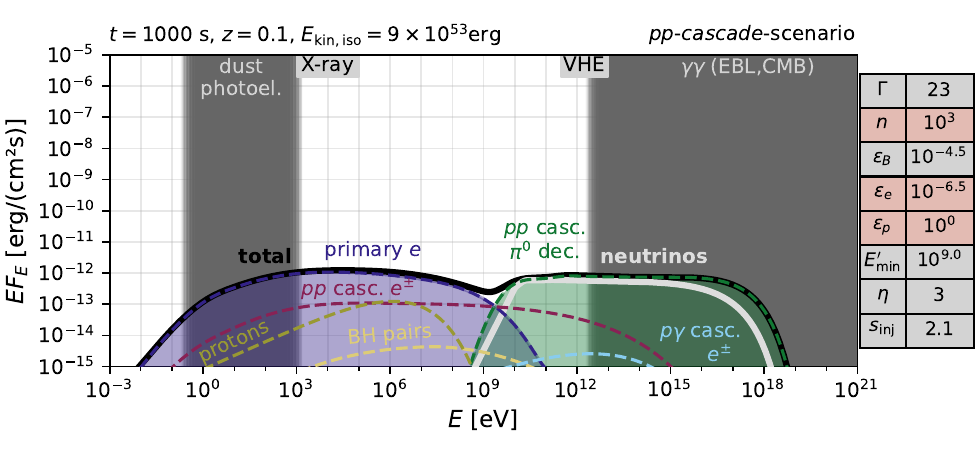}
    \includegraphics[width=0.9\linewidth]{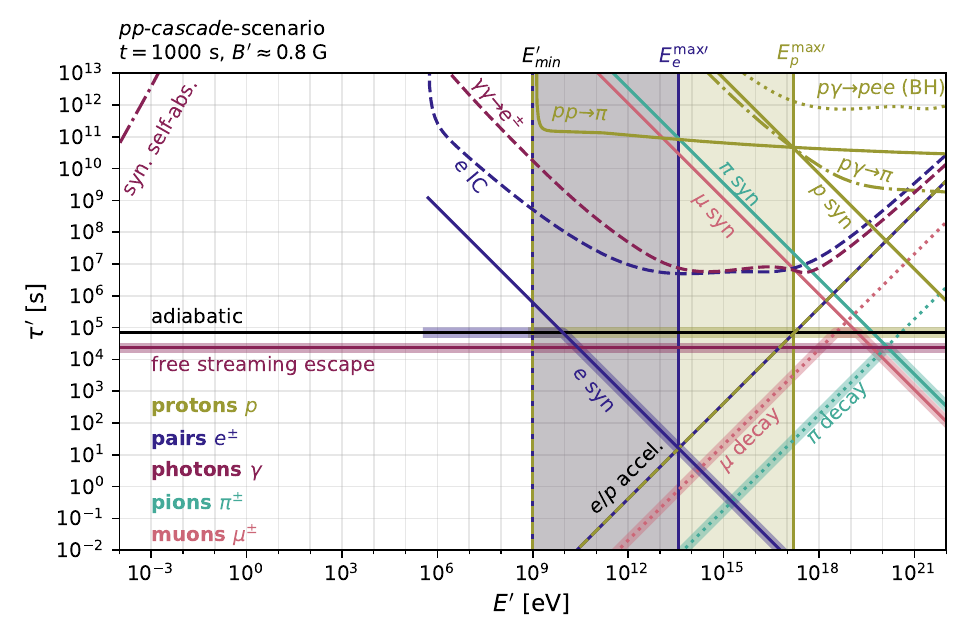}
    \caption{\textbf{\ppcasc scenario (\timedep method) --- Top:} Observed energy flux as a function of observed energy for a representative parameter choice (for a detailed description of the figure see \reffig\ref{fig:SED_SSC} and \ref{fig:psyn_scenario}, note $n$ in cm$^{-3}$ and $\Eminp$ in eV). We highlighted the density $n$ and the parameters $\varepsilon_e, \varepsilon_p$ due to their extreme combination of values.\\
    \textbf{Bottom:} Comoving timescales (for a detailed description of the figure, see \reffig\ref{fig:SSC_timescales} and \ref{fig:psyn_scenario}).}
    \label{fig:ppcascade_scenario}
\end{figure*}

In this scenario, the \vheband band is dominated by the $\pi^0$ decays from $pp$ interactions, whereas the \xrayband band is dominated by the synchrotron radiation from primary electrons. \reffig\ref{fig:ppcascade_scenario} shows the SED and timescale for a representative set of parameters. The constant energy flux range across all SED figures helps to highlight a limitation of this scenario. Despite the rather high upstream density of $n=10^3 ~\mathrm{cm}^{-3}$ (highlighted next to plot), the total flux level is noted to be rather dim\footnote{This is consistent with the intuition that $pp$ interactions are typically important in extended sources, such as star-forming/burst galaxies or galaxy clusters \citep[e.g.,][]{2015ApJ...806...24S,2018ApJ...857...50Y}.}.

At low energies, the primary electron synchrotron component is in the slow cooling regime and can be understood as in the \ssc scenario.
The proton synchrotron component in this scenario is subdominant, peaking at around $E_\gamma^{\mathrm{syn,max},p}\approx 10^6~\eV$. This is consistent with the \psyn scenario, where the magnetic field is higher and therefore the adiabatic-cooling limited maximum energy too (note also that $\eta=3$).

More relevant in this scenario are the components from the $pp$-cascade. In particular, the emission observed above $\approx 1~\mathrm{GeV}$ originates from the decay of the produced, neutral pions. The turn-on energy is related to the lowest energy protons and limited to the threshold at the pion mass $E_\gamma^{\pi^0, \mathrm{turn-on}} \approx \Gamma m_{\pi^0} c^2 / 2 \approx 10^{9}~\eV$. Since the protons cool dominantly adiabatically, their flat spectrum also yields a flat photon spectrum above the turn-on energy. Especially, this extends up to $E_\gamma^{\pi^0, \mathrm{max}} \approx \Gamma \Epmaxp/10\approx 10^{18}~\eV$. This is an interesting feature in the context of the recent flat VHE observations. 

It is worth noting that $pp$-interactions are less efficient compared to the electron synchrotron mechanism, such that a large baryonic loading $\varepsilon_p/\varepsilon_e \approx 10^6$ (highlighted next to plot) is needed for both components to be at a similar energy flux level. Our tilted criterion ($\Delta \photonindex>0$, see \reffig\ref{fig:conceptual_sketch}) only mildly relaxes this extreme requirement. We note that very large baryonic loadings $\gtrsim 10^4$ are frequently considered in the modelling of AGN blazars in the context of neutrino observations; see, e.g., \citet{AM3_paper,2021ApJ...912...54R}. 
Compared to the other scenarios of similar $\Ekiniso$, the inefficiency of the $pp$-interactions also leads to an overall lower energy flux level as dictated by the $pp \to \pi^0 \to \gamma \gamma$ component. Conceptually the flux level could be increased to higher values, but only at the cost of extreme values of $\Ekiniso \gg 10^{55}~\mathrm{erg}$ and only without the $p \gamma$-cascade becoming dominant in the \vheband band (discussed in \refsec \ref{sec:pgammacascade}).

In addition, the synchrotron radiation from secondary electrons/positrons generated from the $pp$-cascade is shown in the SED figure ($pp$-casc. $e^\pm$). In principle, the secondary injection follows the parent proton spectral index, such that efficient electron cooling (see timescales) leads to a flat synchrotron spectrum. The efficient cooling also distributes the energy over a large range of energies ($>10$ orders of magnitude), making this component subdominant.

Lastly, the secondary neutrino spectrum follows the parent proton spectrum approximately, similar to that of the other secondaries, with a slightly lower maximum energy. The neutrino flux is below detection limits (see \refap\ref{ap:neutrinos}).

We also note, that the extreme value of $\varepsilon_p = 1$, relevant for the brightness of this scenario, is already in conflict with the assumption that the non-thermal proton pressure downstream is negligible in the hydrodynamic shock approximation.

Furthermore, we point out that the density of material in the vicinity of GRB explosions is hard to constrain observationally. However, for environments with high star formation, such as molecular clouds, densities of $10^3~\mathrm{cm}^{-3}$ are not uncommon. 

We conclude that a combination of the primary electron synchrotron and the $pp$-cascade induced $\pi^0 \to \gamma \gamma$ emission can yield flat photon spectra when extremely large values of the baryonic loading ($\varepsilon_p/\varepsilon_e \approx 10^6$) and high densities are chosen. We highlight the natural emergence of a hard spectral component extending significantly beyond TeV energies allowed by this scenario despite the comparably dim energy fluxes due to the inefficient $pp$-interactions.

\subsection{The $p\gamma$ cascade scenario} \label{sec:pgammacascade}

\begin{figure*}
    \centering
    \includegraphics[width=\linewidth]{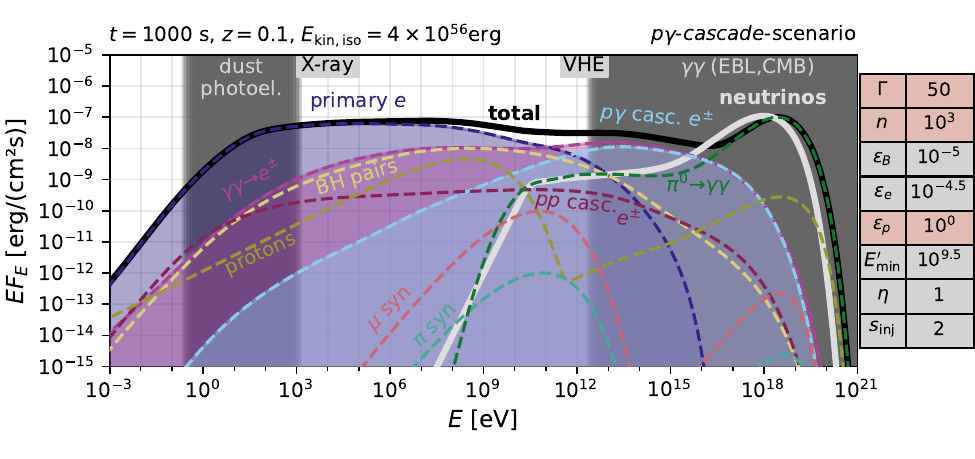}
    \includegraphics[width=0.9\linewidth]{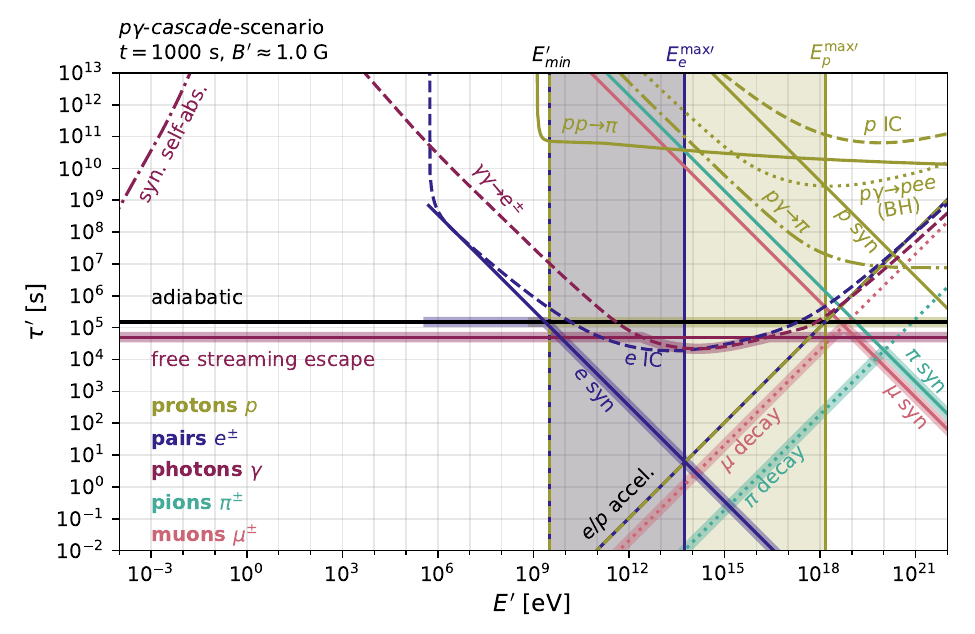}
    \caption{\textbf{\pgcasc scenario (\timedep method) --- Top:} Observed energy flux as a function of observed energy for a representative parameter choice (for a detailed description of the figure see \reffig\ref{fig:SED_SSC} and \ref{fig:psyn_scenario}, note $n$ in cm$^{-3}$ and $\Eminp$ in eV). We highlighted the density $n$, the shock Lorentz factor $\Gamma$, and the parameter $\varepsilon_p$ due to their extreme combination of values.\\
    \textbf{Bottom:} Comoving timescales (for a detailed description of the figure, see \reffig\ref{fig:SSC_timescales} and \ref{fig:psyn_scenario}).}
    \label{fig:pgammacascade_scenario}
\end{figure*}

We finally show a scenario in which the photo-pion-induced cascade synchrotron emission dominates the \vheband band. The SED and timescales are shown in \reffig\ref{fig:pgammacascade_scenario}.

Again, the \xrayband band is dominated by synchrotron emission from primary injected electrons, and the cooling regimes can be understood analogously to the \extsyn scenario. 

The SED at higher energies is dominated by the different emission components from the $ p\gamma$ cascade. They can be understood in a similar manner to the \psyn scenario. Also here, the peak energy of the neutral pion bump can be estimated to $\Eg^{\pi^0\to\gamma\gamma}\approx \Gamma \Epmaxp / 10 \approx 10^{19}~\eV$. Due to the rather abrupt turn-on of the fast cooling synchrotron spectrum at $\Eg \approx 3\times 10^2~\eV$, the $\pi^0 \to \gamma \gamma$ bump does not trigger an efficient pair annihilation cascade.

Different to the \psyn scenario, here, the lower magnetic field $B'\approx 1~\mathrm{G}$ results in the secondary muons and pions decaying before cooling significantly via synchrotron emission. We can understand the shape of the resulting $p\gamma$ cascade electron (positron) synchrotron component by simple scaling arguments. In the $\delta$-approximation, the source term of pions roughly scales as 
\begin{equation}
    q_{E,\pi}' \propto \frac{n_{E,p}'}{t_{p\gamma}'} \propto \frac{q_{E,p}' \tadi}{t_{p\gamma}'} \propto \qty(E_\pi')^{-\injindex+1} \; ,
\end{equation}
since $t_{p\gamma}' \propto 1/\Epp$ in the relevant energies (flat photon target, compare timescale plot). The pions and muons directly decay ($q_{E,e}' \approx q_{E,\mu}' \approx q_{E,\pi}'$), such that the electron spectrum is approximately
\begin{equation}
    \hat{n}_{E,e}' \propto \tsyn \hat{q}_{E,\pi}'  \propto \qty(\Eep)^{-\injindex} \; ,
\end{equation}
since the electrons are fast cooling. This results in the rising synchrotron photon source term with photon index $\photonindex = (\injindex+1)/2 =1.5$ in this scenario. As can be seen from the photon timescale, the energy-independent free streaming escape dominates up to $\Egp\approx 10^{12}~\eV$, with a break feature from the transition to $\gamma \gamma \to e^\pm$ annihilation. This feature imprints itself as a softening in the photon spectrum above $\Eg \approx 10^{13}~\eV$, which results in the effective flat photon SED above TeV energies.
This annihilation subsequently triggers an electromagnetic cascade ($\gamma \gamma \to e^\pm$ component in SED plot), which propagates with an approximately flat spectrum down to at least the MeV energy range.

Due to the high downstream density, the sub-dominant radiating particle components (protons, pions and muons), show besides their synchrotron peaks, an additional (sub-dominant) IC peak.

Compared to the other scenarios, the neutrino flux in this scenario is highest, although only touching the sensitivity of future instruments (see \refap\ref{ap:neutrinos} for further details).

We note that similar parameters can be found to yield slightly softer spectra, fulfilling our selection criteria for $\Delta \photonindex \gtrsim0$.

The difference in the \pgcasc scenario compared to the others is, in particular, the comparatively large energy requirement, coming from the combination of high circum-burst densities and high $\Gamma$. We discuss the feasibility of these conditions in \refsec\ref{sec:energetics}.
We note that the extreme assumption of $\varepsilon_p=1$ (also highlighted next to the plot) is already too high to be compatible with a purely hydrodynamical shock. 
Additionally, the baryonic loading, $\varepsilon_p/\varepsilon_e = 10^{4.5}$, is rather extreme but comparable to that considered in AGN blazar models (see earlier discussion for \ppcasc scenario). This is required to compensate for the low $p\gamma$ efficiency, such that the proton population gives only a small fraction of the energy to the pion population (not calorimetric).

We conclude that this scenario can result in a flat spectrum extending up to energies $\gg$~TeV but is energetically more challenging than the other scenarios that we found.

\section{Discussion} \label{sec:discussion}

\subsection{Limitations of our Results} \label{sec:limitations} 

In this paper, we seek to reproduce the observed spectral properties of GRB afterglows, namely flat (i.e. $\photonindex\approx 2-2.2$) spectra in our \xrayband (keV) and \vheband (TeV) energy bands, that additionally connect to a single power-law, in a single-zone, lepto-hadronic, relativistic-shock model. We provide a robust, comprehensive, and conceptual answer to this question that is based on a large parameter scan. We find five different families of solutions. However, none of these solutions appears to satisfy all the above model selection criteria in a convincing and exclusive manner. In the following, we discuss the range of applicability of our results.

\paragraph{Modelling accuracy} A limitation of our method is the accuracy of our results, which we demonstrate hold to $\Ounity$. However, we consider this sufficient to identify different scenario types, given the current level of uncertainties in both the observations and the modelling. We remind the reader of the uncertain hydrodynamic treatment of the blast wave and relativistic shock, neglecting any feedback from magnetic fields and non-thermal particles, as well as the uncertain treatment of the particle acceleration to a simple power-law injection spectrum.

\paragraph{Thermal particles} The description of the bulk thermal particle population and its transition to the non-thermal population at the injection (energy) scale are still subject to large uncertainty \citep{WarrenEtAl2018}. In light of current particle-in-cell (PIC) simulations a prominent\footnote{It should be noted that a hydrodynamic shock treatment is valid only for $\varepsilon\st{thermal} \gg \varepsilon_{e}, \varepsilon_{p}$, in which case the thermal peak sticks out clearly over the non-thermal power-law.} thermal peak would be expected in the proton and electron (and other secondary) spectral energy distributions \citep[e.g.,][]{MarcowithEtAl2016}. Besides the direct effect of signatures in the synchrotron spectrum at lower energies, this thermal particle population would also impact the SSC spectrum \citep{WarrenEtAl_SSC22}, as well as the proton target density of this paper's $pp$-cascade scenario. An exploration of these signatures is beyond the scope of this paper.

\paragraph{Surrounding density profile} As a representative situation, we focus our work on the case of a constant density profile for the surrounding gas. However, even for a steeper wind density profile, for which $n\propto r_\star^{-2}$, the recently swept-up volume scales as $\approx r_\star^3$, such that the spectra are still expected to be dominated by the most recently injected particles (as they are for the constant density case). It is, therefore, expected that the emission spectra are only affected at the $\Ounity$ level by such a modification to the injection history. We leave a more detailed investigation in this direction for future work.

\paragraph{Observed time} We focus on $t=1000$~s as a representative time of \vheband observations of GRB afterglows. However, the observed time only impacts the Lorentz factor $\Gamma \propto t^{-(3-w)/(8-2w)}$, and the density $n \propto r_\star^{-w} \propto t^{-w/(4-w)}$ \citep{Zhang_GRBbook}. For values of $0\leq w \leq 2$, the scaling indices are at most 1 (or rather -1). This changes the location of our scenarios in the parameter space moderately, but we would expect no new types of solutions to arise from a change in this observation time focus.

\paragraph{Parameter scan and model selection} We performed a grid scan of the parameter space, exploring a finite volume with finite grid spacing. Given the large covered volume of the parameter space, it seems reasonable to consider it sufficiently exhaustive. The limited grid resolution and our simplified model selection process do not impact our qualitative conclusions, which focus on possible scenarios.

\subsection{Energetics} \label{sec:energetics} 
In terms of required energy, in particular the \pgcasc scenario's combination of $\Gamma=50$ and $n=10^3~\mathrm{cm}^{-3}$ requires at $t=10^3~\mathrm{s}$ a seemingly large $\Ekiniso \approx 4\times 10^{56}~\mathrm{erg}$ compared to observed photon fluences. 
However, from a theoretical perspective, we place this in the context of the energy released by the core collapse of a massive rotating star by making the following simplifying assumptions: From the accreted mass $M\approx 10 M_\odot$ a fraction $\varepsilon\st{kin}\approx 0.1$ can be converted to kinetic energy, which is directed into a cone with opening angle $\theta\approx 3^{\circ}$, giving:
\begin{equation}
    \Ekiniso \approx 10^{57}~\mathrm{erg} \qty(\frac{M}{10M_\odot}) \qty(\frac{\varepsilon\st{kin}}{0.1}) \qty(\frac{3^{\circ}}{\theta})^2  \; .
\end{equation}
This estimate aligns with findings of values up to $\Ekiniso \approx  10^{56}~\mathrm{erg}$ in prompt emission internal shock models if GRBs are to power the UHECRs  \citep{2020MNRAS.498.5990H}, for which even larger values of $\Ekiniso$ are still consistent given the uncertainties in the energy dissipation efficiencies in the prompt phase.
Furthermore, if the engine is a newly formed accreting black hole or magnetar, the rotational energy that can be extracted from the system can be estimated as $\Ekiniso \simeq 10^{56}~\mathrm{erg}$ for a solar-mass black hole and a jet opening angle of $3.5^\circ$, see the discussion in~\citet{2023ApJ...944L..34R}. We conclude that the energetic requirements to make hadronic components relevant to the SED are optimistic but not impossible.

\subsection{Ultra-High-Energy Cosmic Rays} \label{sec:uhecrs}
In particular, the \psyn and the \pgcasc scenarios lead to the creation of UHECRs above observed energies of $E_p \approx 10^{18}~\eV$. One can assume that a fraction of the highest-energy particles is no longer efficiently linked to the plasma and, instead of adiabatically cooling, escapes from the blast wave at a similar timescale $\tesc$. One may especially expect that UHECRs can efficiently escape as soon as the Larmor radius becomes comparable to the size of the region; see, e.g., the discussion in \citet{2013ApJ...768..186B}. We note that this transition could lead to a spectral signature for the protons at the highest energies and, therefore, also for the proton-synchrotron photons and secondary cascade particles.

We can parameterise this escape with an uncertain efficiency factor $\varepsilon\st{esc} \lesssim 1$ and a bolometric factor $f_{\mathrm{bol}} \lesssim 1$ that depends on the energy range where escape dominates. In the upstream or progenitor rest frame, we estimate the total injected energy into the UHECR population simply as 
\begin{equation}
    \Euhecriso \approx \varepsilon\st{esc} \, f_{\mathrm{bol}}  \, \varepsilon_p \, \Ekiniso  \; .
\end{equation}
We highlight that the isotropic equivalent energy is considered here since GRB jets not pointing at us also contribute to the UHECR flux at Earth. 

The required energy output in UHECRs per GRB is about $\Euhecriso \simeq 10^{53} \, \mathrm{erg}$ if GRBs are to be the sources of the UHECRs (see \citet{2015APh....62...66B} for a more detailed discussion). Therefore, if all GRBs were alike our prototype and happened at a rate of $1~\mathrm{Gpc}^{-3}\mathrm{yr}^{-1}$, GRBs could comfortably power the UHECRs in the \psyn and \pgcasc scenarios. 

However, the assumption $\varepsilon_p \simeq 1$, suggesting that most of the kinetic energy would be dissipated into UHECRs, requires further investigation as of feedback on the background thermal shock assumed.
Furthermore, we note that current observations favour a UHECR composition including masses larger than just pure proton as simplified in this work \citep[e.g.,][]{2019ApJ...873...88H}.

\subsection{Model Distinction}

The presented scenarios are, by construction, hard to distinguish in the keV--TeV energy range. Full observational coverage of the keV--TeV range remains still crucial, and contemporaneous data is needed to confirm the emerging single power-law nature of the spectrum at a statistically significant level. 

The scenarios which we find could be distinguished via an accurate measurement of the photon index at the highest observable energies ($\approx 10~$TeV), where the \ssc and \psyn scenarios are very soft ($\photonindex \gg 2$), whereas the other models continue hard ($\photonindex\approx 2$). However, the EBL absorption would allow the observation of a GRB at such high energies only for a very close event ($z\lesssim0.01$).

In principle, an additional distinction could be based on the neutrino flux (high in \pgcasc scenario), although this seems challenging with even future instruments (compare \reffig\ref{fig:neutrino_fluence}).

Besides their spectra and possible neutrino signatures, the other observational dimension of GRB afterglows is their evolution with time. In principle, the temporal evolution of the relative ratio between low (electron synchrotron) and high-energy components (differs per scenario) is another way of distinguishing between these scenarios. However, the deceleration of the emitting blast wave mixes the temporal evolution of the flux at a fixed emitted energy with an additional drift of the spectrum towards lower energies. A robust determination of the scaling of both effects is challenging due to the uncertainties in the density profile around the GRB and, observational, the fast decaying light curves (temporal power-law index $\alpha\lesssim -1$, with $EF_E\propto t^\alpha$).

\section{Conclusions} \label{sec:conclusions}

The emerging observational picture of VHE detected GRBs (especially GRB~190114C, GRB~190829A and GRB~221009A) indicates a flat (i.e. $\photonindex\approx 2-2.2$), single power-law spectrum from keV to TeV energies, extending even beyond 10~TeV for GRB~221009A. 
In this paper, we perform a systematic exploration of the possibilities of a one-zone, lepto-hadronic, relativistic-shock model to reproduce this behaviour. 

We find five possible scenarios (see \reftab\ref{tab:summary}): \ssc, \extsyn, \psyn, \ppcasc and \pgcasc. No scenario matches all of the observational criteria in a convincing and exclusive manner.
While for all of the scenarios, the energy flux levels at keV and TeV energies are comparable, the modelled photon index at TeV energies does not reproduce the observed hard values ($\photonindex\approx 2-2.2$) in the \ssc and \psyn scenarios, which exhibit significant curvature. We find that this is different for the \ppcasc and \pgcasc scenarios. These are embedded in higher density environments ($n\approx 10^2-10^4~\mathrm{cm}^{-3}$), such as typical for molecular clouds. Finally, the \extsyn scenario also reproduces flat spectra to the highest energies in a plausible and simple manner, although the small value of $\eta$ needed needs justifying (e.g., in an effective two-zone model).

From the observational perspective, we find the need for improved coordinated multi-wavelength observations of the afterglow phase. 
Specifically, contemporaneous spectral coverage, logarithmically sampled in time and energy, is required to unveil the nature of the VHE emission. Future observational focus should be given: 1) to the keV-TeV range to test the flatness hypothesis; 2) to the highest energies (tens of TeV) to test the origin of the VHE emission. We find the power of neutrino constraints to be limited.

\begin{acknowledgements}
    The authors would like to thank Pavlo Plotko for helpful input on the grid scan using the \texttt{PriNCe analysis tool} and Sylvia Zhu for helpful feedback on the paper. This work was supported by the International Helmholtz-Weizmann Research School for Multimessenger Astronomy, largely funded through the Initiative and Networking Fund of the Helmholtz Association. All figures by the authors under a \href{https://creativecommons.org/licenses/by/4.0/}{CC BY 4.0 license}.
\end{acknowledgements}

\software{NumPy \citep{NumPy_ref}, Matplotlib \citep{Matplotlib_ref}, Astropy \citep{Astropy_ref}, pybind11 \citep{pybind11}, Eigen \citep{eigenweb}, AM$^3$} \citep{AM3_paper}, PriNCe analysis tool \citep{PrinceAnalysisTools}, color schemes \citep{colorSchemes}

\FloatBarrier
\appendix
\section{Quasi-steady-state approximation for hadronic scenarios} \label{ap:qss_approx}

\begin{figure}
    \centering
    \includegraphics[width=0.8\linewidth]{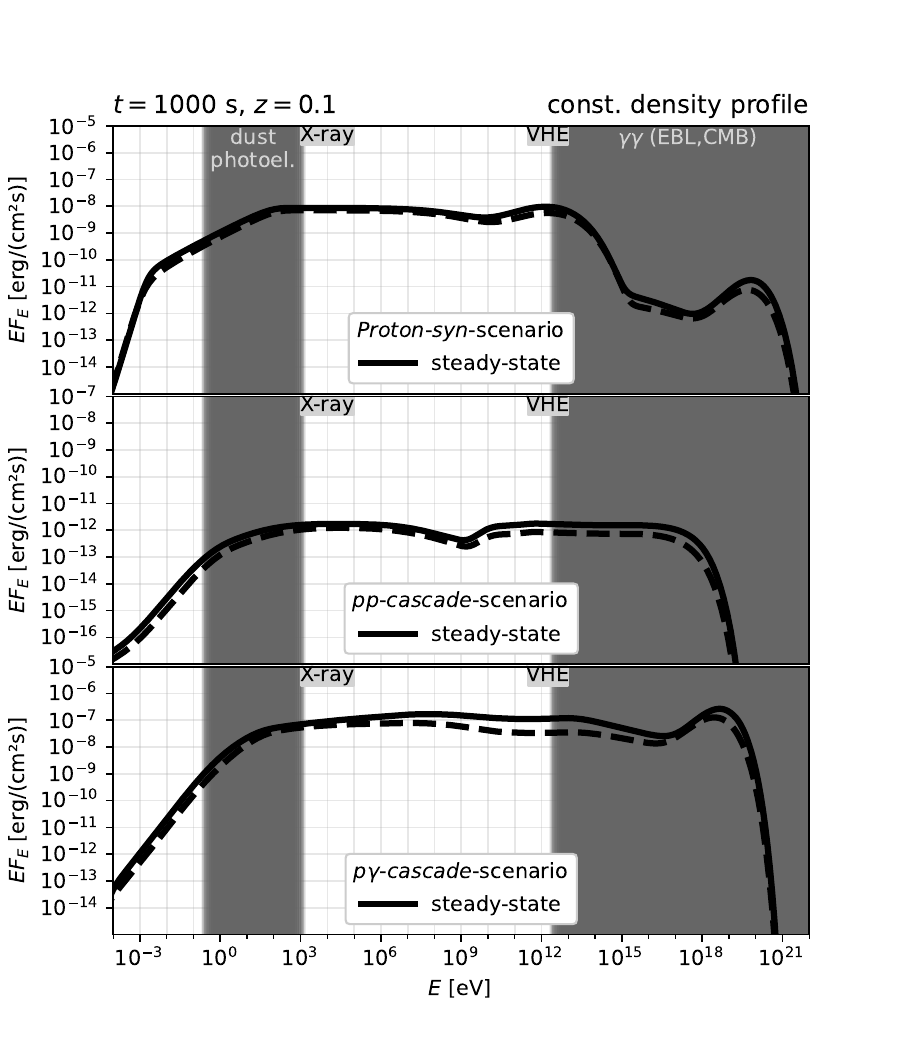}
    \caption{Comparison of the SEDs for the hadronic scenarios in the \steadystate and \timedep method, analogous to \reffig\ref{fig:QSS_comparison}.}
    \label{fig:qss_approx_had}
\end{figure}

We present in \reffig\ref{fig:qss_approx_had} a comparison of the \timedep and \steadystate methods applied to our three hadronic cases. Both methods agree also for these cases well up to $\Ounity$. We note that for the \extsyn scenario, the curves agree similarly well as the synchrotron components in all scenarios; see, e.g., \reffig\ref{fig:QSS_comparison}.

\FloatBarrier
\section{Exemplary particle distribution for \psyn scenario} \label{ap:psyn_particles}
\begin{figure*}
    \centering
    \includegraphics[width=0.7\linewidth]{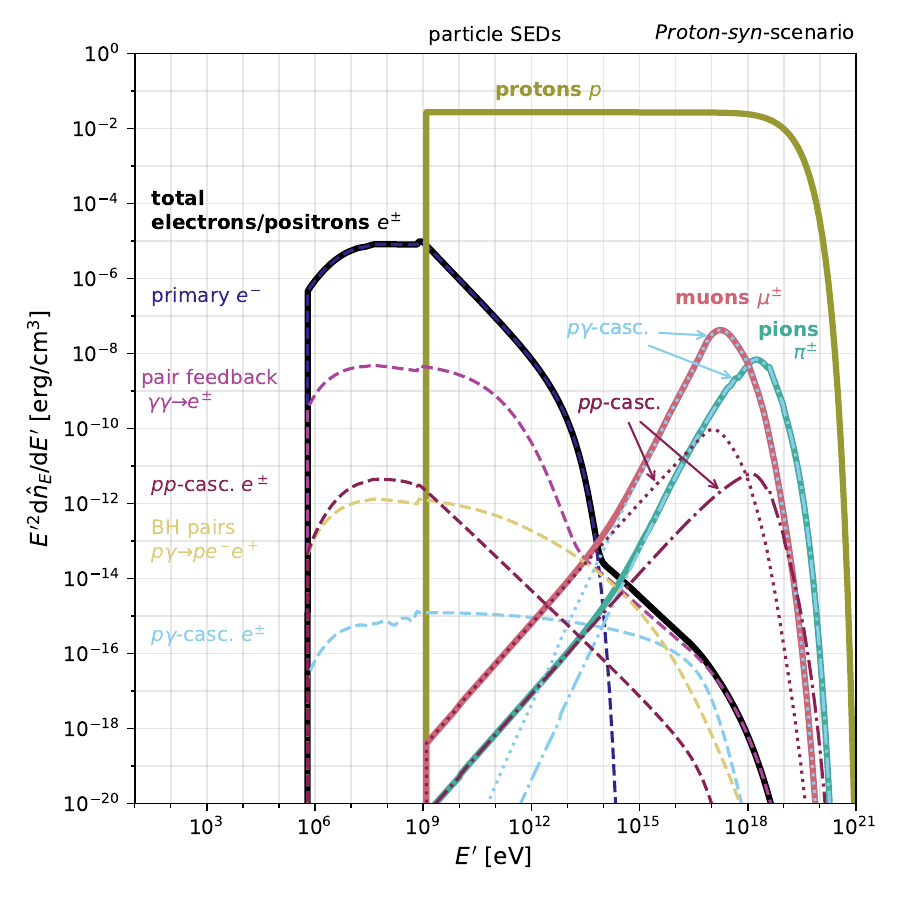}
    \caption{Comoving particle distributions for protons, electrons (including positrons), pions and muons for the \psyn scenario at observed time $t=1000~\mathrm{s}$.}
    \label{fig:psyn_particles}
\end{figure*}

As an example, and for intuition, we give in \reffig\ref{fig:psyn_particles} the comoving energy spectra of the radiating particles at $t=1000~\mathrm{s}$, belonging to the \psyn scenario in \refsec\ref{sec:psyn}.

\FloatBarrier
\section{Neutrino fluences} \label{ap:neutrinos}
\begin{figure}
    \centering
    \includegraphics[width=0.7\linewidth]{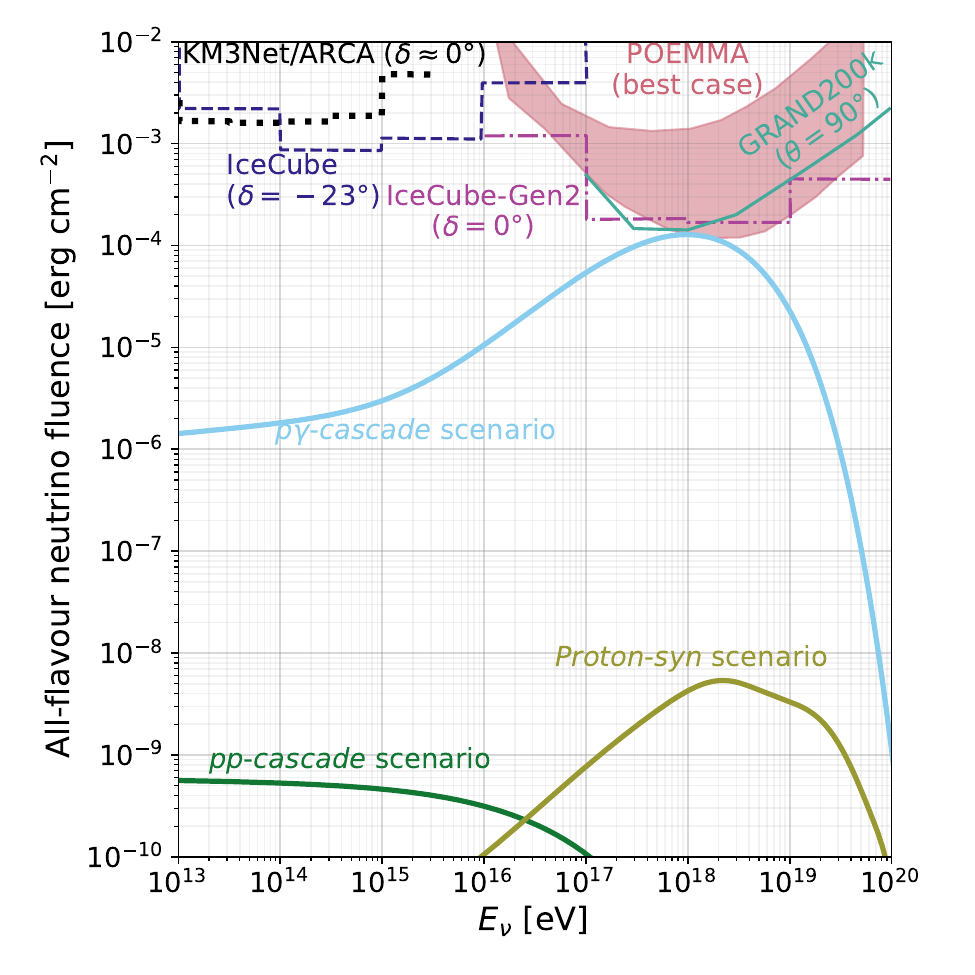}
    \caption{All-flavor ($\nu + \bar{\nu}$) neutrino fluences of the \psyn, \ppcasc and \pgcasc scenarios approximated by the product of the neutrino energy flux $E_\nu F_{E_\nu}$ multiplied with the observed time $t=10^3~\mathrm{s}$. For comparison, we show the 
 90\% CL sensitivity curves of current and future neutrino detectors, such as IceCube, IceCube-Gen2 \citep{IceCube-Gen2:2021rkf}, GRAND \citep{2020SCPMA..6319501A}, POEMMA \citep{2020PhRvD.102l3013V}, and KM3NeT/ARCA230 \citep{KM3NeTARCA_sensitivity}.
    \label{fig:neutrino_fluence}}
\end{figure}

It is useful to discuss the detectability of these scenarios with current and future neutrino telescopes. We show in \reffig\ref{fig:neutrino_fluence} a simple approximation for the neutrino fluence, $t \times E_\nu F_{E_\nu}$. The comparison to existing and future detectors shows that the neutrinos for the \psyn and \ppcasc scenarios are far below the sensitivity, whereas the extreme \pgcasc scenario touches the IceCube-Gen2 sensitivity curve, which implies that the neutrino fluences from GRBs with parameters resembling our \pgcasc scenario could reach the detectable level for IceCube-Gen2. However, this conclusion depends on many factors, such as the duration and the distance to the observers.

\FloatBarrier
\bibliography{references}{}

\begin{thebibliography}{}
\expandafter\ifx\csname natexlab\endcsname\relax\def\natexlab#1{#1}\fi
\providecommand{\url}[1]{\href{#1}{#1}}
\providecommand{\dodoi}[1]{doi:~\href{http://doi.org/#1}{\nolinkurl{#1}}}
\providecommand{\doeprint}[1]{\href{http://ascl.net/#1}{\nolinkurl{http://ascl.net/#1}}}
\providecommand{\doarXiv}[1]{\href{https://arxiv.org/abs/#1}{\nolinkurl{https://arxiv.org/abs/#1}}}

\bibitem[{Abbasi {et~al.}(2021)}]{IceCube-Gen2:2021rkf}
Abbasi, R., {et~al.} 2021, PoS, ICRC2021, 1183, \dodoi{10.22323/1.395.1183}

\bibitem[{{Ackermann} {et~al.}(2014){Ackermann}, {Ajello}, {Asano}, {Atwood}, {Axelsson}, {Baldini}, {Ballet}, {Barbiellini}, {Baring}, {Bastieri}, {Bechtol}, {Bellazzini}, {Bissaldi}, {Bonamente}, {Bregeon}, {Brigida}, {Bruel}, {Buehler}, {Burgess}, {Buson}, {Caliandro}, {Cameron}, {Caraveo}, {Cecchi}, {Chaplin}, {Charles}, {Chekhtman}, {Cheung}, {Chiang}, {Chiaro}, {Ciprini}, {Claus}, {Cleveland}, {Cohen-Tanugi}, {Collazzi}, {Cominsky}, {Connaughton}, {Conrad}, {Cutini}, {D'Ammando}, {de Angelis}, {DeKlotz}, {de Palma}, {Dermer}, {Desiante}, {Diekmann}, {Di Venere}, {Drell}, {Drlica-Wagner}, {Favuzzi}, {Fegan}, {Ferrara}, {Finke}, {Fitzpatrick}, {Focke}, {Franckowiak}, {Fukazawa}, {Funk}, {Fusco}, {Gargano}, {Gehrels}, {Germani}, {Gibby}, {Giglietto}, {Giles}, {Giordano}, {Giroletti}, {Godfrey}, {Granot}, {Grenier}, {Grove}, {Gruber}, {Guiriec}, {Hadasch}, {Hanabata}, {Harding}, {Hayashida}, {Hays}, {Horan}, {Hughes}, {Inoue}, {Jogler}, {J{\'o}hannesson}, {Johnson}, {Kawano}, {Kn{\"o}dlseder}, {Kocevski}, {Kuss}, {Lande}, {Larsson}, {Latronico}, {Longo}, {Loparco}, {Lovellette}, {Lubrano}, {Mayer}, {Mazziotta}, {McEnery}, {Michelson}, {Mizuno}, {Moiseev}, {Monzani}, {Moretti}, {Morselli}, {Moskalenko}, {Murgia}, {Nemmen}, {Nuss}, {Ohno}, {Ohsugi}, {Okumura}, {Omodei}, {Orienti}, {Paneque}, {Pelassa}, {Perkins}, {Pesce-Rollins}, {Petrosian}, {Piron}, {Pivato}, {Porter}, {Racusin}, {Rain{\`o}}, {Rando}, {Razzano}, {Razzaque}, {Reimer}, {Reimer}, {Ritz}, {Roth}, {Ryde}, {Sartori}, {Parkinson}, {Scargle}, {Schulz}, {Sgr{\`o}}, {Siskind}, {Sonbas}, {Spandre}, {Spinelli}, {Tajima}, {Takahashi}, {Thayer}, {Thayer}, {Thompson}, {Tibaldo}, {Tinivella}, {Torres}, {Tosti}, {Troja}, {Usher}, {Vandenbroucke}, {Vasileiou}, {Vianello}, {Vitale}, {Winer}, {Wood}, {Yamazaki}, {Younes}, {Yu}, {Zhu}, {Bhat}, {Briggs}, {Byrne}, {Foley}, {Goldstein}, {Jenke}, {Kippen}, {Kouveliotou}, {McBreen}, {Meegan}, {Paciesas}, {Preece}, {Rau}, {Tierney}, {van der Horst}, {von Kienlin}, {Wilson-Hodge}, {Xiong}, {Cusumano}, {La Parola}, \& {Cummings}}]{Fermi_130427_14}
{Ackermann}, M., {Ajello}, M., {Asano}, K., {et~al.} 2014, Science, 343, 42, \dodoi{10.1126/science.1242353}

\bibitem[{{Aiello} {et~al.}(2024){Aiello}, {Albert}, {Alshamsi}, {Alves Garre}, {Aly}, {Ambrosone}, {Ameli}, {Andre}, {Androutsou}, {Anguita}, {Aphecetche}, {Ardid}, {Ardid}, {Atmani}, {Aublin}, {Badaracco}, {Bailly-Salins}, {Barda{\v{c}}ov{\'a}}, {Baret}, {Bariego-Quintana}, {Basegmez du Pree}, {Becherini}, {Bendahman}, {Benfenati}, {Benhassi}, {Benoit}, {Berbee}, {Bertin}, {Biagi}, {Boettcher}, {Bonanno}, {Boumaaza}, {Bouta}, {Bouwhuis}, {Bozza}, {Bozza}, {Br{\^a}nza{\c{s}}}, {Bretaudeau}, {Breuhaus}, {Bruijn}, {Brunner}, {Bruno}, {Buis}, {Buompane}, {Busto}, {Caiffi}, {Calvo}, {Campion}, {Capone}, {Carenini}, {Carretero}, {Cartraud}, {Castaldi}, {Cecchini}, {Celli}, {Cerisy}, {Chabab}, {Chadolias}, {Chen}, {Cherubini}, {Chiarusi}, {Circella}, {Cocimano}, {Coelho}, {Coleiro}, {Coniglione}, {Coyle}, {Creusot}, {Cuttone}, {Dallier}, {Darras}, {De Benedittis}, {De Martino}, {Decoene}, {Del Burgo}, {Del Rosso}, {Di Mauro}, {Di Palma}, {D{\'\i}az}, {Diaz}, {Diego-Tortosa}, {Distefano}, {Domi}, {Donzaud}, {Dornic}, {D{\"o}rr}, {Drakopoulou}, {Drouhin}, {Dvornick{\'y}}, {Eberl}, {Eckerov{\'a}}, {Eddymaoui}, {van Eeden}, {Eff}, {van Eijk}, {El Bojaddaini}, {El Hedri}, {Enzenh{\"o}fer}, {Ferrara}, {Filipovi{\'c}}, {Filippini}, {Franciotti}, {Fusco}, {Gabriel}, {Gagliardini}, {Gal}, {Garc{\'\i}a M{\'e}ndez}, {Garcia Soto}, {Gatius Oliver}, {Gei{\ss}elbrecht}, {Ghaddari}, {Gialanella}, {Gibson}, {Giorgio}, {Goos}, {Goswami}, {Goupilliere}, {Gozzini}, {Gracia}, {Graf}, {Guidi}, {Guillon}, {Guti{\'e}rrez}, {van Haren}, {Heijboer}, {Hekalo}, {Hennig}, {Hern{\'a}ndez-Rey}, {Ibnsalih}, {Illuminati}, {de Jong}, {de Jong}, {Jung}, {Kalaczy{\'n}ski}, {Kalekin}, {Katz}, {Kistauri}, {Kopper}, {Kouchner}, {Kueviakoe}, {Kulikovskiy}, {Kvatadze}, {Labalme}, {Lahmann}, {Larosa}, {Lastoria}, {Lazo}, {Le Stum}, {Lehaut}, {Leonora}, {Lessing}, {Levi}, {Clark}, {Longhitano}, {Magnani}, {Majumdar}, {Malerba}, {Mamedov}, {Ma{\'n}czak}, {Manfreda}, {Manzaneda}, {Marconi}, {Margiotta}, {Marinelli}, {Markou}, {Martin}, {Mart{\'\i}nez-Mora}, {Marzaioli}, {Mastrodicasa}, {Mastroianni}, {Miccich{\`e}}, {Miele}, {Migliozzi}, {Migneco}, {Mitsou}, {Mollo}, {Morales-Gallegos}, {Morga}, {Moussa}, {Mateo}, {Muller}, {Musone}, {Musumeci}, {Navas}, {Nayerhoda}, {Nicolau}, {Nkosi}, {Fearraigh}, {Oliviero}, {Orlando}, {Oukacha}, {Paesani}, {Palacios Gonz{\'a}lez}, {Papalashvili}, {Parisi}, {Gomez}, {P{\u{a}}un}, {P{\u{a}}v{\u{a}}la{\c{s}}}, {Pe{\~n}a Mart{\'\i}nez}, {Perrin-Terrin}, {Perronnel}, {Pestel}, {Pestes}, {Piattelli}, {Poir{\`e}}, {Popa}, {Pradier}, {Prado}, {Pulvirenti}, {Qu{\'e}m{\'e}ner}, {Quiroz-Rangel}, {Rahaman}, {Randazzo}, {Randriatoamanana}, {Razzaque}, {Rea}, {Real}, {Riccobene}, {Robinson}, {Romanov}, {{\v{S}}aina}, {Salesa Greus}, {Samtleben}, {S{\'a}nchez Losa}, {Sanfilippo}, {Sanguineti}, {Santonastaso}, {Santonocito}, {Sapienza}, {Schnabel}, {Schumann}, {Schutte}, {Seneca}, {Sennan}, {Setter}, {Sgura}, {Shanidze}, {Sharma}, {Shitov}, {{\v{S}}imkovic}, {Simonelli}, {Sinopoulou}, {Smirnov}, {Spisso}, {Spurio}, {Stavropoulos}, {{\v{S}}tekl}, {Taiuti}, {Tayalati}, {Thiersen}, {Tosta e Melo}, {Tragia}, {Trocm{\'e}}, {Tsourapis}, {Tudorache}, {Tzamariudaki}, {Vacheret}, {Valer Melchor}, {Valsecchi}, {Van Elewyck}, {Vannoye}, {Vasileiadis}, {Vazquez de Sola}, {Verilhac}, {Veutro}, {Viola}, {Vivolo}, {Wilms}, {de Wolf}, {Yepes-Ramirez}, {Zarpapis}, {Zavatarelli}, {Zegarelli}, {Zito}, {Zornoza}, {Z{\'u}{\~n}iga}, \& {Zywucka}}]{KM3NeTARCA_sensitivity}
{Aiello}, S., {Albert}, A., {Alshamsi}, M., {et~al.} 2024, Astroparticle Physics, 162, 102990, \dodoi{10.1016/j.astropartphys.2024.102990}

\bibitem[{{Ajello} {et~al.}(2018){Ajello}, {Baldini}, {Barbiellini}, {Bastieri}, {Bellazzini}, {Bissaldi}, {Blandford}, {Bonino}, {Bottacini}, {Bregeon}, {Bruel}, {Buehler}, {Cameron}, {Caputo}, {Caraveo}, {Chiaro}, {Ciprini}, {Cohen-Tanugi}, {Costantin}, {D'Ammando}, {de Palma}, {Di Lalla}, {Di Mauro}, {Di Venere}, {Dom{\'\i}nguez}, {Favuzzi}, {Franckowiak}, {Fukazawa}, {Funk}, {Fusco}, {Gargano}, {Gasparrini}, {Giglietto}, {Giordano}, {Giroletti}, {Green}, {Grenier}, {Guiriec}, {Holt}, {Horan}, {J{\'o}hannesson}, {Kocevski}, {Kuss}, {La Mura}, {Larsson}, {Li}, {Longo}, {Loparco}, {Lubrano}, {Magill}, {Maldera}, {Manfreda}, {Mazziotta}, {Michelson}, {Mizuno}, {Monzani}, {Morselli}, {Negro}, {Nuss}, {Omodei}, {Orienti}, {Orlando}, {Paliya}, {Perkins}, {Persic}, {Pesce-Rollins}, {Piron}, {Principe}, {Racusin}, {Rain{\`o}}, {Rando}, {Razzano}, {Razzaque}, {Reimer}, {Reimer}, {Sgr{\`o}}, {Siskind}, {Spandre}, {Spinelli}, {Tak}, {Thayer}, {Torres}, {Tosti}, {Valverde}, {Vogel}, \& {Wood}}]{SwiftFermi18}
{Ajello}, M., {Baldini}, L., {Barbiellini}, G., {et~al.} 2018, \apj, 863, 138, \dodoi{10.3847/1538-4357/aad000}

\bibitem[{{Ajello} {et~al.}(2019){Ajello}, {Arimoto}, {Axelsson}, {Baldini}, {Barbiellini}, {Bastieri}, {Bellazzini}, {Bhat}, {Bissaldi}, {Blandford}, {Bonino}, {Bonnell}, {Bottacini}, {Bregeon}, {Bruel}, {Buehler}, {Cameron}, {Caputo}, {Caraveo}, {Cavazzuti}, {Chen}, {Cheung}, {Chiaro}, {Ciprini}, {Costantin}, {Crnogorcevic}, {Cutini}, {Dainotti}, {D'Ammando}, {de la Torre Luque}, {de Palma}, {Desai}, {Desiante}, {Di Lalla}, {Di Venere}, {Fana Dirirsa}, {Fegan}, {Franckowiak}, {Fukazawa}, {Funk}, {Fusco}, {Gargano}, {Gasparrini}, {Giglietto}, {Giordano}, {Giroletti}, {Green}, {Grenier}, {Grove}, {Guiriec}, {Hays}, {Hewitt}, {Horan}, {J{\'o}hannesson}, {Kocevski}, {Kuss}, {Latronico}, {Li}, {Longo}, {Loparco}, {Lovellette}, {Lubrano}, {Maldera}, {Manfreda}, {Mart{\'\i}-Devesa}, {Mazziotta}, {Mereu}, {Meyer}, {Michelson}, {Mirabal}, {Mitthumsiri}, {Mizuno}, {Monzani}, {Moretti}, {Morselli}, {Moskalenko}, {Negro}, {Nuss}, {Ohno}, {Omodei}, {Orienti}, {Orlando}, {Palatiello}, {Paliya}, {Paneque}, {Persic}, {Pesce-Rollins}, {Petrosian}, {Piron}, {Poolakkil}, {Poon}, {Porter}, {Principe}, {Racusin}, {Rain{\`o}}, {Rando}, {Razzano}, {Razzaque}, {Reimer}, {Reimer}, {Reposeur}, {Ryde}, {Serini}, {Sgr{\`o}}, {Siskind}, {Sonbas}, {Spandre}, {Spinelli}, {Suson}, {Tajima}, {Takahashi}, {Tak}, {Thayer}, {Torres}, {Troja}, {Valverde}, {Veres}, {Vianello}, {von Kienlin}, {Wood}, {Yassine}, {Zhu}, \& {Zimmer}}]{LAT_GRB_cat_19}
{Ajello}, M., {Arimoto}, M., {Axelsson}, M., {et~al.} 2019, \apj, 878, 52, \dodoi{10.3847/1538-4357/ab1d4e}

\bibitem[{{Ajello} {et~al.}(2020){Ajello}, {Arimoto}, {Axelsson}, {Baldini}, {Barbiellini}, {Bastieri}, {Bellazzini}, {Berretta}, {Bissaldi}, {Blandford}, {Bonino}, {Bottacini}, {Bregeon}, {Bruel}, {Buehler}, {Burns}, {Buson}, {Cameron}, {Caputo}, {Caraveo}, {Cavazzuti}, {Chen}, {Chiaro}, {Ciprini}, {Cohen-Tanugi}, {Costantin}, {Cutini}, {D'Ammando}, {DeKlotz}, {de la Torre Luque}, {de Palma}, {Desai}, {Di Lalla}, {Di Venere}, {Fana Dirirsa}, {Fegan}, {Franckowiak}, {Fukazawa}, {Funk}, {Fusco}, {Gargano}, {Gasparrini}, {Giglietto}, {Gill}, {Giordano}, {Giroletti}, {Granot}, {Green}, {Grenier}, {Grondin}, {Guiriec}, {Hays}, {Horan}, {J{\'o}hannesson}, {Kocevski}, {Kovac'evic'}, {Kuss}, {Larsson}, {Latronico}, {Lemoine-Goumard}, {Li}, {Liodakis}, {Longo}, {Loparco}, {Lovellette}, {Lubrano}, {Maldera}, {Malyshev}, {Manfreda}, {Mart{\'\i}-Devesa}, {Mazziotta}, {McEnery}, {Mereu}, {Meyer}, {Michelson}, {Mitthumsiri}, {Mizuno}, {Monzani}, {Moretti}, {Morselli}, {Moskalenko}, {Negro}, {Nuss}, {Omodei}, {Orienti}, {Orlando}, {Palatiello}, {Paliya}, {Paneque}, {Pei}, {Persic}, {Pesce-Rollins}, {Petrosian}, {Piron}, {Poon}, {Porter}, {Principe}, {Racusin}, {Rain{\`o}}, {Rando}, {Rani}, {Razzano}, {Razzaque}, {Reimer}, {Reimer}, {Ryde}, {Saz Parkinson}, {Serini}, {Sgr{\`o}}, {Siskind}, {Spandre}, {Spinelli}, {Tajima}, {Takagi}, {Takahashi}, {Tak}, {Thayer}, {Thompson}, {Torres}, {Troja}, {Valverde}, {Van Klaveren}, {Wood}, {Yassine}, {Zaharijas}, {Mailyan}, {Bhat}, {Briggs}, {Cleveland}, {Giles}, {Goldstein}, {Hui}, {Malacaria}, {Preece}, {Roberts}, {Veres}, {Wilson-Hodge}, {Kienlin}, {Cenko}, {O'Brien}, {Beardmore}, {Lien}, {Osborne}, {Tohuvavohu}, {D'Elia}, {D'A{\`\i}}, {Perri}, {Gropp}, {Klingler}, {Capalbi}, {Tagliaferri}, {Stamatikos}, \& {De Pasquale}}]{FermiSwift_190114_20}
---. 2020, \apj, 890, 9, \dodoi{10.3847/1538-4357/ab5b05}

\bibitem[{{{\'A}lvarez-Mu{\~n}iz} {et~al.}(2020){{\'A}lvarez-Mu{\~n}iz}, {Alves Batista}, {Balagopal V.}, {Bolmont}, {Bustamante}, {Carvalho}, {Charrier}, {Cognard}, {Decoene}, {Denton}, {De Jong}, {De Vries}, {Engel}, {Fang}, {Finley}, {Gabici}, {Gou}, {Gu}, {Gu{\'e}pin}, {Hu}, {Huang}, {Kotera}, {Le Coz}, {Lenain}, {L{\"u}}, {Martineau-Huynh}, {Mostaf{\'a}}, {Mottez}, {Murase}, {Niess}, {Oikonomou}, {Pierog}, {Qian}, {Qin}, {Ran}, {Renault-Tinacci}, {Roth}, {Schr{\"o}der}, {Sch{\"u}ssler}, {Tasse}, {Timmermans}, {Tueros}, {Wu}, {Zarka}, {Zech}, {Zhang}, {Zhang}, {Zhang}, {Zheng}, \& {Zilles}}]{2020SCPMA..6319501A}
{{\'A}lvarez-Mu{\~n}iz}, J., {Alves Batista}, R., {Balagopal V.}, A., {et~al.} 2020, Science China Physics, Mechanics, and Astronomy, 63, 219501, \dodoi{10.1007/s11433-018-9385-7}

\bibitem[{{Asano} {et~al.}(2020){Asano}, {Murase}, \& {Toma}}]{AsanoEtAl2020}
{Asano}, K., {Murase}, K., \& {Toma}, K. 2020, \apj, 905, 105, \dodoi{10.3847/1538-4357/abc82c}

\bibitem[{{Astropy Collaboration} {et~al.}(2022){Astropy Collaboration}, {Price-Whelan}, {Lim}, {Earl}, {Starkman}, {Bradley}, {Shupe}, {Patil}, {Corrales}, {Brasseur}, {N{\"o}the}, {Donath}, {Tollerud}, {Morris}, {Ginsburg}, {Vaher}, {Weaver}, {Tocknell}, {Jamieson}, {van Kerkwijk}, {Robitaille}, {Merry}, {Bachetti}, {G{\"u}nther}, {Aldcroft}, {Alvarado-Montes}, {Archibald}, {B{\'o}di}, {Bapat}, {Barentsen}, {Baz{\'a}n}, {Biswas}, {Boquien}, {Burke}, {Cara}, {Cara}, {Conroy}, {Conseil}, {Craig}, {Cross}, {Cruz}, {D'Eugenio}, {Dencheva}, {Devillepoix}, {Dietrich}, {Eigenbrot}, {Erben}, {Ferreira}, {Foreman-Mackey}, {Fox}, {Freij}, {Garg}, {Geda}, {Glattly}, {Gondhalekar}, {Gordon}, {Grant}, {Greenfield}, {Groener}, {Guest}, {Gurovich}, {Handberg}, {Hart}, {Hatfield-Dodds}, {Homeier}, {Hosseinzadeh}, {Jenness}, {Jones}, {Joseph}, {Kalmbach}, {Karamehmetoglu}, {Ka{\l}uszy{\'n}ski}, {Kelley}, {Kern}, {Kerzendorf}, {Koch}, {Kulumani}, {Lee}, {Ly}, {Ma}, {MacBride}, {Maljaars}, {Muna}, {Murphy}, {Norman}, {O'Steen}, {Oman}, {Pacifici}, {Pascual}, {Pascual-Granado}, {Patil}, {Perren}, {Pickering}, {Rastogi}, {Roulston}, {Ryan}, {Rykoff}, {Sabater}, {Sakurikar}, {Salgado}, {Sanghi}, {Saunders}, {Savchenko}, {Schwardt}, {Seifert-Eckert}, {Shih}, {Jain}, {Shukla}, {Sick}, {Simpson}, {Singanamalla}, {Singer}, {Singhal}, {Sinha}, {Sip{\H{o}}cz}, {Spitler}, {Stansby}, {Streicher}, {{\v{S}}umak}, {Swinbank}, {Taranu}, {Tewary}, {Tremblay}, {de Val-Borro}, {Van Kooten}, {Vasovi{\'c}}, {Verma}, {de Miranda Cardoso}, {Williams}, {Wilson}, {Winkel}, {Wood-Vasey}, {Xue}, {Yoachim}, {Zhang}, {Zonca}, \& {Astropy Project Contributors}}]{Astropy_ref}
{Astropy Collaboration}, {Price-Whelan}, A.~M., {Lim}, P.~L., {et~al.} 2022, \apj, 935, 167, \dodoi{10.3847/1538-4357/ac7c74}

\bibitem[{{Baerwald} {et~al.}(2013){Baerwald}, {Bustamante}, \& {Winter}}]{2013ApJ...768..186B}
{Baerwald}, P., {Bustamante}, M., \& {Winter}, W. 2013, \apj, 768, 186, \dodoi{10.1088/0004-637X/768/2/186}

\bibitem[{{Baerwald} {et~al.}(2015){Baerwald}, {Bustamante}, \& {Winter}}]{2015APh....62...66B}
---. 2015, Astroparticle Physics, 62, 66, \dodoi{10.1016/j.astropartphys.2014.07.007}

\bibitem[{{Bell} {et~al.}(2018){Bell}, {Araudo}, {Matthews}, \& {Blundell}}]{Bell_RelShock18}
{Bell}, A.~R., {Araudo}, A.~T., {Matthews}, J.~H., \& {Blundell}, K.~M. 2018, \mnras, 473, 2364, \dodoi{10.1093/mnras/stx2485}

\bibitem[{{Beloborodov}(2002)}]{Beloborodov_pairloading_02}
{Beloborodov}, A.~M. 2002, \apj, 565, 808, \dodoi{10.1086/324195}

\bibitem[{{Beloborodov} \& {M{\'e}sz{\'a}ros}(2017)}]{BeloborodovMeszaros17}
{Beloborodov}, A.~M., \& {M{\'e}sz{\'a}ros}, P. 2017, \ssr, 207, 87, \dodoi{10.1007/s11214-017-0348-6}

\bibitem[{{Berezinsky} \& {Kalashev}(2016)}]{Berezinsky_cascade_16}
{Berezinsky}, V., \& {Kalashev}, O. 2016, \prd, 94, 023007, \dodoi{10.1103/PhysRevD.94.023007}

\bibitem[{{Blandford} \& {McKee}(1976)}]{BlandfordMcKee1976}
{Blandford}, R.~D., \& {McKee}, C.~F. 1976, Physics of Fluids, 19, 1130, \dodoi{10.1063/1.861619}

\bibitem[{Bohm(1949)}]{Bohm_1949}
Bohm, D. 1949, Qualitative Description of the Arc Plasma in a Magnetic Field.
\newblock \url{https://cir.nii.ac.jp/crid/1570009750450419968}

\bibitem[{{Cao} {et~al.}(2023){Cao}, {Aharonian}, {An}, {Axikegu}, {Bai}, {Bao}, {Bastieri}, {Bi}, {Bi}, {Cai}, {Cao}, {Cao}, {Cao}, {Chang}, {Chang}, {Chen}, {Chen}, {Chen}, {Chen}, {Chen}, {Chen}, {Chen}, {Chen}, {Chen}, {Chen}, {Chen}, {Chen}, {Cheng}, {Cheng}, {Cui}, {Cui}, {Cui}, {Cui}, {Dai}, {Dai}, {Dai}, {Danzengluobu}, {della Volpe}, {Dong}, {Duan}, {Fan}, {Fan}, {Fang}, {Fang}, {Feng}, {Feng}, {Feng}, {Feng}, {Feng}, {Gabici}, {Gao}, {Gao}, {Gao}, {Gao}, {Gao}, {Gao}, {Ge}, {Geng}, {Giacinti}, {Gong}, {Gou}, {Gu}, {Guo}, {Guo}, {Guo}, {Guo}, {Han}, {He}, {He}, {He}, {He}, {He}, {Heller}, {Hor}, {Hou}, {Hou}, {Hou}, {Hu}, {Hu}, {Hu}, {Huang}, {Huang}, {Huang}, {Huang}, {Huang}, {Huang}, {Huang}, {Ji}, {Jia}, {Jia}, {Jiang}, {Jiang}, {Jiang}, {Jin}, {Kang}, {Ke}, {Kuleshov}, {Kurinov}, {Li}, {Li}, {Li}, {Li}, {Li}, {Li}, {Li}, {Li}, {Li}, {Li}, {Li}, {Li}, {Li}, {Li}, {Li}, {Li}, {Li}, {Li}, {Li}, {Liang}, {Liang}, {Lin}, {Liu}, {Liu}, {Liu}, {Liu}, {Liu}, {Liu}, {Liu}, {Liu}, {Liu}, {Liu}, {Liu}, {Liu}, {Liu}, {Liu}, {Lu}, {Luo}, {Lv}, {Ma}, {Ma}, {Ma}, {Mao}, {Min}, {Mitthumsiri}, {Mu}, {Nan}, {Neronov}, {Ou}, {Pang}, {Pattarakijwanich}, {Pei}, {Qi}, {Qi}, {Qiao}, {Qin}, {Ruffolo}, {S{\'a}iz}, {Semikoz}, {Shao}, {Shao}, {Shchegolev}, {Sheng}, {Shu}, {Song}, {Stenkin}, {Stepanov}, {Su}, {Sun}, {Sun}, {Sun}, {Tam}, {Tang}, {Tang}, {Tian}, {Wang}, {Wang}, {Wang}, {Wang}, {Wang}, {Wang}, {Wang}, {Wang}, {Wang}, {Wang}, {Wang}, {Wang}, {Wang}, {Wang}, {Wang}, {Wang}, {Wang}, {Wang}, {Wang}, {Wang}, {Wang}, {Wei}, {Wei}, {Wei}, {Wen}, {Wu}, {Wu}, {Wu}, {Wu}, {Wu}, {Xi}, {Xia}, {Xia}, {Xiang}, {Xiao}, {Xiao}, {Xin}, {Xin}, {Xing}, {Xiong}, {Xu}, {Xu}, {Xu}, {Xu}, {Xue}, {Yan}, {Yan}, {Yan}, {Yang}, {Yang}, {Yang}, {Yang}, {Yang}, {Yang}, {Yang}, {Yang}, {Yang}, {Yao}, {Yao}, {Ye}, {Yin}, {Yin}, {You}, {You}, {Yu}, {Yuan}, {Yue}, {Zeng}, {Zeng}, {Zeng}, {Zha}, {Zhang}, {Zhang}, {Zhang}, {Zhang}, {Zhang}, {Zhang}, {Zhang}, {Zhang}, {Zhang}, {Zhang}, {Zhang}, {Zhang}, {Zhang}, {Zhang}, {Zhang}, {Zhang}, {Zhang}, {Zhang}, {Zhao}, {Zhao}, {Zhao}, {Zhao}, {Zhao}, {Zheng}, {Zhou}, {Zhou}, {Zhou}, {Zhou}, {Zhou}, {Zhou}, {Zhou}, {Zhu}, {Zhu}, {Zhu}, {Zhu}, \& {Zuo}}]{LHAASO_221009_KM2}
{Cao}, Z., {Aharonian}, F., {An}, Q., {et~al.} 2023, Science Advances, 9, eadj2778, \dodoi{10.1126/sciadv.adj2778}

\bibitem[{{Chiang} \& {Dermer}(1999)}]{ChiangDermer99}
{Chiang}, J., \& {Dermer}, C.~D. 1999, \apj, 512, 699, \dodoi{10.1086/306789}

\bibitem[{{Das} \& {Razzaque}(2023)}]{Das_intergalCasc_221009_23}
{Das}, S., \& {Razzaque}, S. 2023, \aap, 670, L12, \dodoi{10.1051/0004-6361/202245377}

\bibitem[{{Derishev} \& {Piran}(2021)}]{DerishevPiran21}
{Derishev}, E., \& {Piran}, T. 2021, \apj, 923, 135, \dodoi{10.3847/1538-4357/ac2dec}

\bibitem[{{Derishev} \& {Piran}(2016)}]{Derishev_pairbalance16}
{Derishev}, E.~V., \& {Piran}, T. 2016, \mnras, 460, 2036, \dodoi{10.1093/mnras/stw1175}

\bibitem[{{Fan} {et~al.}(2008){Fan}, {Piran}, {Narayan}, \& {Wei}}]{FanPiran08}
{Fan}, Y.-Z., {Piran}, T., {Narayan}, R., \& {Wei}, D.-M. 2008, \mnras, 384, 1483, \dodoi{10.1111/j.1365-2966.2007.12765.x}

\bibitem[{{Fukushima} {et~al.}(2017){Fukushima}, {To}, {Asano}, \& {Fujita}}]{FukushimaToAsano17}
{Fukushima}, T., {To}, S., {Asano}, K., \& {Fujita}, Y. 2017, \apj, 844, 92, \dodoi{10.3847/1538-4357/aa7b83}

\bibitem[{{Gao} {et~al.}(2017){Gao}, {Pohl}, \& {Winter}}]{GaoEtAl_AM3_17}
{Gao}, S., {Pohl}, M., \& {Winter}, W. 2017, \apj, 843, 109, \dodoi{10.3847/1538-4357/aa7754}

\bibitem[{{Ghisellini} {et~al.}(2020){Ghisellini}, {Ghirlanda}, {Oganesyan}, {Ascenzi}, {Nava}, {Celotti}, {Salafia}, {Ravasio}, \& {Ronchi}}]{Ghisellini_ProtonSynPrompt20}
{Ghisellini}, G., {Ghirlanda}, G., {Oganesyan}, G., {et~al.} 2020, \aap, 636, A82, \dodoi{10.1051/0004-6361/201937244}

\bibitem[{{Giannios} \& {Spruit}(2005)}]{GianniosSpruit_prompt_pynting05}
{Giannios}, D., \& {Spruit}, H.~C. 2005, \aap, 430, 1, \dodoi{10.1051/0004-6361:20047033}

\bibitem[{{Gill} \& {Granot}(2023)}]{GillGranot}
{Gill}, R., \& {Granot}, J. 2023, \mnras, 524, L78, \dodoi{10.1093/mnrasl/slad075}

\bibitem[{{Gro{\v{s}}elj} {et~al.}(2022){Gro{\v{s}}elj}, {Sironi}, \& {Beloborodov}}]{GroseljEtAl2022}
{Gro{\v{s}}elj}, D., {Sironi}, L., \& {Beloborodov}, A.~M. 2022, \apj, 933, 74, \dodoi{10.3847/1538-4357/ac713e}

\bibitem[{{Guarini} {et~al.}(2023){Guarini}, {Tamborra}, {B{\'e}gu{\'e}}, \& {Rudolph}}]{Guarini_neutrinos23}
{Guarini}, E., {Tamborra}, I., {B{\'e}gu{\'e}}, D., \& {Rudolph}, A. 2023, \mnras, 523, 149, \dodoi{10.1093/mnras/stad1421}

\bibitem[{Guennebaud {et~al.}(2010)Guennebaud, Jacob, {et~al.}}]{eigenweb}
Guennebaud, G., Jacob, B., {et~al.} 2010, Eigen v3, http://eigen.tuxfamily.org

\bibitem[{{H.~E.~S.~S. Collaboration} {et~al.}(2017){H.~E.~S.~S. Collaboration}, {Abdalla}, {Abramowski}, {Aharonian}, {Ait Benkhali}, {Akhperjanian}, {Andersson}, {Ang{\"u}ner}, {Arakawa}, {Arrieta}, {Aubert}, {Backes}, {Balzer}, {Barnard}, {Becherini}, {Tjus}, {Berge}, {Bernhard}, {Bernl{\"o}hr}, {Blackwell}, {B{\"o}ttcher}, {Boisson}, {Bolmont}, {Bonnefoy}, {Bordas}, {Bregeon}, {Brun}, {Brun}, {Bryan}, {B{\"u}chele}, {Bulik}, {Capasso}, {Carr}, {Casanova}, {Cerruti}, {Chakraborty}, {Chaves}, {Chen}, {Chevalier}, {Coffaro}, {Colafrancesco}, {Cologna}, {Condon}, {Conrad}, {Cui}, {Davids}, {Decock}, {Degrange}, {Deil}, {Devin}, {de Wilt}, {Dirson}, {Djannati-Ata{\"\i}}, {Domainko}, {Donath}, {Drury}, {Dutson}, {Dyks}, {Edwards}, {Egberts}, {Eger}, {Ernenwein}, {Eschbach}, {Farnier}, {Fegan}, {Fernandes}, {Fiasson}, {Fontaine}, {F{\"o}rster}, {Funk}, {F{\"u}{\ss}ling}, {Gabici}, {Gallant}, {Garrigoux}, {Giavitto}, {Giebels}, {Glicenstein}, {Gottschall}, {Goyal}, {Grondin}, {Hahn}, {Haupt}, {Hawkes}, {Heinzelmann}, {Henri}, {Hermann}, {Hinton}, {Hofmann}, {Hoischen}, {Holch}, {Holler}, {Horns}, {Ivascenko}, {Iwasaki}, {Jacholkowska}, {Jamrozy}, {Janiak}, {Jankowsky}, {Jankowsky}, {Jingo}, {Jogler}, {Jouvin}, {Jung-Richardt}, {Kastendieck}, {Katarzy{\'n}ski}, {Katsuragawa}, {Katz}, {Kerszberg}, {Khangulyan}, {Kh{\'e}lifi}, {King}, {Klepser}, {Klochkov}, {Klu{\'z}niak}, {Kolitzus}, {Komin}, {Kosack}, {Krakau}, {Kraus}, {Kr{\"u}ger}, {Laffon}, {Lamanna}, {Lau}, {Lees}, {Lefaucheur}, {Lefranc}, {Lemi{\`e}re}, {Lemoine-Goumard}, {Lenain}, {Leser}, {Lohse}, {Lorentz}, {Liu}, {L{\'o}pez-Coto}, {Lypova}, {Marandon}, {Marcowith}, {Mariaud}, {Marx}, {Maurin}, {Maxted}, {Mayer}, {Meintjes}, {Meyer}, {Mitchell}, {Moderski}, {Mohamed}, {Mohrmann}, {Mor{\r{a}}}, {Moulin}, {Murach}, {Nakashima}, {de Naurois}, {Niederwanger}, {Niemiec}, {Oakes}, {O'Brien}, {Odaka}, {Ohm}, {Ostrowski}, {Oya}, {Padovani}, {Panter}, {Parsons}, {Pekeur}, {Pelletier}, {Perennes}, {Petrucci}, {Peyaud}, {Piel}, {Pita}, {Poon}, {Prokhorov}, {Prokoph}, {P{\"u}hlhofer}, {Punch}, {Quirrenbach}, {Raab}, {Rauth}, {Reimer}, {Reimer}, {Renaud}, {de los Reyes}, {Richter}, {Rieger}, {Romoli}, {Rowell}, {Rudak}, {Rulten}, {Sahakian}, {Saito}, {Salek}, {Sanchez}, {Santangelo}, {Sasaki}, {Schlickeiser}, {Sch{\"u}ssler}, {Schulz}, {Schwanke}, {Schwemmer}, {Seglar-Arroyo}, {Settimo}, {Seyffert}, {Shafi}, {Shilon}, {Simoni}, {Sol}, {Spanier}, {Spengler}, {Spies}, {Stawarz}, {Steenkamp}, {Stegmann}, {Stycz}, {Sushch}, {Takahashi}, {Tavernet}, {Tavernier}, {Taylor}, {Terrier}, {Tibaldo}, {Tiziani}, {Tluczykont}, {Trichard}, {Tsuji}, {Tuffs}, {Uchiyama}, {van der Walt}, {van Eldik}, {van Rensburg}, {van Soelen}, {Vasileiadis}, {Veh}, {Venter}, {Viana}, {Vincent}, {Vink}, {Voisin}, {V{\"o}lk}, {Vuillaume}, {Wadiasingh}, {Wagner}, {Wagner}, {Wagner}, {White}, {Wierzcholska}, {Willmann}, {W{\"o}rnlein}, {Wouters}, {Yang}, {Zaborov}, {Zacharias}, {Zanin}, {Zdziarski}, {Zech}, {Zefi}, {Ziegler}, \& {{\.Z}ywucka}}]{HESS_EBL17}
{H.~E.~S.~S. Collaboration}, {Abdalla}, H., {Abramowski}, A., {et~al.} 2017, \aap, 606, A59, \dodoi{10.1051/0004-6361/201731200}

\bibitem[{{H.~E.~S.~S. Collaboration} {et~al.}(2021){H.~E.~S.~S. Collaboration}, {Abdalla}, {Aharonian}, {Ait Benkhali}, {Ang{\"u}ner}, {Arcaro}, {Armand}, {Armstrong}, {Ashkar}, {Backes}, {Baghmanyan}, {Barbosa Martins}, {Barnacka}, {Barnard}, {Becherini}, {Berge}, {Bernl{\"o}hr}, {Bi}, {Bissaldi}, {B{\"o}ttcher}, {Boisson}, {Bolmont}, {de Bony de Lavergne}, {Breuhaus}, {Brun}, {Brun}, {Bryan}, {B{\"u}chele}, {Bulik}, {Bylund}, {Caroff}, {Carosi}, {Casanova}, {Chand}, {Chandra}, {Chen}, {Cotter}, {Cury{\l}o}, {Damascene Mbarubucyeye}, {Davids}, {Davies}, {Deil}, {Devin}, {Dirson}, {Djannati-Ata{\"\i}}, {Dmytriiev}, {Donath}, {Doroshenko}, {Dreyer}, {Duffy}, {Dyks}, {Egberts}, {Eichhorn}, {Einecke}, {Emery}, {Ernenwein}, {Feijen}, {Fegan}, {Fiasson}, {Fichet de Clairfontaine}, {Fontaine}, {Funk}, {F{\"u}{\ss}ling}, {Gabici}, {Gallant}, {Giavitto}, {Giunti}, {Glawion}, {Glicenstein}, {Grondin}, {Hahn}, {Haupt}, {Hermann}, {Hinton}, {Hofmann}, {Hoischen}, {Holch}, {Holler}, {H{\"o}rbe}, {Horns}, {Huber}, {Jamrozy}, {Jankowsky}, {Jankowsky}, {Jardin-Blicq}, {Joshi}, {Jung-Richardt}, {Kasai}, {Kastendieck}, {Katarzy{\'n}ski}, {Katz}, {Khangulyan}, {Kh{\'e}lifi}, {Klepser}, {Klu{\'z}niak}, {Komin}, {Konno}, {Kosack}, {Kostunin}, {Kreter}, {Lamanna}, {Lemi{\`e}re}, {Lemoine-Goumard}, {Lenain}, {Leuschner}, {Levy}, {Lohse}, {Lypova}, {Mackey}, {Majumdar}, {Malyshev}, {Malyshev}, {Marandon}, {Marchegiani}, {Marcowith}, {Mares}, {Mart{\'\i}-Devesa}, {Marx}, {Maurin}, {Meintjes}, {Meyer}, {Mitchell}, {Moderski}, {Mohrmann}, {Montanari}, {Moore}, {Morris}, {Moulin}, {Muller}, {Murach}, {Nakashima}, {Nayerhoda}, {de Naurois}, {Ndiyavala}, {Niemiec}, {Oakes}, {O'Brien}, {Odaka}, {Ohm}, {Olivera-Nieto}, {de Ona Wilhelmi}, {Ostrowski}, {Panny}, {Panter}, {Parsons}, {Peron}, {Peyaud}, {Piel}, {Pita}, {Poireau}, {Priyana Noel}, {Prokhorov}, {Prokoph}, {P{\"u}hlhofer}, {Punch}, {Quirrenbach}, {Raab}, {Rauth}, {Reichherzer}, {Reimer}, {Reimer}, {Remy}, {Renaud}, {Rieger}, {Rinchiuso}, {Romoli}, {Rowell}, {Rudak}, {Ruiz-Velasco}, {Sahakian}, {Sailer}, {Salzmann}, {Sanchez}, {Santangelo}, {Sasaki}, {Scalici}, {Sch{\"a}fer}, {Sch{\"u}ssler}, {Schutte}, {Schwanke}, {Seglar-Arroyo}, {Senniappan}, {Seyffert}, {Shafi}, {Shapopi}, {Shiningayamwe}, {Simoni}, {Sinha}, {Sol}, {Specovius}, {Spencer}, {Spir-Jacob}, {Stawarz}, {Sun}, {Steenkamp}, {Stegmann}, {Steinmassl}, {Steppa}, {Takahashi}, {Tam}, {Tavernier}, {Taylor}, {Terrier}, {Thiersen}, {Tiziani}, {Tluczykont}, {Tomankova}, {Tsirou}, {Tuffs}, {Uchiyama}, {van der Walt}, {van Eldik}, {van Rensburg}, {van Soelen}, {Vasileiadis}, {Veh}, {Venter}, {Vincent}, {Vink}, {V{\"o}lk}, {Wadiasingh}, {Wagner}, {Watson}, {Werner}, {White}, {Wierzcholska}, {Wong}, {Yusafzai}, {Zacharias}, {Zanin}, {Zargaryan}, {Zdziarski}, {Zech}, {Zhu}, {Zorn}, {Zouari}, {{\.Z}ywucka}, {Evans}, \& {Page}}]{HESS_190829}
{H.~E.~S.~S. Collaboration}, {Abdalla}, H., {Aharonian}, F., {et~al.} 2021, Science, 372, 1081, \dodoi{10.1126/science.abe8560}

\bibitem[{Harris {et~al.}(2020)Harris, Millman, van~der Walt, Gommers, Virtanen, Cournapeau, Wieser, Taylor, Berg, Smith, Kern, Picus, Hoyer, van Kerkwijk, Brett, Haldane, del R{\'{i}}o, Wiebe, Peterson, G{\'{e}}rard-Marchant, Sheppard, Reddy, Weckesser, Abbasi, Gohlke, \& Oliphant}]{NumPy_ref}
Harris, C.~R., Millman, K.~J., van~der Walt, S.~J., {et~al.} 2020, Nature, 585, 357, \dodoi{10.1038/s41586-020-2649-2}

\bibitem[{Heinze(2020)}]{PrinceAnalysisTools}
Heinze, J. 2020, PriNCe analysis tool

\bibitem[{{Heinze} {et~al.}(2020){Heinze}, {Biehl}, {Fedynitch}, {Boncioli}, {Rudolph}, \& {Winter}}]{2020MNRAS.498.5990H}
{Heinze}, J., {Biehl}, D., {Fedynitch}, A., {et~al.} 2020, \mnras, 498, 5990, \dodoi{10.1093/mnras/staa2751}

\bibitem[{{Heinze} {et~al.}(2019){Heinze}, {Fedynitch}, {Boncioli}, \& {Winter}}]{2019ApJ...873...88H}
{Heinze}, J., {Fedynitch}, A., {Boncioli}, D., \& {Winter}, W. 2019, \apj, 873, 88, \dodoi{10.3847/1538-4357/ab05ce}

\bibitem[{{H.E.S.S. Collaboration} {et~al.}(2019){H.E.S.S. Collaboration}, {Abdalla}, {Adam}, {Aharonian}, {Ait Benkhali}, {Ang{\"u}ner}, {et~al.}}]{HESS_180720B}
{H.E.S.S. Collaboration}, {Abdalla}, H., {Adam}, R., {et~al.} 2019, \nat, 575, 464, \dodoi{10.1038/s41586-019-1743-9}

\bibitem[{{Huang} {et~al.}(2022){Huang}, {Kirk}, {Giacinti}, \& {Reville}}]{HuangReville22}
{Huang}, Z.-Q., {Kirk}, J.~G., {Giacinti}, G., \& {Reville}, B. 2022, \apj, 925, 182, \dodoi{10.3847/1538-4357/ac3f38}

\bibitem[{Hunter(2007)}]{Matplotlib_ref}
Hunter, J.~D. 2007, Computing in Science \& Engineering, 9, 90, \dodoi{10.1109/MCSE.2007.55}

\bibitem[{{Isravel} {et~al.}(2023{\natexlab{a}}){Isravel}, {B{\'e}gu{\'e}}, \& {Pe'er}}]{Isravel_221009_psyn_23}
{Isravel}, H., {B{\'e}gu{\'e}}, D., \& {Pe'er}, A. 2023{\natexlab{a}}, \apj, 956, 12, \dodoi{10.3847/1538-4357/acefcd}

\bibitem[{{Isravel} {et~al.}(2023{\natexlab{b}}){Isravel}, {Pe'er}, \& {B{\'e}gu{\'e}}}]{Isravel_190114_psyn_23}
{Isravel}, H., {Pe'er}, A., \& {B{\'e}gu{\'e}}, D. 2023{\natexlab{b}}, \apj, 955, 70, \dodoi{10.3847/1538-4357/acec73}

\bibitem[{Jakob {et~al.}(2017)Jakob, Rhinelander, \& Moldovan}]{pybind11}
Jakob, W., Rhinelander, J., \& Moldovan, D. 2017, pybind11 -- Seamless operability between C++11 and Python

\bibitem[{{Khangulyan} {et~al.}(2021){Khangulyan}, {Aharonian}, {Romoli}, \& {Taylor}}]{KhangulyanEtAl_clumpyB_21}
{Khangulyan}, D., {Aharonian}, F., {Romoli}, C., \& {Taylor}, A. 2021, \apj, 914, 76, \dodoi{10.3847/1538-4357/abfcbf}

\bibitem[{{Khangulyan} {et~al.}(2023){Khangulyan}, {Taylor}, \& {Aharonian}}]{KhangulyanEtAl_2zoneSSC_23}
{Khangulyan}, D., {Taylor}, A.~M., \& {Aharonian}, F. 2023, \apj, 947, 87, \dodoi{10.3847/1538-4357/acc24e}

\bibitem[{{Klinger} {et~al.}(2023){Klinger}, {Tak}, {Taylor}, \& {Zhu}}]{KlingerEtAl_GRB190114C}
{Klinger}, M., {Tak}, D., {Taylor}, A.~M., \& {Zhu}, S.~J. 2023, \mnras, 520, 839, \dodoi{10.1093/mnras/stad142}

\bibitem[{{Klinger} {et~al.}(2024){Klinger}, {Taylor}, {Parsotan}, {Beardmore}, {Heinz}, \& {Zhu}}]{Klinger_221009_24}
{Klinger}, M., {Taylor}, A.~M., {Parsotan}, T., {et~al.} 2024, \mnras, 529, L47, \dodoi{10.1093/mnrasl/slad185}

\bibitem[{Klinger {et~al.}(2024)Klinger, Yuan, Taylor, \& Winter}]{zenodo_ref}
Klinger, M., Yuan, C., Taylor, A.~M., \& Winter, W. 2024, {Code from: Lepto-Hadronic Scenarios for TeV Extensions of Gamma-Ray Burst Afterglow Spectra}, v1.1.0,  Zenodo, \dodoi{10.5281/zenodo.13961284}

\bibitem[{{Klinger} {et~al.}(2024){Klinger}, {Rudolph}, {Rodrigues}, {Yuan}, {Fichet de Clairfontaine}, {Fedynitch}, {Winter}, {Pohl}, \& {Gao}}]{AM3_paper}
{Klinger}, M., {Rudolph}, A., {Rodrigues}, X., {et~al.} 2024, \apjs, 275, 4, \dodoi{10.3847/1538-4365/ad725c}

\bibitem[{{Kobayashi} {et~al.}(1997){Kobayashi}, {Piran}, \& {Sari}}]{Kobayashi_internalshocks97}
{Kobayashi}, S., {Piran}, T., \& {Sari}, R. 1997, \apj, 490, 92, \dodoi{10.1086/512791}

\bibitem[{{Kouveliotou} {et~al.}(2013){Kouveliotou}, {Granot}, {Racusin}, {Bellm}, {Vianello}, {Oates}, {Fryer}, {Boggs}, {Christensen}, {Craig}, {Dermer}, {Gehrels}, {Hailey}, {Harrison}, {Melandri}, {McEnery}, {Mundell}, {Stern}, {Tagliaferri}, \& {Zhang}}]{Kouveliotou_130427_13}
{Kouveliotou}, C., {Granot}, J., {Racusin}, J.~L., {et~al.} 2013, \apjl, 779, L1, \dodoi{10.1088/2041-8205/779/1/L1}

\bibitem[{{Lesage} {et~al.}(2023){Lesage}, {Veres}, {Briggs}, {Goldstein}, {Kocevski}, {Burns}, {Wilson-Hodge}, {Bhat}, {Huppenkothen}, {Fryer}, {Hamburg}, {Racusin}, {Bissaldi}, {Cleveland}, {Dalessi}, {Fletcher}, {Giles}, {Hristov}, {Hui}, {Mailyan}, {Malacaria}, {Poolakkil}, {Roberts}, {von Kienlin}, {Wood}, {Ajello}, {Arimoto}, {Baldini}, {Ballet}, {Baring}, {Bastieri}, {Gonzalez}, {Bellazzini}, {Bissaldi}, {Blandford}, {Bonino}, {Bruel}, {Buson}, {Cameron}, {Caputo}, {Caraveo}, {Cavazzuti}, {Chiaro}, {Cibrario}, {Ciprini}, {Orestano}, {Crnogorcevic}, {Cuoco}, {Cutini}, {D'Ammando}, {De Gaetano}, {Di Lalla}, {Di Venere}, {Dom{\'\i}nguez}, {Fegan}, {Ferrara}, {Fleischhack}, {Fukazawa}, {Funk}, {Fusco}, {Galanti}, {Gammaldi}, {Gargano}, {Gasbarra}, {Gasparrini}, {Germani}, {Giacchino}, {Giglietto}, {Gill}, {Giroletti}, {Granot}, {Green}, {Grenier}, {Guiriec}, {Gustafsson}, {Hays}, {Hewitt}, {Horan}, {Hou}, {Kuss}, {Latronico}, {Laviron}, {Lemoine-Goumard}, {Li}, {Liodakis}, {Longo}, {Loparco}, {Lorusso}, {Lovellette}, {Lubrano}, {Maldera}, {Manfreda}, {Mart{\'\i}-Devesa}, {Mazziotta}, {McEnery}, {Mereu}, {Meyer}, {Michelson}, {Mizuno}, {Monzani}, {Morselli}, {Moskalenko}, {Negro}, {Nuss}, {Omodei}, {Orlando}, {Ormes}, {Paneque}, {Panzarini}, {Persic}, {Pesce-Rollins}, {Pillera}, {Piron}, {Poon}, {Porter}, {Principe}, {Rain{\`o}}, {Rando}, {Rani}, {Razzano}, {Razzaque}, {Reimer}, {Reimer}, {Ryde}, {S{\'a}nchez-Conde}, {Parkinson}, {Scotton}, {Serini}, {Sgr{\`o}}, {Sharma}, {Siskind}, {Spandre}, {Spinelli}, {Tajima}, {Torres}, {Valverde}, {Venters}, {Wadiasingh}, {Wood}, \& {Zaharijas}}]{FermiGBM_grb221009a}
{Lesage}, S., {Veres}, P., {Briggs}, M.~S., {et~al.} 2023, \apjl, 952, L42, \dodoi{10.3847/2041-8213/ace5b4}

\bibitem[{{LHAASO Collaboration} {et~al.}(2023){LHAASO Collaboration}, {Cao}, {Aharonian}, {An}, {Axikegu}, {Bai}, {Bai}, {Bao}, {Bastieri}, {Bi}, {Bi}, {Cai}, {Cao}, {Cao}, {Cao}, {Chang}, {Chang}, {Chen}, {Chen}, {Chen}, {Chen}, {Chen}, {Chen}, {Chen}, {Chen}, {Chen}, {Chen}, {Chen}, {Cheng}, {Cheng}, {Cheng}, {Cui}, {Cui}, {Cui}, {Dai}, {Dai}, {Danzengluobu}, {Della Volpe}, {Dong}, {Duan}, {Fan}, {Fan}, {Fang}, {Fang}, {Feng}, {Feng}, {Feng}, {Feng}, {Feng}, {Gao}, {Gao}, {Gao}, {Gao}, {Gao}, {Gao}, {Ge}, {Geng}, {Gong}, {Gou}, {Gu}, {Guo}, {Guo}, {Guo}, {Guo}, {Han}, {He}, {He}, {He}, {He}, {He}, {Heller}, {Hor}, {Hou}, {Hou}, {Hou}, {Hu}, {Hu}, {Hu}, {Huang}, {Huang}, {Huang}, {Huang}, {Huang}, {Ji}, {Jia}, {Jia}, {Jiang}, {Jiang}, {Jiang}, {Jin}, {Kang}, {Ke}, {Kuleshov}, {Kurinov}, {Li}, {Li}, {Li}, {Li}, {Li}, {Li}, {Li}, {Li}, {Li}, {Li}, {Li}, {Li}, {Li}, {Li}, {Li}, {Li}, {Li}, {Li}, {Li}, {Liang}, {Liang}, {Lin}, {Liu}, {Liu}, {Liu}, {Liu}, {Liu}, {Liu}, {Liu}, {Liu}, {Liu}, {Liu}, {Liu}, {Liu}, {Liu}, {Liu}, {Liu}, {Liu}, {Long}, {Lu}, {Luo}, {Lv}, {Ma}, {Ma}, {Ma}, {Mao}, {Min}, {Mitthumsiri}, {Nan}, {Ou}, {Pang}, {Pattarakijwanich}, {Pei}, {Qi}, {Qi}, {Qiao}, {Qin}, {Ruffolo}, {Saiz}, {Shao}, {Shao}, {Shchegolev}, {Sheng}, {Song}, {Stenkin}, {Stepanov}, {Su}, {Sun}, {Sun}, {Sun}, {Tam}, {Tang}, {Tian}, {Wang}, {Wang}, {Wang}, {Wang}, {Wang}, {Wang}, {Wang}, {Wang}, {Wang}, {Wang}, {Wang}, {Wang}, {Wang}, {Wang}, {Wang}, {Wang}, {Wang}, {Wang}, {Wang}, {Wei}, {Wei}, {Wei}, {Wen}, {Wu}, {Wu}, {Wu}, {Wu}, {Wu}, {Xi}, {Xia}, {Xia}, {Xiang}, {Xiao}, {Xiao}, {Xin}, {Xin}, {Xing}, {Xiong}, {Xu}, {Xu}, {Xu}, {Xue}, {Yan}, {Yan}, {Yan}, {Yang}, {Yang}, {Yang}, {Yang}, {Yang}, {Yang}, {Yang}, {Yang}, {Yang}, {Yao}, {Ye}, {Yin}, {Yin}, {You}, {You}, {Yu}, {Yuan}, {Yue}, {Zeng}, {Zeng}, {Zeng}, {Zeng}, {Zhang}, {Zhang}, {Zhang}, {Zhang}, {Zhang}, {Zhang}, {Zhang}, {Zhang}, {Zhang}, {Zhang}, {Zhang}, {Zhang}, {Zhang}, {Zhang}, {Zhang}, {Zhang}, {Zhang}, {Zhang}, {Zhang}, {Zhao}, {Zhao}, {Zhao}, {Zhao}, {Zhao}, {Zheng}, {Zhou}, {Zhou}, {Zhou}, {Zhou}, {Zhou}, {Zhou}, {Zhu}, {Zhu}, {Zhu}, {Zhu}, \& {Zuo}}]{LHAASO_221009_WCDA}
{LHAASO Collaboration}, {Cao}, Z., {Aharonian}, F., {et~al.} 2023, Science, 380, 1390, \dodoi{10.1126/science.adg9328}

\bibitem[{{Liu} {et~al.}(2023){Liu}, {Zhang}, \& {Wang}}]{Liu_LAT_221009_23}
{Liu}, R.-Y., {Zhang}, H.-M., \& {Wang}, X.-Y. 2023, \apjl, 943, L2, \dodoi{10.3847/2041-8213/acaf5e}

\bibitem[{{MAGIC Collaboration} {et~al.}(2019{\natexlab{a}}){MAGIC Collaboration}, {Acciari}, {Ansoldi}, {Antonelli}, {Arbet Engels}, {Baack}, {Babi{\'c}}, {Banerjee}, {Barres de Almeida}, {Barrio}, {Becerra Gonz{\'a}lez}, {Bednarek}, {Bellizzi}, {Bernardini}, {Berti}, {Besenrieder}, {Bhattacharyya}, {Bigongiari}, {Biland}, {Blanch}, {Bonnoli}, {Bo{\v{s}}njak}, {Busetto}, {Carosi}, {Carosi}, {Ceribella}, {Chai}, {Chilingaryan}, {Cikota}, {Colak}, {Colin}, {Colombo}, {Contreras}, {Cortina}, {Covino}, {D'Amico}, {D'Elia}, {da Vela}, {Dazzi}, {de Angelis}, {de Lotto}, {Delfino}, {Delgado}, {Depaoli}, {di Pierro}, {di Venere}, {Do Souto Espi{\~n}eira}, {Dominis Prester}, {Donini}, {Dorner}, {Doro}, {Elsaesser}, {Fallah Ramazani}, {Fattorini}, {Fern{\'a}ndez-Barral}, {Ferrara}, {Fidalgo}, {Foffano}, {Fonseca}, {Font}, {Fruck}, {Fukami}, {Gallozzi}, {Garc{\'\i}a L{\'o}pez}, {Garczarczyk}, {Gasparyan}, {Gaug}, {Giglietto}, {Giordano}, {Godinovi{\'c}}, {Green}, {Guberman}, {Hadasch}, {Hahn}, {Herrera}, {Hoang}, {Hrupec}, {H{\"u}tten}, {Inada}, {Inoue}, {Ishio}, {Iwamura}, {Jouvin}, {Kerszberg}, {Kubo}, {Kushida}, {Lamastra}, {Lelas}, {Leone}, {Lindfors}, {Lombardi}, {Longo}, {L{\'o}pez}, {L{\'o}pez-Coto}, {L{\'o}pez-Oramas}, {Loporchio}, {Machado de Oliveira Fraga}, {Maggio}, {Majumdar}, {Makariev}, {Mallamaci}, {Maneva}, {Manganaro}, {Mannheim}, {Maraschi}, {Mariotti}, {Mart{\'\i}nez}, {Masuda}, {Mazin}, {Mi{\'c}anovi{\'c}}, {Miceli}, {Minev}, {Miranda}, {Mirzoyan}, {Molina}, {Moralejo}, {Morcuende}, {Moreno}, {Moretti}, {Munar-Adrover}, {Neustroev}, {Nigro}, {Nilsson}, {Ninci}, {Nishijima}, {Noda}, {Nogu{\'e}s}, {N{\"o}the}, {Nozaki}, {Paiano}, {Palacio}, {Palatiello}, {Paneque}, {Paoletti}, {Paredes}, {Pe{\~n}il}, {Peresano}, {Persic}, {Prada Moroni}, {Prandini}, {Puljak}, {Rhode}, {Rib{\'o}}, {Rico}, {Righi}, {Rugliancich}, {Saha}, {Sahakyan}, {Saito}, {Sakurai}, {Satalecka}, {Schmidt}, {Schweizer}, {Sitarek}, {{\v{S}}nidari{\'c}}, {Sobczynska}, {Somero}, {Stamerra}, {Strom}, {Strzys}, {Suda}, {Suri{\'c}}, {Takahashi}, {Tavecchio}, {Temnikov}, {Terzi{\'c}}, {Teshima}, {Torres-Alb{\`a}}, {Tosti}, {Tsujimoto}, {Vagelli}, {van Scherpenberg}, {Vanzo}, {Vazquez Acosta}, {Vigorito}, {Vitale}, {Vovk}, {Will}, {Zari{\'c}}, \& {Nava}}]{MAGIC_190114_data19}
{MAGIC Collaboration}, {Acciari}, V.~A., {Ansoldi}, S., {et~al.} 2019{\natexlab{a}}, \nat, 575, 455, \dodoi{10.1038/s41586-019-1750-x}

\bibitem[{{MAGIC Collaboration} {et~al.}(2019{\natexlab{b}}){MAGIC Collaboration}, {Acciari}, {Ansoldi}, {Antonelli}, {Engels}, {Baack}, {Babi{\'c}}, {Banerjee}, {Barres de Almeida}, {Barrio}, {Becerra Gonz{\'a}lez}, {Bednarek}, {Bellizzi}, {Bernardini}, {Berti}, {Besenrieder}, {Bhattacharyya}, {Bigongiari}, {Biland}, {Blanch}, {Bonnoli}, {Bo{\v{s}}njak}, {Busetto}, {Carosi}, {Ceribella}, {Chai}, {Chilingaryan}, {Cikota}, {Colak}, {Colin}, {Colombo}, {Contreras}, {Cortina}, {Covino}, {D'Elia}, {da Vela}, {Dazzi}, {de Angelis}, {de Lotto}, {Delfino}, {Delgado}, {Depaoli}, {di Pierro}, {di Venere}, {Do Souto Espi{\~n}eira}, {Dominis Prester}, {Donini}, {Dorner}, {Doro}, {Elsaesser}, {Fallah Ramazani}, {Fattorini}, {Ferrara}, {Fidalgo}, {Foffano}, {Fonseca}, {Font}, {Fruck}, {Fukami}, {Garc{\'\i}a L{\'o}pez}, {Garczarczyk}, {Gasparyan}, {Gaug}, {Giglietto}, {Giordano}, {Godinovi{\'c}}, {Green}, {Guberman}, {Hadasch}, {Hahn}, {Herrera}, {Hoang}, {Hrupec}, {H{\"u}tten}, {Inada}, {Inoue}, {Ishio}, {Iwamura}, {Jouvin}, {Kerszberg}, {Kubo}, {Kushida}, {Lamastra}, {Lelas}, {Leone}, {Lindfors}, {Lombardi}, {Longo}, {L{\'o}pez}, {L{\'o}pez-Coto}, {L{\'o}pez-Oramas}, {Loporchio}, {Machado de Oliveira Fraga}, {Maggio}, {Majumdar}, {Makariev}, {Mallamaci}, {Maneva}, {Manganaro}, {Mannheim}, {Maraschi}, {Mariotti}, {Mart{\'\i}nez}, {Mazin}, {Mi{\'c}anovi{\'c}}, {Miceli}, {Minev}, {Miranda}, {Mirzoyan}, {Molina}, {Moralejo}, {Morcuende}, {Moreno}, {Moretti}, {Munar-Adrover}, {Neustroev}, {Nigro}, {Nilsson}, {Ninci}, {Nishijima}, {Noda}, {Nogu{\'e}s}, {Nozaki}, {Paiano}, {Palatiello}, {Paneque}, {Paoletti}, {Paredes}, {Pe{\~n}il}, {Peresano}, {Persic}, {Moroni}, {Prandini}, {Puljak}, {Rhode}, {Rib{\'o}}, {Rico}, {Righi}, {Rugliancich}, {Saha}, {Sahakyan}, {Saito}, {Sakurai}, {Satalecka}, {Schmidt}, {Schweizer}, {Sitarek}, {{\v{S}}nidari{\'c}}, {Sobczynska}, {Somero}, {Stamerra}, {Strom}, {Strzys}, {Suda}, {Suri{\'c}}, {Takahashi}, {Tavecchio}, {Temnikov}, {Terzi{\'c}}, {Teshima}, {Torres-Alb{\`a}}, {Tosti}, {Vagelli}, {van Scherpenberg}, {Vanzo}, {Vazquez Acosta}, {Vigorito}, {Vitale}, {Vovk}, {Will}, {Zari{\'c}}, {Nava}, {Veres}, {Bhat}, {Briggs}, {Cleveland}, {Hamburg}, {Hui}, {Mailyan}, {Preece}, {Roberts}, {von Kienlin}, {Wilson-Hodge}, {Kocevski}, {Arimoto}, {Tak}, {Asano}, {Axelsson}, {Barbiellini}, {Bissaldi}, {Dirirsa}, {Gill}, {Granot}, {McEnery}, {Omodei}, {Razzaque}, {Piron}, {Racusin}, {Thompson}, {Campana}, {Bernardini}, {Kuin}, {Siegel}, {Cenko}, {O'Brien}, {Capalbi}, {Da{\i}}, {de Pasquale}, {Gropp}, {Klingler}, {Osborne}, {Perri}, {Starling}, {Tagliaferri}, {Tohuvavohu}, {Ursi}, {Tavani}, {Cardillo}, {Casentini}, {Piano}, {Evangelista}, {Verrecchia}, {Pittori}, {Lucarelli}, {Bulgarelli}, {Parmiggiani}, {Anderson}, {Anderson}, {Bernardi}, {Bolmer}, {Caballero-Garc{\'\i}a}, {Carrasco}, {Castell{\'o}n}, {Castro Segura}, {Castro-Tirado}, {Cherukuri}, {Cockeram}, {D'Avanzo}, {di Dato}, {Diretse}, {Fender}, {Fern{\'a}ndez-Garc{\'\i}a}, {Fynbo}, {Fruchter}, {Greiner}, {Gromadzki}, {Heintz}, {Heywood}, {van der Horst}, {Hu}, {Inserra}, {Izzo}, {Jaiswal}, {Jakobsson}, {Japelj}, {Kankare}, {Kann}, {Kouveliotou}, {Klose}, {Levan}, {Li}, {Lotti}, {Maguire}, {Malesani}, {Manulis}, {Marongiu}, {Martin}, {Melandri}, {Micha{\l}owski}, {Miller-Jones}, {Misra}, {Moin}, {Mooley}, {Nasri}, {Nicholl}, {Noschese}, {Novara}, {Pandey}, {Peretti}, {P{\'e}rez Del Pulgar}, {P{\'e}rez-Torres}, {Perley}, {Piro}, {Ragosta}, {Resmi}, {Ricci}, {Rossi}, {S{\'a}nchez-Ram{\'\i}rez}, {Selsing}, {Schulze}, {Smartt}, {Smith}, {Sokolov}, {Stevens}, {Tanvir}, {Th{\"o}ne}, {Tiengo}, {Tremou}, {Troja}, {de Ugarte Postigo}, {Valeev}, {Vergani}, {Wieringa}, {Woudt}, {Xu}, {Yaron}, \& {Young}}]{MAGIC_190114C_MWL19}
---. 2019{\natexlab{b}}, \nat, 575, 459, \dodoi{10.1038/s41586-019-1754-6}

\bibitem[{{Marcowith} {et~al.}(2016){Marcowith}, {Bret}, {Bykov}, {Dieckman}, {O'C Drury}, {Lemb{\`e}ge}, {Lemoine}, {Morlino}, {Murphy}, {Pelletier}, {Plotnikov}, {Reville}, {Riquelme}, {Sironi}, \& {Stockem Novo}}]{MarcowithEtAl2016}
{Marcowith}, A., {Bret}, A., {Bykov}, A., {et~al.} 2016, Reports on Progress in Physics, 79, 046901, \dodoi{10.1088/0034-4885/79/4/046901}

\bibitem[{{M{\'e}sz{\'a}ros}(2006)}]{2006RPPh...69.2259M}
{M{\'e}sz{\'a}ros}, P. 2006, Reports on Progress in Physics, 69, 2259, \dodoi{10.1088/0034-4885/69/8/R01}

\bibitem[{{Meszaros} {et~al.}(1994){Meszaros}, {Rees}, \& {Papathanassiou}}]{MeszarosReesPapathanassiou_94}
{Meszaros}, P., {Rees}, M.~J., \& {Papathanassiou}, H. 1994, \apj, 432, 181, \dodoi{10.1086/174559}

\bibitem[{{Misra} {et~al.}(2021){Misra}, {Resmi}, {Kann}, {Marongiu}, {Moin}, {Klose}, {Bernardi}, {de Ugarte Postigo}, {Jaiswal}, {Schulze}, {Perley}, {Ghosh}, {Dimple}, {Kumar}, {Gupta}, {Micha{\l}owski}, {Mart{\'\i}n}, {Cockeram}, {Cherukuri}, {Bhalerao}, {Anderson}, {Pandey}, {Anupama}, {Th{\"o}ne}, {Barway}, {Wieringa}, {Fynbo}, \& {Habeeb}}]{MisraEtAl2021}
{Misra}, K., {Resmi}, L., {Kann}, D.~A., {et~al.} 2021, \mnras, 504, 5685, \dodoi{10.1093/mnras/stab1050}

\bibitem[{{Mochkovitch} \& {Daigne}(1998)}]{MochkovitchDaigne98}
{Mochkovitch}, R., \& {Daigne}, F. 1998, in American Institute of Physics Conference Series, Vol. 428, Gamma-Ray Bursts, 4th Hunstville Symposium, ed. C.~A. {Meegan}, R.~D. {Preece}, \& T.~M. {Koshut}, 667--671, \dodoi{10.1063/1.55396}

\bibitem[{{Murase} {et~al.}(2018){Murase}, {Toomey}, {Fang}, {Oikonomou}, {Kimura}, {Hotokezaka}, {Kashiyama}, {Ioka}, \& {M{\'e}sz{\'a}ros}}]{2018ApJ...854...60M}
{Murase}, K., {Toomey}, M.~W., {Fang}, K., {et~al.} 2018, \apj, 854, 60, \dodoi{10.3847/1538-4357/aaa48a}

\bibitem[{{O'Connor} {et~al.}(2023){O'Connor}, {Troja}, {Ryan}, {Beniamini}, {van Eerten}, {Granot}, {Dichiara}, {Ricci}, {Lipunov}, {Gillanders}, {Gill}, {Moss}, {Anand}, {Andreoni}, {Becerra}, {Buckley}, {Butler}, {Cenko}, {Chasovnikov}, {Durbak}, {Francile}, {Hammerstein}, {van der Horst}, {Kasliwal}, {Kouveliotou}, {Kutyrev}, {Lee}, {Srinivasaragavan}, {Topolev}, {Watson}, {Yang}, \& {Zhirkov}}]{OConnor_structuredjet_221009_23}
{O'Connor}, B., {Troja}, E., {Ryan}, G., {et~al.} 2023, Science Advances, 9, eadi1405, \dodoi{10.1126/sciadv.adi1405}

\bibitem[{{Pe'er} \& {Ryde}(2018)}]{PeerRyde18}
{Pe'er}, A., \& {Ryde}, F. 2018, in Fourteenth Marcel Grossmann Meeting - MG14, ed. M.~{Bianchi}, R.~T. {Jansen}, \& R.~{Ruffini}, 806--840, \dodoi{10.1142/9789813226609_0044}

\bibitem[{{Pe'er} \& {Waxman}(2005)}]{PeerWaxman05}
{Pe'er}, A., \& {Waxman}, E. 2005, \apj, 628, 857, \dodoi{10.1086/431139}

\bibitem[{{Pei}(1992)}]{Pei92}
{Pei}, Y.~C. 1992, \apj, 395, 130, \dodoi{10.1086/171637}

\bibitem[{{Pennanen} {et~al.}(2014){Pennanen}, {Vurm}, \& {Poutanen}}]{PennanenVurmPoutanen14}
{Pennanen}, T., {Vurm}, I., \& {Poutanen}, J. 2014, \aap, 564, A77, \dodoi{10.1051/0004-6361/201322520}

\bibitem[{{Petropoulou} \& {Mastichiadis}(2009)}]{PetropoulouMastichiadis09}
{Petropoulou}, M., \& {Mastichiadis}, A. 2009, \aap, 507, 599, \dodoi{10.1051/0004-6361/200912970}

\bibitem[{{Piran}(2004)}]{Piran_review04}
{Piran}, T. 2004, Reviews of Modern Physics, 76, 1143, \dodoi{10.1103/RevModPhys.76.1143}

\bibitem[{{Planck Collaboration} {et~al.}(2020){Planck Collaboration}, {Aghanim}, {Akrami}, {Ashdown}, {Aumont}, {Baccigalupi}, {Ballardini}, {Banday}, {Barreiro}, {Bartolo}, {Basak}, {Battye}, {Benabed}, {Bernard}, {Bersanelli}, {Bielewicz}, {Bock}, {Bond}, {Borrill}, {Bouchet}, {Boulanger}, {Bucher}, {Burigana}, {Butler}, {Calabrese}, {Cardoso}, {Carron}, {Challinor}, {Chiang}, {Chluba}, {Colombo}, {Combet}, {Contreras}, {Crill}, {Cuttaia}, {de Bernardis}, {de Zotti}, {Delabrouille}, {Delouis}, {Di Valentino}, {Diego}, {Dor{\'e}}, {Douspis}, {Ducout}, {Dupac}, {Dusini}, {Efstathiou}, {Elsner}, {En{\ss}lin}, {Eriksen}, {Fantaye}, {Farhang}, {Fergusson}, {Fernandez-Cobos}, {Finelli}, {Forastieri}, {Frailis}, {Fraisse}, {Franceschi}, {Frolov}, {Galeotta}, {Galli}, {Ganga}, {G{\'e}nova-Santos}, {Gerbino}, {Ghosh}, {Gonz{\'a}lez-Nuevo}, {G{\'o}rski}, {Gratton}, {Gruppuso}, {Gudmundsson}, {Hamann}, {Handley}, {Hansen}, {Herranz}, {Hildebrandt}, {Hivon}, {Huang}, {Jaffe}, {Jones}, {Karakci}, {Keih{\"a}nen}, {Keskitalo}, {Kiiveri}, {Kim}, {Kisner}, {Knox}, {Krachmalnicoff}, {Kunz}, {Kurki-Suonio}, {Lagache}, {Lamarre}, {Lasenby}, {Lattanzi}, {Lawrence}, {Le Jeune}, {Lemos}, {Lesgourgues}, {Levrier}, {Lewis}, {Liguori}, {Lilje}, {Lilley}, {Lindholm}, {L{\'o}pez-Caniego}, {Lubin}, {Ma}, {Mac{\'\i}as-P{\'e}rez}, {Maggio}, {Maino}, {Mandolesi}, {Mangilli}, {Marcos-Caballero}, {Maris}, {Martin}, {Martinelli}, {Mart{\'\i}nez-Gonz{\'a}lez}, {Matarrese}, {Mauri}, {McEwen}, {Meinhold}, {Melchiorri}, {Mennella}, {Migliaccio}, {Millea}, {Mitra}, {Miville-Desch{\^e}nes}, {Molinari}, {Montier}, {Morgante}, {Moss}, {Natoli}, {N{\o}rgaard-Nielsen}, {Pagano}, {Paoletti}, {Partridge}, {Patanchon}, {Peiris}, {Perrotta}, {Pettorino}, {Piacentini}, {Polastri}, {Polenta}, {Puget}, {Rachen}, {Reinecke}, {Remazeilles}, {Renzi}, {Rocha}, {Rosset}, {Roudier}, {Rubi{\~n}o-Mart{\'\i}n}, {Ruiz-Granados}, {Salvati}, {Sandri}, {Savelainen}, {Scott}, {Shellard}, {Sirignano}, {Sirri}, {Spencer}, {Sunyaev}, {Suur-Uski}, {Tauber}, {Tavagnacco}, {Tenti}, {Toffolatti}, {Tomasi}, {Trombetti}, {Valenziano}, {Valiviita}, {Van Tent}, {Vibert}, {Vielva}, {Villa}, {Vittorio}, {Wandelt}, {Wehus}, {White}, {White}, {Zacchei}, \& {Zonca}}]{Planck18}
{Planck Collaboration}, {Aghanim}, N., {Akrami}, Y., {et~al.} 2020, \aap, 641, A6, \dodoi{10.1051/0004-6361/201833910}

\bibitem[{{Ressler} \& {Laskar}(2017)}]{ResslerLaskar2017}
{Ressler}, S.~M., \& {Laskar}, T. 2017, \apj, 845, 150, \dodoi{10.3847/1538-4357/aa8268}

\bibitem[{{Rodrigues} {et~al.}(2021){Rodrigues}, {Garrappa}, {Gao}, {Paliya}, {Franckowiak}, \& {Winter}}]{2021ApJ...912...54R}
{Rodrigues}, X., {Garrappa}, S., {Gao}, S., {et~al.} 2021, \apj, 912, 54, \dodoi{10.3847/1538-4357/abe87b}

\bibitem[{{Rudolph} {et~al.}(2023){Rudolph}, {Petropoulou}, {Winter}, \& {Bo{\v{s}}njak}}]{2023ApJ...944L..34R}
{Rudolph}, A., {Petropoulou}, M., {Winter}, W., \& {Bo{\v{s}}njak}, {\v{Z}}. 2023, \apjl, 944, L34, \dodoi{10.3847/2041-8213/acb6d7}

\bibitem[{{Ryan} {et~al.}(2020){Ryan}, {van Eerten}, {Piro}, \& {Troja}}]{Ryan_Afterglowpy20}
{Ryan}, G., {van Eerten}, H., {Piro}, L., \& {Troja}, E. 2020, \apj, 896, 166, \dodoi{10.3847/1538-4357/ab93cf}

\bibitem[{{Sari} \& {Esin}(2001)}]{SariEsin01}
{Sari}, R., \& {Esin}, A.~A. 2001, \apj, 548, 787, \dodoi{10.1086/319003}

\bibitem[{{Sari} {et~al.}(1996){Sari}, {Narayan}, \& {Piran}}]{SarietAl96}
{Sari}, R., {Narayan}, R., \& {Piran}, T. 1996, \apj, 473, 204, \dodoi{10.1086/178136}

\bibitem[{{Sato} {et~al.}(2023){Sato}, {Murase}, {Ohira}, \& {Yamazaki}}]{SatoEtAl}
{Sato}, Y., {Murase}, K., {Ohira}, Y., \& {Yamazaki}, R. 2023, \mnras, 522, L56, \dodoi{10.1093/mnrasl/slad038}

\bibitem[{{Senno} {et~al.}(2015){Senno}, {M{\'e}sz{\'a}ros}, {Murase}, {Baerwald}, \& {Rees}}]{2015ApJ...806...24S}
{Senno}, N., {M{\'e}sz{\'a}ros}, P., {Murase}, K., {Baerwald}, P., \& {Rees}, M.~J. 2015, \apj, 806, 24, \dodoi{10.1088/0004-637X/806/1/24}

\bibitem[{{Tavani} {et~al.}(2023){Tavani}, {Piano}, {Bulgarelli}, {Foffano}, {Ursi}, {Verrecchia}, {Pittori}, {Casentini}, {Giuliani}, {Longo}, {Panebianco}, {Di Piano}, {Baroncelli}, {Fioretti}, {Parmiggiani}, {Argan}, {Trois}, {Vercellone}, {Cardillo}, {Antonelli}, {Barbiellini}, {Caraveo}, {Cattaneo}, {Chen}, {Costa}, {Del Monte}, {Di Cocco}, {Donnarumma}, {Evangelista}, {Feroci}, {Gianotti}, {Labanti}, {Lazzarotto}, {Lipari}, {Lucarelli}, {Marisaldi}, {Mereghetti}, {Morselli}, {Pacciani}, {Pellizzoni}, {Perotti}, {Picozza}, {Pilia}, {Rapisarda}, {Rappoldi}, {Rubini}, {Soffitta}, {Trifoglio}, {Vittorini}, \& {D'Amico}}]{Tavani_AGILE_221009_23}
{Tavani}, M., {Piano}, G., {Bulgarelli}, A., {et~al.} 2023, \apjl, 956, L23, \dodoi{10.3847/2041-8213/acfaff}

\bibitem[{Tol(2021)}]{colorSchemes}
Tol, P. 2021, {INTRODUCTION TO COLOUR SCHEMES}, \url{https://personal.sron.nl/~pault/}, {Accessed: 2024-03-12}

\bibitem[{{van der Horst} {et~al.}(2014){van der Horst}, {Paragi}, {de Bruyn}, {Granot}, {Kouveliotou}, {Wiersema}, {Starling}, {Curran}, {Wijers}, {Rowlinson}, {Anderson}, {Fender}, {Yang}, \& {Strom}}]{VDHorstEtAl2014}
{van der Horst}, A.~J., {Paragi}, Z., {de Bruyn}, A.~G., {et~al.} 2014, \mnras, 444, 3151, \dodoi{10.1093/mnras/stu1664}

\bibitem[{{van Eerten} {et~al.}(2010){van Eerten}, {Zhang}, \& {MacFadyen}}]{VanEertenZhangMacFadyen10}
{van Eerten}, H., {Zhang}, W., \& {MacFadyen}, A. 2010, \apj, 722, 235, \dodoi{10.1088/0004-637X/722/1/235}

\bibitem[{{Vanthieghem} {et~al.}(2020){Vanthieghem}, {Lemoine}, {Plotnikov}, {Grassi}, {Grech}, {Gremillet}, \& {Pelletier}}]{Vanthieghem_Micro_20}
{Vanthieghem}, A., {Lemoine}, M., {Plotnikov}, I., {et~al.} 2020, Galaxies, 8, 33, \dodoi{10.3390/galaxies8020033}

\bibitem[{{Venters} {et~al.}(2020){Venters}, {Reno}, {Krizmanic}, {Anchordoqui}, {Gu{\'e}pin}, \& {Olinto}}]{2020PhRvD.102l3013V}
{Venters}, T.~M., {Reno}, M.~H., {Krizmanic}, J.~F., {et~al.} 2020, \prd, 102, 123013, \dodoi{10.1103/PhysRevD.102.123013}

\bibitem[{{Vietri}(1995)}]{Vietri95}
{Vietri}, M. 1995, \apj, 453, 883, \dodoi{10.1086/176448}

\bibitem[{{Wang} {et~al.}(2023){Wang}, {Tang}, {Zhang}, {Zheng}, {Xiong}, {Ren}, \& {Zhang}}]{Wang_pgammacasc_23}
{Wang}, K., {Tang}, Q.-W., {Zhang}, Y.-Q., {et~al.} 2023, arXiv e-prints, arXiv:2310.11821, \dodoi{10.48550/arXiv.2310.11821}

\bibitem[{{Warren} {et~al.}(2018){Warren}, {Barkov}, {Ito}, {Nagataki}, \& {Laskar}}]{WarrenEtAl2018}
{Warren}, D.~C., {Barkov}, M.~V., {Ito}, H., {Nagataki}, S., \& {Laskar}, T. 2018, \mnras, 480, 4060, \dodoi{10.1093/mnras/sty2138}

\bibitem[{{Warren} {et~al.}(2022){Warren}, {Dainotti}, {Barkov}, {Ahlgren}, {Ito}, \& {Nagataki}}]{WarrenEtAl_SSC22}
{Warren}, D.~C., {Dainotti}, M., {Barkov}, M.~V., {et~al.} 2022, \apj, 924, 40, \dodoi{10.3847/1538-4357/ac2f43}

\bibitem[{{Warren} {et~al.}(2015){Warren}, {Ellison}, {Bykov}, \& {Lee}}]{WarrenEtAl2015}
{Warren}, D.~C., {Ellison}, D.~C., {Bykov}, A.~M., \& {Lee}, S.-H. 2015, \mnras, 452, 431, \dodoi{10.1093/mnras/stv1304}

\bibitem[{{Waxman}(1995)}]{Waxman_GRB_CR_95}
{Waxman}, E. 1995, \prl, 75, 386, \dodoi{10.1103/PhysRevLett.75.386}

\bibitem[{{Willingale} {et~al.}(2013){Willingale}, {Starling}, {Beardmore}, {Tanvir}, \& {O'Brien}}]{Willingale_abs13}
{Willingale}, R., {Starling}, R.~L.~C., {Beardmore}, A.~P., {Tanvir}, N.~R., \& {O'Brien}, P.~T. 2013, \mnras, 431, 394, \dodoi{10.1093/mnras/stt175}

\bibitem[{{Wilms} {et~al.}(2000){Wilms}, {Allen}, \& {McCray}}]{Wilms2000}
{Wilms}, J., {Allen}, A., \& {McCray}, R. 2000, \apj, 542, 914, \dodoi{10.1086/317016}

\bibitem[{{Yuan} {et~al.}(2018){Yuan}, {M{\'e}sz{\'a}ros}, {Murase}, \& {Jeong}}]{2018ApJ...857...50Y}
{Yuan}, C., {M{\'e}sz{\'a}ros}, P., {Murase}, K., \& {Jeong}, D. 2018, \apj, 857, 50, \dodoi{10.3847/1538-4357/aab774}

\bibitem[{{Yuan} {et~al.}(2022){Yuan}, {Murase}, {Guetta}, {Pe'er}, {Bartos}, \& {M{\'e}sz{\'a}ros}}]{2022ApJ...932...80Y}
{Yuan}, C., {Murase}, K., {Guetta}, D., {et~al.} 2022, \apj, 932, 80, \dodoi{10.3847/1538-4357/ac6ddf}

\bibitem[{{Zhang}(2018)}]{Zhang_GRBbook}
{Zhang}, B. 2018, {The Physics of Gamma-Ray Bursts}, \dodoi{10.1017/9781139226530}

\bibitem[{{Zhang} \& {Yan}(2011)}]{Zhang_ICMART11}
{Zhang}, B., \& {Yan}, H. 2011, \apj, 726, 90, \dodoi{10.1088/0004-637X/726/2/90}

\bibitem[{{Zhang} {et~al.}(2023{\natexlab{a}}){Zhang}, {Murase}, {Ioka}, {Song}, {Yuan}, \& {M{\'e}sz{\'a}ros}}]{Zhang_reverseShockproton_23}
{Zhang}, B.~T., {Murase}, K., {Ioka}, K., {et~al.} 2023{\natexlab{a}}, \apjl, 947, L14, \dodoi{10.3847/2041-8213/acc79f}

\bibitem[{{Zhang} {et~al.}(2021){Zhang}, {Murase}, {Yuan}, {Kimura}, \& {M{\'e}sz{\'a}ros}}]{2021ApJ...908L..36Z}
{Zhang}, B.~T., {Murase}, K., {Yuan}, C., {Kimura}, S.~S., \& {M{\'e}sz{\'a}ros}, P. 2021, \apjl, 908, L36, \dodoi{10.3847/2041-8213/abe0b0}

\bibitem[{{Zhang} {et~al.}(2020){Zhang}, {Christie}, {Petropoulou}, {Rueda-Becerril}, \& {Giannios}}]{ZhangPetropoulou_190114C_ecternalIC20}
{Zhang}, H., {Christie}, I.~M., {Petropoulou}, M., {Rueda-Becerril}, J.~M., \& {Giannios}, D. 2020, \mnras, 496, 974, \dodoi{10.1093/mnras/staa1583}

\bibitem[{{Zhang} {et~al.}(2023{\natexlab{b}}){Zhang}, {Huang}, {Liu}, \& {Wang}}]{Zhang_GBM_221009_23}
{Zhang}, H.-M., {Huang}, Y.-Y., {Liu}, R.-Y., \& {Wang}, X.-Y. 2023{\natexlab{b}}, \apjl, 956, L21, \dodoi{10.3847/2041-8213/acfcab}

\end{thebibliography}
\bibliographystyle{aasjournal}


\end{CJK*}
\end{document}